\documentclass[12pt,a4paper]{article}
\pdfoutput=1

\usepackage{amsmath,amssymb}
\usepackage[nosort]{cite}
\usepackage[hyperref,bulletsep]{collect}
\usepackage[%
  colorlinks=false,%
  pdfpagelabels,%
  bookmarks=true,%
  bookmarksnumbered=true,%
  plainpages=false,%
  citebordercolor={.7 .7 1},%
  linkbordercolor={.7 .7 1},%
  pdfborder={0 0 1 [3]},%
  pdfstartview=FitH,%
  pdfpagemode=UseOutlines,%
  ]{hyperref}
\def\gfxon{\usepackage[final]{graphicx}}

\gfxon

\ifx\hypersetup\sadfkjashdfkxja\else
\hypersetup{pdftitle={Quantum Stability for the Heisenberg Ferromagnet}}
\hypersetup{pdfauthor={Till Bargheer, Niklas Beisert, Nikolay Gromov}}
\hypersetup{pdfkeywords={Heisenberg Ferromagnet, Spectral Curve, Stability, Elliptic Solution, Two-Cut Solution}}
\fi

\makeatletter \@addtoreset{equation}{section} \makeatother

\allowdisplaybreaks[3]

\makeatletter
\let\old@startsection=\@startsection
\let\oldl@section=\l@section
\renewcommand{\@startsection}[6]{\old@startsection{#1}{#2}{#3}{#4}{#5}{#6\mathversion{bold}}}
\renewcommand{\l@section}[2]{%
\vspace{-0.5em}%
\oldl@section{\mathversion{bold}#1}{#2}}
\makeatother

\makeatletter
\let\old@makecaption=\@makecaption
\def\@makecaption{\small\old@makecaption}
\makeatother

\usepackage{subfig}
\let\oldPhi=\Phi
\let\oldPsi=\Psi
\let\oldGamma=\Gamma
\let\oldDelta=\Delta
\let\oldLambda=\Lambda
\let\oldSigma=\Sigma
\let\oldTheta=\Theta
\let\oldPi=\Pi
\renewcommand{\Phi}{\mathnormal{\oldPhi}}
\renewcommand{\Psi}{\mathnormal{\oldPsi}}
\renewcommand{\Gamma}{\mathnormal{\oldGamma}}
\renewcommand{\Lambda}{\mathnormal{\oldLambda}}
\renewcommand{\Sigma}{\mathnormal{\oldSigma}}
\renewcommand{\Delta}{\mathnormal{\oldDelta}}
\renewcommand{\Theta}{\mathnormal{\oldTheta}}
\renewcommand{\Pi}{\mathnormal{\oldPi}}

\newcommand{\superN}{\mathcal{N}}

\newcommand{\arsinh}{\mathop{\mathrm{arsinh}}}
\newcommand{\artanh}{\mathop{\mathrm{artanh}}}
\newcommand{\arcoth}{\mathop{\mathrm{arcoth}}}

\newcommand{\res}{\mathop{\mathrm{res}}\nolimits}
\renewcommand{\Re}{\mathop{\mathrm{Re}}}
\renewcommand{\Im}{\mathop{\mathrm{Im}}}

\newcommand{\order}[1]{\mathcal{O}(#1)}

\newcommand{\Integers}{\mathbb{Z}}
\newcommand{\Reals}{\mathbb{R}}
\newcommand{\Complex}{\mathbb{C}}

\newcommand{\ud}{\mathrm{d}}
\newcommand{\contour}[1]{\mathcal{#1}}

\makeatletter
\newcommand{\ellSN}{\mathop{\operator@font sn}\nolimits}
\newcommand{\ellCN}{\mathop{\operator@font cn}\nolimits}
\newcommand{\ellDN}{\mathop{\operator@font dn}\nolimits}
\newcommand{\ellAM}{\mathop{\operator@font am}\nolimits}
\newcommand{\ellK}{\mathop{\smash{\operator@font K}\vphantom{a}}\nolimits}
\newcommand{\ellF}{\mathop{\smash{\operator@font F}\vphantom{a}}\nolimits}
\newcommand{\ellE}{\mathop{\smash{\operator@font E}\vphantom{a}}\nolimits}
\newcommand{\ellPi}{\mathop{\smash{\operator@font \oldPi}\vphantom{a}}\nolimits}
\newcommand{\elltPi}{\mathop{\smash{\operator@font \tilde{\oldPi}}\vphantom{a}}\nolimits}
\makeatother

\ifx\genfrac\sdflkaj

\else

\fi
\newcommand{\sfrac}[2]{{\textstyle\frac{#1}{#2}}}
\newcommand{\half}{\sfrac{1}{2}}
\newcommand{\ihalf}{\sfrac{i}{2}}
\newcommand{\quarter}{\sfrac{1}{4}}

\newcommand{\indup}[1]{_{\mathrm{#1}}}

\newcommand{\alg}[1]{\mathfrak{#1}}
\newcommand{\grp}[1]{\mathrm{#1}}

\newcommand{\lrbrk}[1]{\left(#1\right)}
\newcommand{\bigbrk}[1]{\bigl(#1\bigr)}

\newcommand{\bigabs}[1]{\bigl|#1\bigr|}

\newcommand{\set}[1]{\{#1\}}

\newcommand{\nn}{\nonumber}
\newcommand{\nln}{\nonumber\\}
\newcommand{\nl}[1][0pt]{\nonumber\\[#1]&\hspace{-4\arraycolsep}&\mathord{}}

\newcommand{\earel}[1]{\mathrel{}&\hspace{-2\arraycolsep}#1\hspace{-2\arraycolsep}&\mathrel{}}
\newcommand{\eq}{\earel{=}}

\def\[{\begin{equation}}
\def\]{\end{equation}}
\def\<{\begin{eqnarray}}
\def\>{\end{eqnarray}}

\makeatletter
\def\mr@ignsp#1 {\ifx\:#1\@empty\else #1\expandafter\mr@ignsp\fi}%
\newcommand{\multiref}[1]{\begingroup
\xdef\mr@no@sparg{\expandafter\mr@ignsp#1 \: }%
\def\mr@comma{}%
\@for\mr@refs:=\mr@no@sparg\do{\mr@comma\def\mr@comma{,}\ref{\mr@refs}}%
\endgroup}
\makeatother

\newcommand{\hypref}[2]{\ifx\href\asklfhas #2\else\href{#1}{#2}\fi}

\newcommand{\secref}[1]{Sec.~\multiref{#1}}

\newcommand{\appref}[1]{App.~\multiref{#1}}

\newcommand{\tabref}[1]{Tab.~\multiref{#1}}

\newcommand{\figref}[1]{Fig.~\multiref{#1}}
\renewcommand{\eqref}[1]{(\multiref{#1})}

\ifx\href\asklfhas\newcommand{\href}[2]{#2}\fi
\newcommand{\arxivno}[1]{\href{http://arxiv.org/abs/#1}{#1}}

\begin{document}

\iftrue

\pagenumbering{roman}
\thispagestyle{empty}
\begin{flushright}\footnotesize
\texttt{\arxivno{arxiv:0804.0324}}\\
\texttt{AEI-2008-010}\\
\texttt{LPTENS 08/19}\\
\texttt{SPhT-t08/051}\\
\end{flushright}
\vspace{0cm}

\begin{center}%
{\Large\textbf{\mathversion{bold}%
Quantum Stability for\\the Heisenberg Ferromagnet}\par}\vspace{1cm}%

\textsc{Till Bargheer$^{\dagger}$, Niklas Beisert$^{\dagger}$
and
Nikolay Gromov$^{\ddag\P\S}$} \vspace{8mm}

\textit{$^{\dag}$Max-Planck-Institut f\"ur Gravitationsphysik\\%
Albert-Einstein-Institut\\%
Am M\"uhlenberg 1, 14476 Potsdam, Germany}%
\vspace{3mm}%

\textit{$^{\ddag}$Service de Physique Th\'eorique, CNRS-URA 2306\\ C.E.A.-Saclay,\\%
91191 Gif-sur-Yvette, France}%
\vspace{3mm}%

\textit{$^{\P}$Laboratoire de Physique Th\'eorique
 de l'Ecole Normale Sup\'erieure\\
24 rue Lhomond, Paris 75231, France}%
\vspace{3mm}%

\textit{$^{\S}$St.\ Petersburg INP\\
Gatchina, 188 300, St.\ Petersburg, Russia}%
\vspace{3mm}%

\texttt{bargheer,nbeisert@aei.mpg.de\quad gromov@thd.pnpi.spb.ru}
\par\vspace{1cm}

\textbf{Abstract}\vspace{7mm}

\begin{minipage}{12.7cm}
Highly spinning classical strings on $\Reals\times S^3$ are 
described by the Landau--Lifshitz model or equivalently by the 
Heisenberg ferromagnet in the thermodynamic limit. The spectrum of
this model can be given in terms of spectral curves. However, it is a
priori not clear whether any given admissible spectral curve can
actually be realised as a solution to the discrete Bethe equations, a
property which can be referred to as \emph{stability}. In order to study the
issue of stability, we find and explore the general two-cut solution
or elliptic curve. It turns out that the moduli space of this elliptic
curve shows a surprisingly rich structure. We present the various
cases with illustrations and thus gain some insight into the features
of multi-cut solutions. It appears that all admissible spectral curves
are indeed stable if the branch cuts are positioned in a suitable,
non-trivial fashion.
\end{minipage}

\end{center}

\tableofcontents
\newpage

\newpage
\setcounter{page}{1}
\pagenumbering{arabic}
\renewcommand{\thefootnote}{\arabic{footnote}}
\setcounter{footnote}{0}

\fi

\section{Introduction}

The Heisenberg magnet \cite{Heisenberg:1928aa} is one of the very
first quantum mechanical models, and it serves as the prototypical
spin chain. Although it was set up almost 80 years ago it still
remains a fascinating subject with many features left to be understood.
Only three years after the discovery of the model and owing to its integrability,
Bethe was able to write a set of equations
\[\label{eq:xxxbethe}
\begin{array}[b]{c}\displaystyle
\lrbrk{\frac{u_k+\ihalf}{u_k-\ihalf}}^L=
\mathop{\prod_{j=1}}_{j\neq k}^{M}\frac{u_k-u_j+i}{u_k-u_j-i}
\quad\mbox{for }k=1,\ldots,M,
\\[28pt]\displaystyle
e^{iP}=\prod_{j=1}^M\frac{u_j+\ihalf}{u_j-\ihalf}\,,\qquad
E=\sum_{j=1}^M\lrbrk{\frac{i}{u_j+\ihalf}-\frac{i}{u_j-\ihalf}},
\end{array}
\]
which determine the complete spectrum \cite{Bethe:1931hc}. The terms
``Bethe equations'' or ``Bethe ansatz'' later became synonymous for the exact
solution for generic integrable spin chain models.

Although the Bethe equations describe the complete and exact spectrum,
it is virtually impossible (and perhaps not very enlightening) to
solve them concretely for generic states somewhere in the middle of
the spectrum. Nevertheless some corners of the spectrum are accessible
(and interesting). In particular these are the low-energy and
high-energy states for very long chains (the thermodynamic limit),
where the Bethe equations are approximated by integral equations. Most
studies have focused on the low-energy spectrum of the antiferromagnet
and this regime is well-understood, see \cite{Korepin:1993aa} for a review.
For example, the antiferromagnetic state is a solution to an integral
equation \cite{Hulthen:1938aa} and its excitation quanta are called
spinons \cite{Takhtajan:1982zz,Faddeev:1981ip}.

\begin{figure}
\centering
  \subfloat[]{\includegraphics{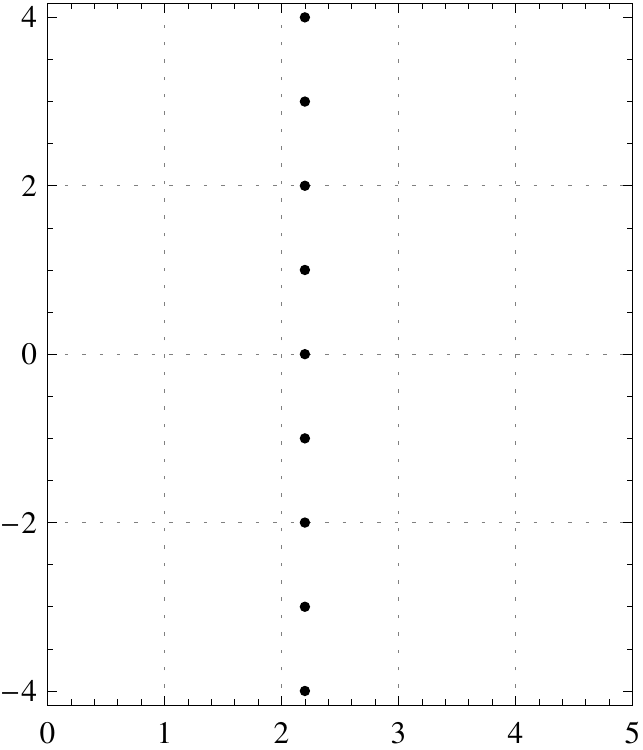}\label{fig:bethestring}}\qquad
  \subfloat[]{\includegraphics{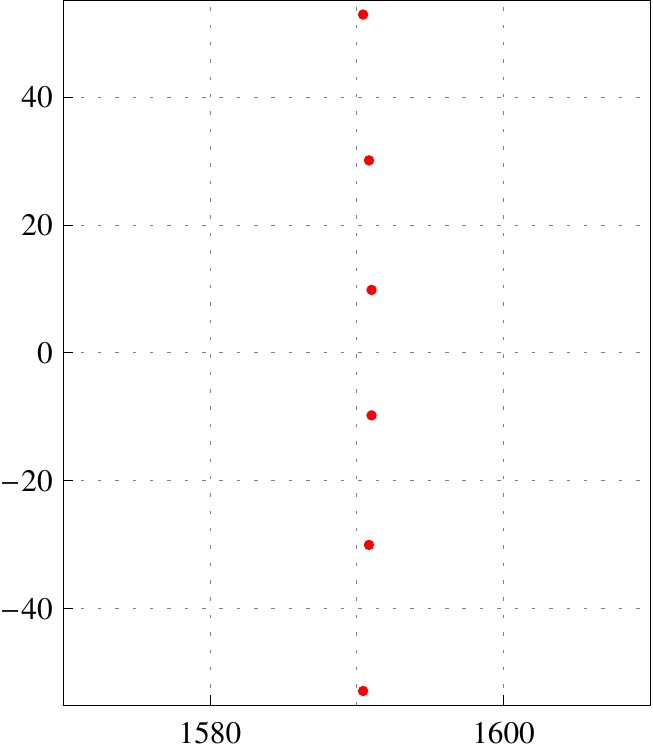}\label{fig:shortstring}}
\caption{A Bethe string \protect\subref{fig:bethestring} and a short
string \protect\subref{fig:shortstring}. Bethe strings have a strictly
regular pattern and only exist on infinitely long chains. For short
strings, the separation of adjacent roots is of order
$\order{\sqrt{L}}$; the string shown here lives on a chain of length
$L=10,000$.}
\label{fig:rootstrings}
\end{figure}

Conversely, the low-energy regime of the ferromagnet (or equivalently
the high-energy regime of the antiferromagnet) is much less explored.
The ferromagnetic ground state coincides with the vacuum of the Bethe ansatz,
it is trivial. The excitation quanta are called magnons, and each magnon
corresponds to a single Bethe root $u$ describing the rapidity of the
magnon. Magnons can form bound states which are usually called Bethe
strings. In a $k$-string centred at rapidity $u$ there are $k$ Bethe
roots $u_{1,\ldots,k}$ arranged in a regular pattern
$u_j=u+\ihalf(k+1-2j)$, i.e.\ the separation of adjacent constituent
magnon rapidities is $i$, see \figref{fig:bethestring}.
Of course this distribution pattern of Bethe roots is not exact when
the length $L$ of the chain is finite and in fact large deviations are
observed. Nevertheless the string hypothesis can be used to perform a
counting of all states which gives the expected exact result even at
finite $L$ \cite{Bethe:1931hc}
(see also \cite{Baxter:2001sx} for a recent account including references).

\begin{figure}
\centering
  \subfloat[$M/L=30/200$]{\includegraphics{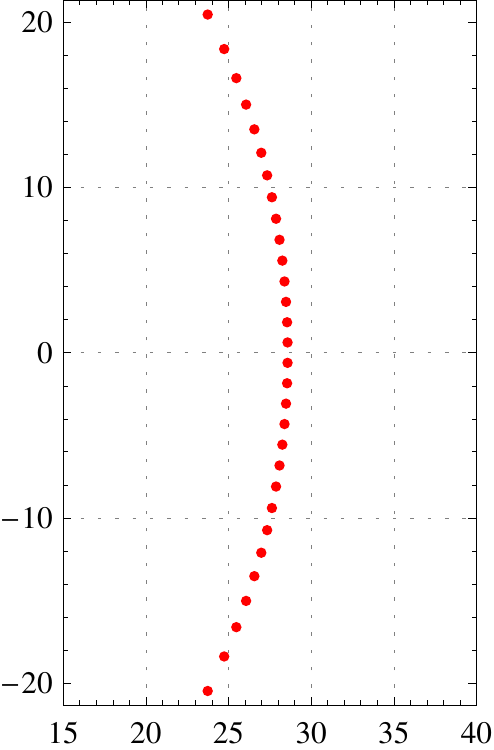}}\hfill
  \subfloat[$M/L=40/162$]{\includegraphics{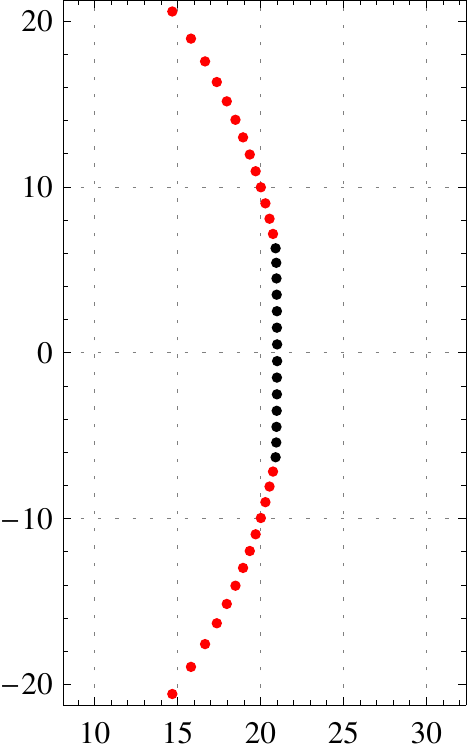}\label{fig:longstringb}}\hfill
  \subfloat[$M/L=40/115$]{\includegraphics{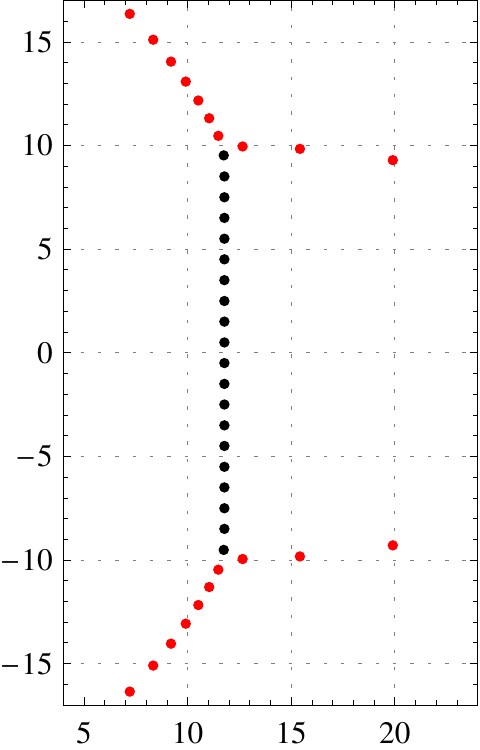}}
\caption{Some long strings with mode number $n=1$ and various fillings $\alpha=M/L$
($L=\text{length}$, $M=\text{number of roots}$).}
\label{fig:longstring}
\end{figure}

An interesting type of ``Bethe string'' which has not found much
attention until recently is one where the rapidity $u$ scales as the
length $L$ of the chain. Here one distinguishes between short and long
strings where the number of constituent magnons is either of
$\order{1}$ or of $\order{L}$. For these strings the regular pattern
is violated strongly, the distance between adjacent Bethe roots
deviates much from $i$. Long strings were first investigated by
Sutherland in \cite{Sutherland:1995aa}. This particular string consists of a
condensate core and two tails, see \figref{fig:longstringb}.
Similar long strings were later considered in \cite{Beisert:2003xu}.
The distance of Bethe roots in the core is very close
to $i$ while in the curved tails the density of roots decreases to $0$
at the ends.
Short strings were considered in \cite{Dhar:2000aa,Minahan:2002ve}.
Unlike for standard strings the distance of Bethe roots is of $\order{\sqrt{L}}$,
see \figref{fig:shortstring}. Short strings can as well be considered as
very short versions of long strings \cite{Beisert:2003ea}.
The distinction of the two types
is nevertheless useful because of their different energy scales in the
limit of large $L$. Furthermore, long strings can be considered as
smooth classical objects for which the discrete magnon constituents
play a minor role. Short strings on the other hand are quantum objects
and the number of constituent magnons must be a positive integer. In
fact a short string is better viewed as a coherent state of bosonic
magnons. In the strict thermodynamic limit the magnons do not interact
with each other, their rapidities are merely influenced by the long strings
which do interact non-trivially among themselves.

Interest in the ferromagnetic regime was recently sparked by the
AdS/CFT correspondence. Minahan and Zarembo showed the equivalence \cite{Minahan:2002ve}
of the spectrum of planar one-loop anomalous dimensions in the
$\alg{su}(2)$ sector of $\superN=4$ supersymmetric gauge theory to the
spectrum of the Heisenberg magnet with zero
overall momentum. Consequently it was observed that the energies of
certain long ultra-relativistic string configurations in
$\Reals\times S^3$ \cite{Frolov:2003qc,Frolov:2003xy}
agree precisely with the energies of
certain long strings of the Heisenberg ferromagnet in the thermodynamic
limit \cite{Beisert:2003xu}, 
see \cite{Tseytlin:2003ii,Plefka:2005bk} for reviews. 
This extended the earlier observation of
\cite{Berenstein:2002jq} that the spectrum of magnons or short string
excitations above the ferromagnetic vacuum agrees with the spectrum of
quantum excitations of a short ultra-relativistic string orbiting the
equator of $S^3$. Some time later Kruczenski related both the
ultra-relativistic limit of strings on $\Reals\times S^3$ and the
thermodynamic limit of the Heisenberg ferromagnet to one and the same
Landau--Lifshitz model on $S^2$ \cite{Kruczenski:2003gt}, see also
\cite{Fradkin:1991aa}.

The general solution for this classical model was constructed in
\cite{Bikbaev:1983aa,Kazakov:2004qf} in terms of spectral curves,
see \cite{Dubrovin:1976aa}.
It was furthermore shown that the moduli space of spectral curves of
any given genus has the expected dimension. The features of the
spectral curves are in one-to-one correspondence to the integral
equations obtained from the above thermodynamic limit of the Bethe equations.
It is therefore clear that
any state of the Heisenberg ferromagnet in the thermodynamic limit is
approximated by a spectral curve. Less obvious is the question if
every admissible spectral curve also has a corresponding solution to
the Bethe equations, and if not, what are the stability criteria? In
\cite{Beisert:2005mq} one such criterion was derived from the
requirement that the mode number of a long string can be defined
unambiguously and self-consistently.
The statement is that the directed density $\rho$ of Bethe roots
should not encircle the points $in$, $n\in\Integers\setminus\set{0}$
when moving along the contour of a string.
This is for example achieved if the density
is bounded by $|\rho(x)|<1$ or $|\Im \rho(x)|<1$,
i.e.\ the distribution of Bethe roots should not be denser
than the ideal Bethe string.
In particular we expect $|\rho(x)|<1$ where the cut crosses the real
axis.
It is in fact a natural bound for the quantum model:
The values $\rho=in$ (along the imaginary axis) are distinguished
because the pairwise interactions of Bethe roots in
\eqref{eq:xxxbethe} become singular at that point. For the strictly classical
model on the other hand the quantity $\rho$ does not have a meaning,
and solutions with density bounded by one are not at all different
from solutions where the density exceeds one. A notion of stability
exists in the semiclassical theory \cite{Frolov:2003tu};
it demands the absence of tachyonic fluctuations. This condition does not
coincide with $|\rho|\leq 1$, it turns out that the latter is a
stronger requirement in the case of a single cut \cite{Beisert:2005bv}.

So the question how to properly define stability remains. In this
paper we would like to gain further insight into this issue by
studying the general one- and two-cut solutions. The moduli space of
two-cut solutions has two discrete parameters and two continuous ones,
and it is sufficiently rich to investigate stability. In particular,
we want to understand the role of condensate cores in the context of
stability. In order to compare our analysis of the one- and two-cut
spectral curves to actual solutions of the discrete Bethe equations
\eqref{eq:xxxbethe}, we further develop a numerical method for the
construction of such solutions with large numbers of Bethe roots. Up
to now, only few solutions that can be compared to corresponding
spectral curves have been constructed \cite{Beisert:2003xu}, and they
only have a comparatively small number of Bethe roots. The numerical
solutions with finite numbers of Bethe roots represent quantum states
on a chain of finite length and hence can also be used to examine
finite-size corrections to the thermodynamic limit. For one-cut
configurations these corrections were studied analytically in
\cite{Beisert:2005mq,Hernandez:2005nf,Beisert:2005bv}, some more
configurations of roots were analyzed analytically and numerically in
\cite{Gromov:2007ky}. In this work we
examine numerically the leading order finite-size effects explicitly for
configurations beyond the one-cut case.

This paper is organised as follows. We start in
\secref{sec:spectral_curves} with a review of the general solution in
the thermodynamic limit by means of spectral curves. In order to
illustrate our procedure on a simple example we will first reconsider
and discuss the general one-cut solution in \secref{sec:onecut}.
The corresponding construction within the Landau--Lifshitz model
is given in \appref{sec:ll-model}.
\secref{sec:twocut} contains the derivation of the general two-cut
solution and its properties. In \secref{sec:consecutive} we will
continue by applying the two-cut solution to study the issue of
stability. To that end we have to determine the physical shape of the
branch cuts in various regions of the parameter space. Finally in
\secref{sec:numerics} we test our predictions against
solutions of the Bethe equations with large but finite length,
constructed numerically, in order to substantiate our claims. We
conclude in \secref{sec:conc}.

\section{Spectral Curves for the Heisenberg Ferromagnet}
\label{sec:spectral_curves}

We start by reviewing the spectral curves
which describe solutions of the Bethe equations
\eqref{eq:xxxbethe} in the thermodynamic limit
\cite{Kazakov:2004qf}.

\subsection{Baxter Equation}

To derive the properties of the spectral curves
it seems convenient to consider the Baxter equation
as advertised in \cite{Gromov:2005gp}
\[
T(u)Q(u)=
(u+\ihalf)^L Q(u-i)
+
(u-\ihalf)^L Q(u+i).
\]
This equation is fully equivalent to the Bethe equations
in the following way:
Let $Q$ and $T$ be polynomials of degree $M$ and $L$, respectively.
Then the set of solutions to the Bethe equation
is equivalent to the set of Baxter equations.
The Bethe roots $u_k$ are simply the roots
of the Baxter Q-function $Q(u)=\prod_{k=1}^M(u-u_k)$.
Furthermore the transfer matrix eigenvalue $T(u)$
encodes the momentum eigenvalue $P$ and the energy eigenvalue $E$
in two equivalent ways as follows
\<\label{eq:momeng}
\frac{T(u+\ihalf)}{(u+i)^L}\eq
\exp\bigbrk{+iP+iuE+\order{u^2}}
+\order{u^L},
\nln
\frac{T(u-\ihalf)}{(u-i)^L}\eq
\exp\bigbrk{-iP-iuE+\order{u^2}}
+\order{u^L}.
\>
In terms of the Q-function the momentum and energy read (unless $L<2$)
\[\label{eq:momeng2}
e^{iP}=\frac{Q(-\ihalf)}{Q(+\ihalf)}\,,
\qquad
E=\frac{iQ'(+\ihalf)}{Q(+\ihalf)}-\frac{iQ'(-\ihalf)}{Q(-\ihalf)}\,.
\]
Note that for gauge theory and AdS/CFT the states are required
to be cyclic, i.e.\ the net momentum must be zero, $e^{iP}=1$.
Here we will however not restrict to cyclic states but consider
states with arbitrary momentum.

Finally, the number of magnons $M$ can be read off from the
transfer matrix eigenvalue expanded around $u=\infty$
\[\label{eq:magnonnum}
\frac{T(u)}{u^L}=2+\bigbrk{\sfrac{3}{4}L-\quarter(L-2M)(L-2M+2)}\frac{1}{u^{2}}+\order{1/u^3}.
\]
Note that $\quarter(L-2M)(L-2M+2)$
is the eigenvalue of the quadratic Casimir
for a representation with spin $\half L-M$.%
\footnote{Note the ambiguity $M\leftrightarrow L+1-M$
which is related to the existence of
mirror solutions of the Bethe/Baxter equations with $M>\half L$.
These solutions have zero norm and thus are unphysical.
Let us here restrict to physical states with $M\leq \half L$.}

\subsection{Thermodynamic Limit}

Before we take the thermodynamic limit, we make the following
substitutions:
Define
\[\label{eq:qmom}
\exp\bigbrk{ip(x)}=\exp(i/2x)\frac{Q(xL-\ihalf)}{Q(xL+\ihalf)}
\qquad\mbox{and}\qquad
t(x)=\frac{T(xL)}{(xL)^L}\,,
\]
where $p(x)$ is called the quasi-momentum
and $t(x)$ is the properly rescaled
transfer matrix eigenvalue.
It is clear that the function $p(x)$ has logarithmic singularities with
opposite prefactors at $x=(u_k\pm\ihalf)/L$
and a pole with residue $i/2$ at $x=0$.
Furthermore, we fix the ambiguity of $p(x)$ by shifts of $2\pi$
by setting $p(\infty)=0$.
Finally, $t(x)$ is a polynomial in $1/x$ of degree $L$,
i.e.\ it has an $L$-fold pole at $x=0$
and is analytic everywhere else.
In these variables the Baxter equation reads
\<
t(x)\eq
(1+i/2Lx)^L \exp\lrbrk{-\frac{i/2}{x-i/2L}}\exp\bigbrk{+ip(x-i/2L)}
\nl+
(1-i/2Lx)^L \exp\lrbrk{+\frac{i/2}{x+i/2L}}\exp\bigbrk{-ip(x+i/2L)}.
\>

It is now easy to take the thermodynamic limit $L\to\infty$.
In this limit the magnon number is assumed to scale like $M=L\alpha$
with fixed total filling $\alpha$.
The magnon rapidities scale like $u_k\sim L$
and are assumed to distribute smoothly along certain contours
in the complex plane with distance $\order{L^0}$.
The distance between adjacent magnon rapidities
defines the density $\rho(x)$ of Bethe roots
\[\label{eq:density0}
\rho(x)\approx \frac{1}{u_{k+1}-u_k}\,\qquad
\mbox{for }
u_k\approx u_{k+1}\approx xL.
\]
Note that the contours typically
arrange vertically in the complex plane
(but not necessarily strictly along the imaginary direction).
Therefore the density function $\rho(x)$ is in general complex and
its phase is inversely related to the
direction of the contour at the point $x$.
For definiteness, let us decide that cuts
generally go upwards in the complex plane,
$\Im u_{k+1}> \Im u_k$.
That means that the density will typically have
a negative imaginary part, $\Im \rho(x)< 0$.

The Baxter equation in the thermodynamic limit becomes simply
\[\label{eq:baxtercont}
t(x)=2\cos\bigbrk{p(x)}.
\]
The logarithmic singularities in $p(x)$ move together
to form linear discontinuities.
These discontinuities are interpreted as branch cuts
connecting various Riemann sheets of $p(x)$.
Apart from these and the pole
at $x=0$ with residue $i/2$ the quasi-momentum $p(x)$
is analytic everywhere.
The function $t(x)$ is also analytic
except for an exponential singularity at $x=0$
originating from the pole of degree $L\to\infty$.
Solutions $p(x),t(x)$ to the equation \eqref{eq:baxtercont}
with the above properties describe the spectrum in the thermodynamic limit.
Let us now study the properties of the solutions in more detail.

\subsection{Branch Cuts}
\label{sec:branch_cuts}

The function $p(x)$ has discontinuities along certain branch cuts
whose union shall be denoted by $\mathcal{C}$.

\paragraph{Density.}

The discontinuity of a cut
is proportional to the density $\rho(x)$ of Bethe roots \eqref{eq:density0}
in the vicinity of the point $x$
\[\label{eq:density}
\rho (x) = \frac{1}{2\pi i}\bigbrk{p(x-\varepsilon)-p(x+\varepsilon)} ,
\qquad x\in \mathcal{C}.
\]
We shall take $\varepsilon$ to be an infinitesimally small positive number,
i.e.\ $p(x+\varepsilon)$ denotes the limiting value of $p$ at $x\in \mathcal{C}$
towards the right of the cut.
Note that the combination $\ud x\,\rho(x)$ must be real and positive
which determines the direction of physical branch cuts.
The integrated density along a connected cut $\mathcal{C}_k$ will be denoted by
the filling $\alpha_k$
\[\label{eq:partfill}
\alpha_k=\int_{\mathcal{C}_k}\ud x\,\rho(x).
\]
%

\paragraph{Standard Cut.}

The function $t(x)$ must remain analytic across a branch cut of $p(x)$.
Equation \eqref{eq:baxtercont} tells us that $p(x)$ can change
sign and shift by a multiple of $2\pi$ without causing
a discontinuity in $t(x)$.

If the sign changes across a branch cut
we thus have
\[\label{eq:sqrtcut}
p(x+\epsilon)+p(x-\epsilon)=2\pi n_k\qquad
\mbox{for }x\in\mathcal{C}_k,
\]
where $n_k$ is the (constant) mode number associated to
the connected branch cut $\mathcal{C}_k$.
Using \eqref{eq:density} we can relate the density to the quasi-momentum
\[
\rho (x)
= \frac{1}{\pi i}
\bigbrk{p(x-\varepsilon) - \pi n_k }
= \frac{1}{\pi i}
\bigbrk{\pi n_k-p(x+\varepsilon) }.
\label{eq:rho}
\]
This type of cut can end in a square-root singularity $x_\ast$ of $p(x)$
where it takes the value $p(x_\ast)=\pi n_k$
\[\label{eq:sqrtsing}
p(x_\ast+\epsilon)=\pi n_k+\order{\sqrt{\epsilon}}\qquad
\mbox{for }x_\ast\in\delta\mathcal{C}_k.
\]
Note that this singularity of $p(x)$ is indeed compatible with
analyticity of $t(x)$ at the singularity, cf.\ \eqref{eq:baxtercont}.
At $x_\ast$ the branch cut can be oriented in three different
directions: Consider the radial coordinates $x=x_\ast+re^{i\varphi}$.
The combination $\ud x \rho(x)\sim e^{3i\varphi/2} \sqrt{r}\ud r$ must
be real and therefore the three possible orientations are separated by
$240^\circ$ (where the full rotation is by $720^\circ$ due to the
square root singularity).

\paragraph{Condensate Cut.}

\begin{figure}[tb]
\centering
  \includegraphics{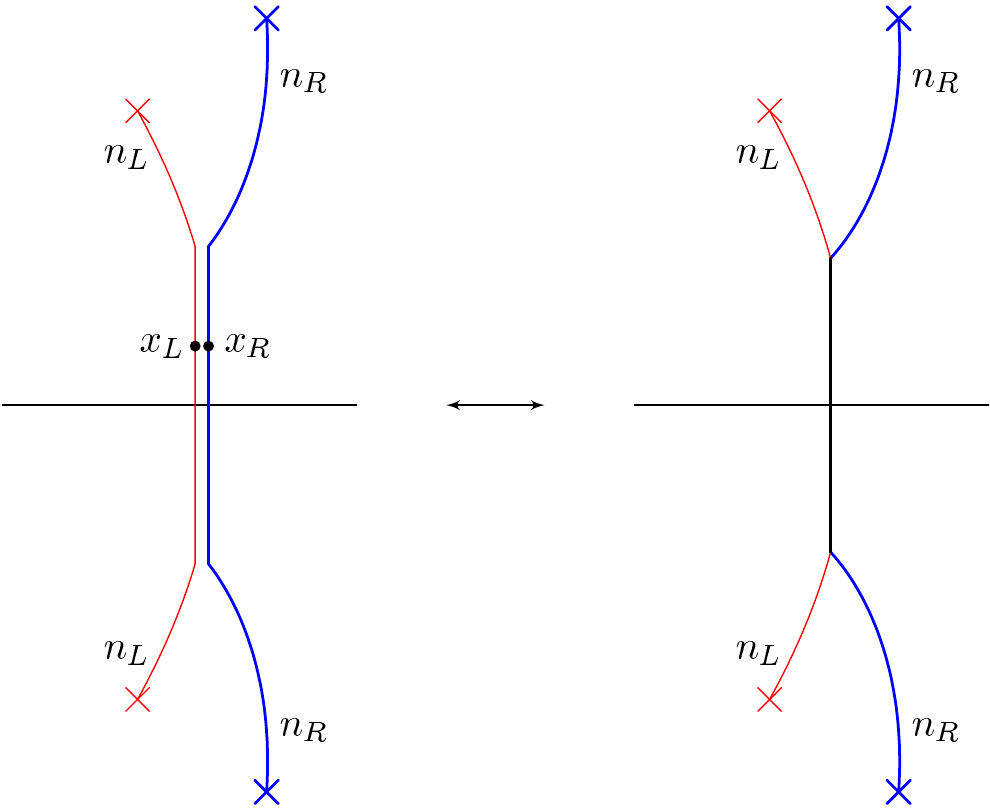}
  \caption[Two standard cuts form a condensate cut]{Two standard
  cuts $\contour{C}_L$ and $\contour{C}_R$ with mode numbers $n_L$
  and $n_R$ whose physical contours (as determined by the condition
  that $\rho(x)\, \ud x$ be real) meet at a common point. The
  standard cut contours start and end at square-root singularities,
  which are indicated by crosses. Beyond their common point, the cut
  contours can be placed on top of each other -- the result is a
  condensate cut as shown in the right picture. The density
  $\rho\indup c$ on the condensate cut can be determined by placing
  the individual contours infinitesimally close and parallel to each
  other, as shown in the left picture. It turns out that the density
  on the condensate cut is constant and purely imaginary:
  $\rho\indup c(x)=i(n_R-n_L)$, see \eqref{eq:logcut}. Consequently,
  the contour of the condensate cut must be vertical.}
\label{fig:condensate_cut}
\end{figure}

Conversely, if the sign does not change across a branch cut, the
quasi-momentum must obey
\[\label{eq:logcut}
p(x-\epsilon)-p(x+\epsilon)=2\pi n'_k\qquad
\mbox{for }x\in\mathcal{C}'_k.
\]
This type of cut would be required to end on a logarithmic singularity
which is not compatible with analyticity of $t(x)$ in \eqref{eq:baxtercont}.
Nevertheless such a can exist if it ends on other cuts:
Indeed, we can view a logarithmic cut \eqref{eq:logcut}
as the union of two parallel standard cuts \eqref{eq:sqrtcut}.
Therefore a logarithmic cut $\mathcal{C}'$ can split up into two
parallel cuts $\mathcal{C}\indup{L}$ (left) and $\mathcal{C}\indup{R}$ (right)
at some point, see \figref{fig:condensate_cut}.
Evidence for this splitting is provided by some numerical solutions to the Bethe equations
for small $L$ \cite{Dhar:2000aa,Beisert:2003xu}.
Compatibility of \eqref{eq:logcut} with \eqref{eq:sqrtcut} requires
$n'=n\indup{L}-n\indup{R}$.
The integer $n'_k$ determines the density
of Bethe roots \eqref{eq:density} to $\rho(x)=-in'_k$ and thus a
logarithmic cut has constant integral density and extends strictly
along the imaginary direction.

\subsection{Observables}

According to \eqref{eq:momeng2}
the momentum $P$ and energy $\tilde E=EL$ appear in the
expansion of the quasi-momentum at $x=0$
\[
p(x)=\frac{1}{2x} + P + \tilde E x + \order{x^2}.
\label{eq:p_exp_0}
\]
The relation between $t(x)$ and the momentum and energy is slightly trickier:
The limit of the expansion \eqref{eq:momeng}
gives two asymptotic expansions at the exponential singularity
of $t(x)$ at
\<
t(x)\exp(-i/2x)\eq
\exp \bigbrk{+iP+ix\tilde E+\order{x^2}}
+\order{\exp(-i/x)},
\nln
t(x)\exp(+i/2x)\eq
\exp \bigbrk{-iP-ix\tilde E+\order{x^2}}
+\order{\exp(+i/x)},
\>
where the terms $\order{u^L}$ in \eqref{eq:momeng}
are interpreted as exponential singularities.
Putting the two expansions together and comparing with \eqref{eq:baxtercont}
is in agreement with \eqref{eq:p_exp_0}.

The total filling $\alpha=M/L$
is easily obtained from the definition \eqref{eq:qmom} of the quasi-momentum
\[
p(x) = \frac{1-2\alpha}{2x} + \order{1/x^{2}} \,.
\label{eq:p_exp_inf}
\]
The expansion of $t(x)$ around $x=\infty$ in \eqref{eq:magnonnum}
yields a compatible expression, which however does not fix the above sign
\[
t(x)=2-\frac{(1-2\alpha)^2}{4x^{2}}+\order{1/x^3}.
\]

One can also derive two useful relations between
the total and partial fillings
\[
\alpha =\sum_k \alpha_k\,,
\label{eq:p_exp_inf2}
\]
and the overall momentum and the mode numbers
\[
P = \sum_k 2\pi n_k \alpha_k \,.
\label{eq:sum_nl_al}
\]

\subsection{Spectral Curve}
\label{sec:spectral_curve}

For the differential $\ud p(x)$ the analytic
structure is somewhat simpler: The shifts of $p(x)$ by constants which
arise when passing through a branch cut are not seen in $dp(x)$. The
differential $\ud p(x)$ thus has merely two Riemann
sheets with opposite signs and it is called the spectral curve. In
particular, the condensate cuts drop out in the spectral curve. Only
the standard branch cuts are seen in $dp(x)$; they change the sheet or
equivalently flip the sign.

\begin{figure}\centering
  \includegraphics{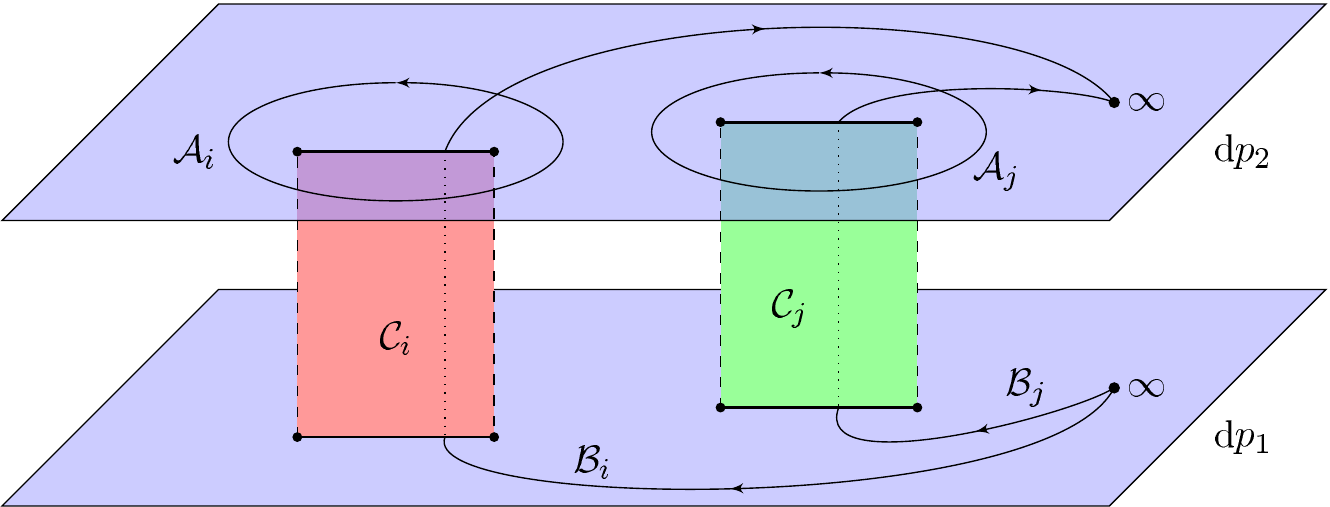}
  \caption{Cycles of the spectral curve $\ud p$. Shown are the two
  sheets $\ud p_1$, $\ud p_2$ of the Riemann surface of the spectral
  curve $\ud p$. The two sheets are connected by cuts $\contour{C}_i$.
  To each cut $\contour{C}_i$ is associated an A-cycle $\contour{A}_i$
  and a B-cycle $\contour{B}_i$. The A-cycle encircles the cut
  contour, staying on the same sheet, while the B-cycle starts at
  $x=\infty$ on one sheet, passes through the cut and ends at
  $x=\infty$ on the other sheet.}
\label{fig:cycles}
\end{figure}

The above parameters of the solution can be read off from the spectral
curve. First we need to introduce A-cycles and B-cycles for the cuts.
A-cycles wind around the cuts while B-cycles extend from $x=\infty$
through a cut and back to $x=\infty$ but on the other Riemann sheet,
see \figref{fig:cycles}.%
\footnote{Note that the B-cycles defined here are not cycles in the
strict sense because they are open curves.
Conventionally, B-cycles are defined as closed curves
that pass through two branch cuts.
A spectral curve with $K$ cuts has genus $K-1$ and hence only
$K-1$ A-cycles and $K-1$ closed B-cycles are independent. The
set of $K$ open B-cycles as defined here is equivalent
to the set of $K-1$ closed B-cycles plus one open B-cycle.}%
For a normal cut the A-cycle does not intersect with any
branch cut of the quasi-momentum. The A-period of
$\ud p(x)$ is therefore zero
\[
\oint_{\mathcal{A}_k} \ud p(x)=0.
\]
The B-period yields the mode number $n_k$ of the cut
\[
\int_{\mathcal{B}_k} \ud p(x)=2\pi n_k.
\label{eq:Bj}
\]
To understand this we split up the contour
into the two parts before and after crossing the cut at point $x$.
The first part yields $p(x+\varepsilon)-p(\infty)=p(x+\varepsilon)$
due to our normalisation $p(\infty)=0$.
On the other sheet the sign of the quasi-momentum is flipped
and consequently the second part yields $-p(\infty)+p(x-\varepsilon)=p(x-\varepsilon)$.
Altogether we obtain $p(x+\varepsilon)+p(x-\varepsilon)$ and by
\eqref{eq:sqrtcut} this equals $2\pi n_k$.
Note that a vanishing A-period of $\ud p$ enables us to be
rather unspecific about how the B-period returns to $x=\infty$
on the second sheet.

Finally, the partial filling \eqref{eq:partfill}
of a cut can be expressed through the
A-period of $x\,\ud p(x)$
\[\label{eq:fillingcurve}
\alpha_k
= \int_{\mathcal{C}_k} \ud x\, \rho(x)
=-\frac{1}{2\pi i}\oint_{\mathcal{A}_k}\ud x\, p(x)
=\frac{1}{2\pi i}\oint_{\mathcal{A}_k} x\, \ud p(x).
\]
This follows by substitution of \eqref{eq:density} and partial integration.
Note that the partial integration assumes that the A-period of $\ud p$ is zero.

The above discussion shows that the cut contours and corresponding cycles
can be deformed continuously without affecting the parameters of the solution.
However, special care has to be taken when a cycle moves into a cut or
passes through a singularity.
In particular, the reality condition $\ud \rho(x)\geq 0$ appears to play no
important role in the purely classical approximation.

\begin{figure}\centering
  \subfloat[]{\includegraphics{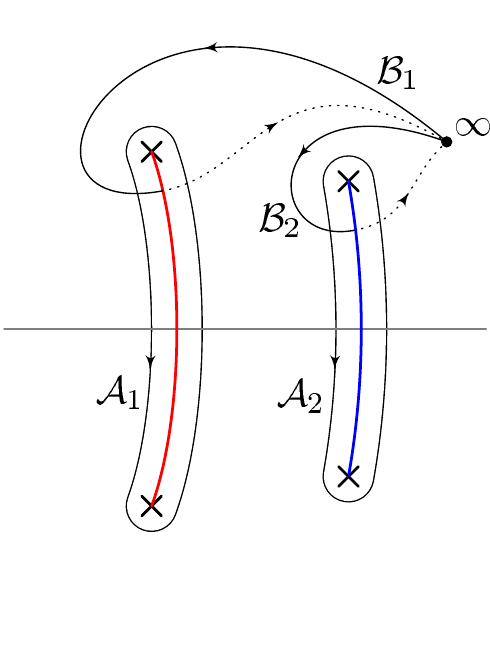}\label{fig:cycles.vertical}}\hfill
  \subfloat[]{\includegraphics{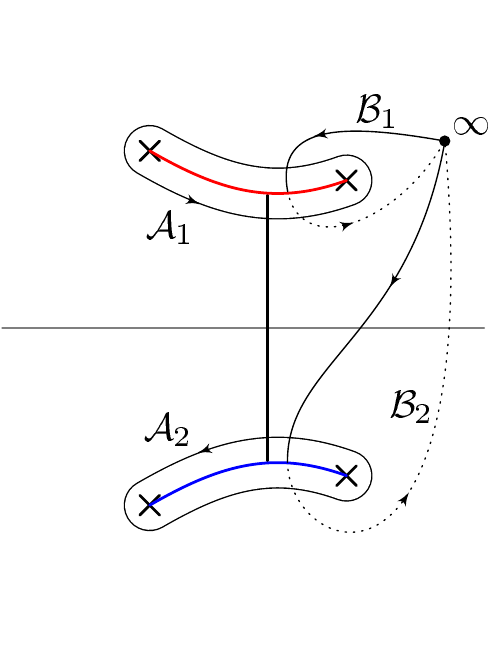}\label{fig:cycles.horizontal}}\hfill
  \subfloat[]{\includegraphics{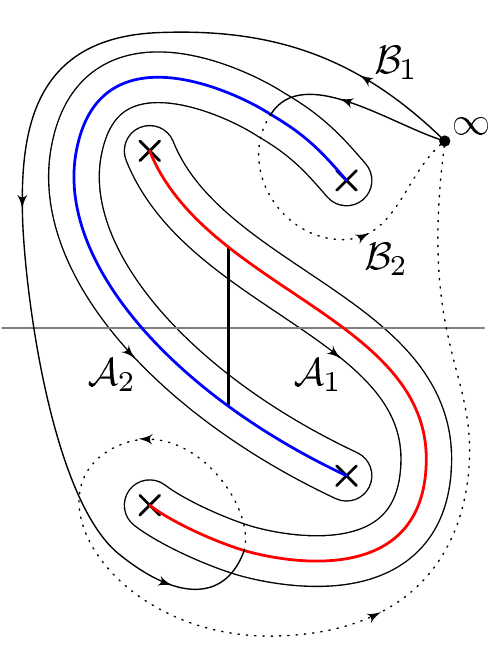}\label{fig:cycles.wild}}
  \caption{Possible cut contours (blue/red,thick) and corresponding
  cycles for a core with four tails. In the standard configuration
  \protect\subref{fig:cycles.vertical}, complex conjugate
  pairs of branch points are connected more or less directly. In this
  case, the A-periods vanish and the B-periods are non-ambiguous, even
  when the central cut segments join and form a condensate core as in
  \protect\figref{fig:condensate_cut}. When the branch points are
  connected horizontally \protect\subref{fig:cycles.horizontal},
  the mode numbers (B-periods) of the cuts are ambiguous, hence the configuration
  requires a condensate (black,vertical) between the two cuts. This
  implies that the A-periods do not vanish. Branch cuts that wind
  around each other \protect\subref{fig:cycles.wild} also
  require a condensate cut, hence complex conjugate branch points
  carry different mode numbers.}
\label{fig:cyclescondensate}
\end{figure}

In the case of many cuts the assignment of cycles
is not necessarily unique anymore.
Let us discuss what happens when two standard cuts
join to form a condensate cut with four tails.
Now there are various ways in which the four branch points
could be connected, see \figref{fig:cyclescondensate}.
The standard method to set up the cuts is to
connect complex conjugate pairs of branch points
more or less directly, as in \figref{fig:cycles.vertical}.
Then the above assignments of parameters works well.
All cuts are of the standard kind and condensate cuts
arise from two standard cuts running parallel for a while,
see \figref{fig:condensate_cut}.

Another option would be to connect the branch points horizontally with
a condensate cut forming between the two standard cuts
(\figref{fig:cycles.horizontal}). This option leads to non-zero
A-periods
\[ \oint_{\mathcal{A}_k} \ud p(x)=\pm 2\pi n'_j, \]
where $n'_j$ is the density of the condensate cut being intersected by
the A-cycles. The non-trivial A-cycles then lead to ambiguities in the
definition of mode numbers $n_k$ and fillings $\alpha_k$. Also
non-technically it is unclear how to associate suitable $n_k$ and
$\alpha_k$ to these two individual cuts. The situation becomes even
worse if the cuts wind around the branch points
(\figref{fig:cycles.wild}). Therefore the standard straight connection
of complex conjugate pairs of branch points appears best to describe
the parameters.

\subsection{Finite Gap Solutions}

The simplest types of spectral curves with the desired properties
are the finite-gap solutions.
These have finitely many branch cuts (``gaps'') and
therefore finitely many branch points.
The general ansatz for $p'(x)$ is given by
\[
p'(x)^2 = \frac{g(x)^2}{x^4 h(x)}\,,
\label{eq:dp_ansatz}
\]
where $g(x)$ and $h(x)$ are polynomials of degree $K$ and $2K$, respectively
\[
g(x) = \sum_{k=0}^{K} c_k x^k,\qquad
h(x) = \prod _{k=1} ^{2 K} (x-x_k) .
\label{eq:y^2}
\]
The roots $x_k$ are the square root branch points which must come in (complex conjugate) pairs.
A pair is typically connected by a branch cut
and therefore $K$ is the number of standard branch cuts.
The genus of the algebraic curve $p'(x)$ equals $g=K-1$.

\subsection{Stability}
\label{sec:stab}

Stability addresses the question which classical spectral curves
with real and positive densities on the cuts can be realised approximately
by solutions of the Bethe equations with large but finite $L$.
An important stability criterion was derived in \cite{Beisert:2005mq}
\[\label{eq:stability}
\int_{\mathcal{C}_k} d\log\frac{n+i\rho(x)}{n-i\rho(x)}=0\quad\mbox{for all }n\in\Integers.
\]
The criterion implies that the branch of the above logarithm can be
uniquely defined on an isolated standard branch cut $\mathcal{C}_k$.
The argument of the logarithm
\[
\frac{n+i\rho(x)}{n-i\rho(x)}\approx
\frac{n(u_{k+1}-u_k)+i}{n(u_{k+1}-u_k)-i}\approx
\frac{u_{k+n}-u_k+i}{u_{k+n}-u_k-i}
\]
approximates the scattering term in the Bethe equation \eqref{eq:xxxbethe}.
In a logarithmic form of the Bethe equations the condition implies
that mode numbers can be unambiguously associated to the cuts,
even in the full quantum theory.
Unlike the classical quantities, here the density appears
independently of the differential $\ud x$ and therefore
it is crucial to use the physical contour of the cut $\mathcal{C}_k$.

The most interesting case is $n=1$; the conditions for $n>1$
appear to be less constraining and perhaps redundant.
It is quite clear that if the absolute density is bounded by unity everywhere,
\[
|\rho(x)|<1,
\label{eq:stabcrit}
\]
the stability criterion will be satisfied. However this condition
is too restrictive in general.
Nevertheless, for a stand-alone cut the density is typically
highest at the centre $x_0$ where the cut crosses the real axis
and where the density becomes purely imaginary.
It is then necessary to obey
\[
|\rho(x_0)|<1,
\]
in order to satisfy the stability condition.
Otherwise the argument of the logarithm in \eqref{eq:stability}
will cross the negative real axis at $x_0$ and wind around the origin once.

What remains obscure at this point is how to interpret the stability condition
for two standard cuts joined by a condensate.
Also it is not clear if \eqref{eq:stability} alone can
ensure that a given configuration of cuts can be realised
by a concrete solution of the Bethe equations.
In the following we shall study the general one-cut and two-cut solutions
in order to shed some light on the various classically allowed
configurations of cuts and when the stability condition is satisfied.

\section{One-Cut Solution}
\label{sec:onecut}

In this section we review the simplest type of spectral curve with
one cut.
It has two parameters, the mode number $n$ and the filling $\alpha$.
Its genus is zero and thus we will only encounter algebraic
and trigonometric functions.
We also review the corresponding construction
for the Landau--Lifshitz model in \appref{sec:ll-model}.

\subsection{Solution}

The general one-cut solution was obtained in \cite{Kazakov:2004qf}.
It has one mode number $n$, one filling $\alpha$ and it takes the form
\[\label{eq:onecutqmom}
p(x)=\pi n+\frac{1-2\pi n x}{2x}\sqrt{1+\frac{8\pi n \alpha x}{(1-2\pi n x)^2}}\,.
\]
The derivative of the quasi-momentum defines the spectral curve
\[
p'(x)^2=
\frac{(1-2\pi n x(1-2\alpha))^2}{4x^4\bigbrk{(1-2\pi n x)^2+8\pi n \alpha x}}
\]
and matches with \eqref{eq:dp_ansatz,eq:y^2}.
The quasi-momentum is in agreement with
the expansions at $x=0,\infty$ \eqref{eq:p_exp_0,eq:p_exp_inf}
and with the condition for the branch points \eqref{eq:sqrtsing}.
The higher terms in the expansion at $x=0$ yield the total momentum and energy
\[
P=2\pi n\alpha,
\qquad
\tilde E=4\pi^2 n^2\alpha(1-\alpha).
\label{eq:onecut_EP}
\]
If we choose to restrict to cyclic states as required for AdS/CFT
we have to set $P=2\pi m$ with integer $m$.
Then the total filling must be a rational number $\alpha=m/n$.
However, all these solutions are unstable.
As we shall see later this is related to the fact that
the total momentum leaves the first Brillouin zone, $|P|>\pi$.

\subsection{Cut Contour}
\label{sec:contour}

\begin{figure}\centering
\includegraphics{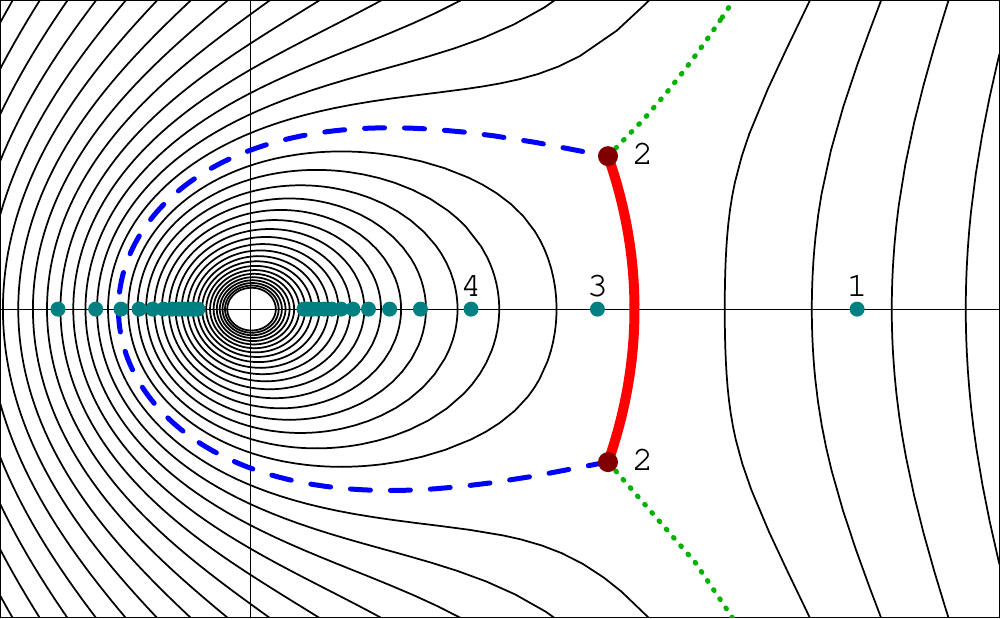}

\caption{Physical position of the cut (solid/red)
for a one-cut solution with $n=2$, $\alpha=0.05$.
Two alternative contours with positive filling originating from the same
branch points are shown: The second (dashed/blue) contour corresponds
to the conjugate one-cut solution with $n=-2$, $\alpha=0.95$.
This contour is unphysical because the filling exceeds $0.5$.
The third contour (dotted/green) extends to infinity,
it is inconsistent because the filling is infinite.
The remaining lines depict possible cuts with real filling
not ending on the branch points.
Fluctuation points are marked by dots.}
\label{fig:onerealcuts}
\end{figure}

The physical contour of the cut is not a simple function.
In particular, it is not given by the natural
branch cut of the square root in $p(x)$;
it lies somewhat closer to the origin.
In order to find the contour, we shall make use of
the identity \eqref{eq:rho} which relates the
density to the quasi-momentum.
The integrated density must be real and positive;
therefore we will need the integral of the quasi-momentum
\<
\Lambda(x)=\int p(x)\,\ud x
\eq
\pi n x
+\frac{1-2\pi n x}{2}
\sqrt{1+\frac{8\pi n \alpha x}{(1-2\pi n x)^2}}
\nl
-\artanh\lrbrk{\frac{1-2\pi n x}{1+2\pi n x}
\sqrt{1+\frac{8\pi n \alpha x}{(1-2\pi n x)^2}}}
\nl
-\alpha\arcoth\lrbrk{\frac{1-2\pi n x}{1-2\pi n x-2\alpha}
\sqrt{1+\frac{8\pi n \alpha x}{(1-2\pi n x)^2}}}.
\label{eq:onecut_Lambda}
\>
On the cut the combination $\Lambda(x)-\pi n x$ must be real.
At the branch points
\[
x_\pm=\frac{1-2\alpha\pm 2i\sqrt{\alpha(1-\alpha)}}{2\pi n}
\label{eq:onecut_bp}
\]
it takes the values $\Lambda(x_\pm)-\pi nx_\pm=\pm \ihalf \pi \alpha$
and thus the contour is defined by $\Lambda(x)-\pi n x\in \ihalf\pi\alpha[-1,1]$.
In \figref{fig:onerealcuts} we have plotted the contour of a sample cut.

In fact as discussed below \eqref{eq:sqrtsing},
a branch cut with positive density can, in principle, originate from a branch point
in three different directions.
The shortest path will turn out to be the only physical choice.

The longer circular path which encircles the origin
is equivalent to the one-cut solution with
opposite mode number $n'=-n$ and conjugate filling $\alpha'=1-\alpha$.
This can be seen as follows:
We deform the shorter cut continuously to the longer cut by rotating it
towards the right by $240^\circ$.%
\footnote{Alternatively one can rotate by $120^\circ$
and flip the sign of $p$.}
At some point the cut has to pass
the point $x=\infty$. This has the following two effects.
Firstly, the A-cycle describing the filling \eqref{eq:fillingcurve}
intersects $x=\infty$ twice. This adds to $\alpha$ the residues $\half-\alpha$
and $\half-\alpha$ (the opposite sheet with opposite circulation)
so we obtain $\alpha'=\alpha+2(\half-\alpha)=1-\alpha$.
Secondly, the point $x=\infty$ now resides on a different sheet.
This implies $p(\infty)=2\pi n$
and we have to subtract the constant $2\pi n$ to normalise to $p(\infty)=0$.
At the branch point that leads to $p(x_\ast)=-\pi n$, i.e.\ the new
mode number is $n'=-n$.
Note that this configuration has a large filling $\alpha\geq \half$
because the shorter cut has $\alpha\leq \half$. It is therefore
unphysical.

For the third choice we rotate the contour towards the left by $240^\circ$.
This contour escapes towards $\pm i\infty$ on both sides.
Moreover the density $\rho$ on the contour approaches a
finite value at infinity. Therefore the total filling
on the contour is infinite, and the configuration is inconsistent.

\subsection{Stability}

\begin{figure}[tbp]
\centering
  \includegraphics{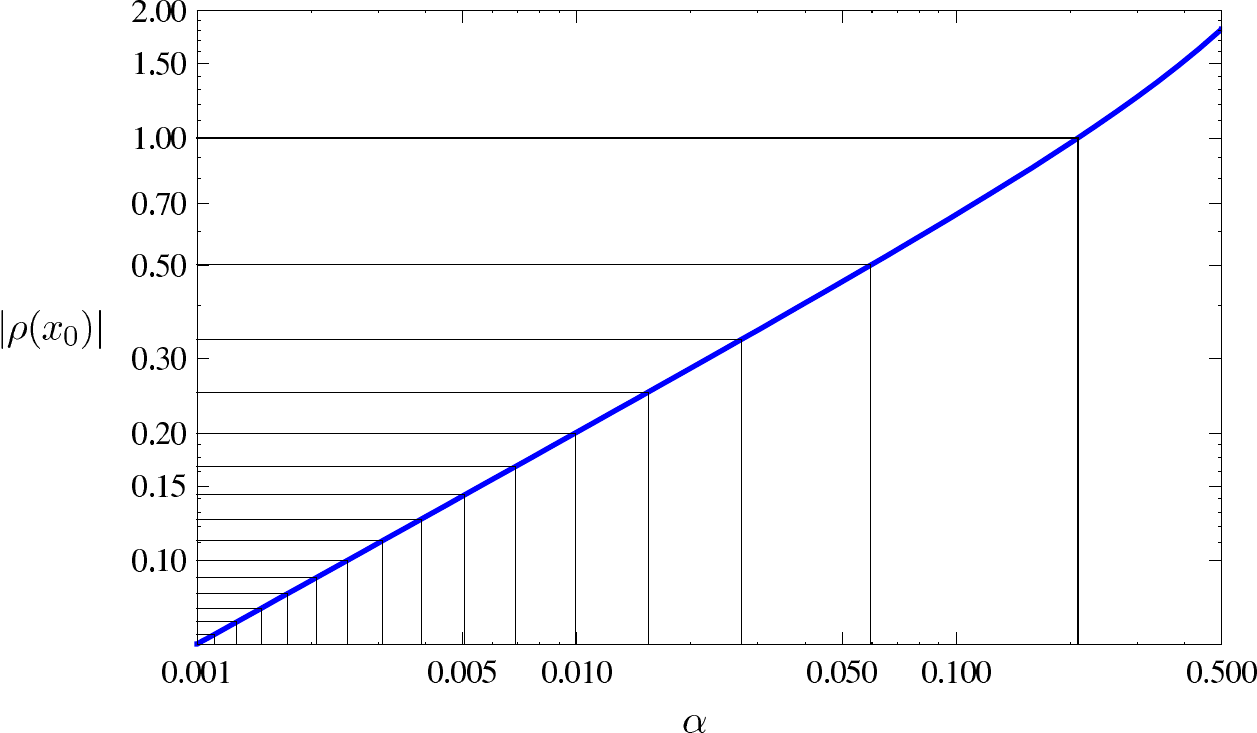}\\
  \caption[Maximal densities of stable one-cut solutions]{Absolute
  density $|\rho(x_0)|$ at the centre of a single cut with mode
  number $|n|=1$ versus the cut's filling fraction $\alpha$
  (blue/thick). Also displayed are the lines $1/m$ for integer $m$.
  According to equation \eqref{eq:maxfill_scaling}, the intersection
  points of these lines with the density mark the maximal fillings
  $\alpha \indup{cond}$ of stable solutions with mode numbers $m$
  (see also \protect\tabref{tab:maxfill}).}
\label{fig:maxfill}
\end{figure}

We are now ready to consider the issue of stability.
The stability condition for the one-cut solutions
implies that the density must be bounded by unity
where the cut crosses the real line.
Assume the cut crosses the real axis at $x_0$
which must be the solution of the equation $\Lambda(x_0)=\pi n x_0$.
This equation is transcendental and we can only solve it numerically
for given values of $n$ and $\alpha$.
Once we have the value $x_0$
the absolute density is given by
\[\label{eq:densityonaxis}
\bigabs{\rho(x_0)}=\frac{1}{\pi}\,\bigabs{p(x_0)-\pi n}.
\]
The density for solutions with $n=1$ is plotted in \figref{fig:maxfill}.
We are interested in the filling $\alpha\indup{cond}$ where the
density reaches the maximum allowed value, $|\rho(x_0)|=1$ or
$p(x_0)=\pi(n+1)$. Let us assume for simplicity that $n>0$. Because
of the way $p(x)$ and $\Lambda(x)$ scale with $x$ and $n$, the
intersection point of a cut with mode number $n$ and filling $a$ and
the density at that point are directly related to the intersection
point and the density of the cut with mode number $1$ and the same
filling:
\[
  x_0 (n,\alpha) = \frac{x_0 (1,\alpha)}{n} \,, \qquad
  \rho_n (x_0(n,\alpha)) = n \rho_1 (x_0(1,\alpha)) \,.
\label{eq:maxfill_scaling}
\]
Hence it is sufficient to obtain the fillings $\alpha$ at which the
absolute density at the centre of the cut with mode number $1$ equals
$1/n$. These fillings are exactly the maximal fillings of the cuts
with mode number $n$ that are allowed by the stability criterion; they
are indicated in \figref{fig:maxfill}.

Alternatively, one can solve the equation $p(x_0) = \pi(n+1)$ for
$x_0$ in terms of $\alpha$ and $n$. The solution is substituted in
$\Lambda(x_0)=\pi n x_0$ which yields the equation
\[
2(q+n)\arcoth(q+2n)-\frac{1-q^2}{2n}\artanh\lrbrk{\frac{1-q^2-2nq}{2n}}=1\,.
\]
Here $q$ is an auxiliary variable that encodes the maximal filling
$\alpha\indup{cond}$ and the intersection point $x_0$
\[ \alpha\indup{cond}=\frac{1-q^2}{4n(q+n)}\,, \qquad
x_0=\frac{1}{2\pi(q+n)}\,.  \]
We list the first few values of $\alpha\indup{cond}$ in
\tabref{tab:maxfill}. Note that $\alpha\indup{cond}<1/2n$, which
implies that stable solutions have a momentum $P<\pi$.
Hence, cyclic one-cut solutions cannot be stable. In fact one can solve
the above equation perturbatively using that $q\sim 1/n$. The
expansion of $\alpha\indup{cond}$ reads
\[
  \alpha\indup{cond}= \frac{1}{4 n^2}-\frac{7}{144
  n^4}+\frac{367}{38880 n^6} -\frac{540373}{293932800n^8}
  +\frac{1895953}{5290790400 n^{10}} +\order{1/n^{12}}.
\label{eq:onecut.alphacond.exp}
\]

\begin{table}
  \centering
  \begin{tabular}{|r|c|c|}
    \hline
    $|n|$&$\alpha\indup{cond}$&$\tilde{E}\indup{cond}/4\pi^2$\\
    \hline
     1&0.2092896452&0.1654874896\\
     2&0.0596024470&0.2241999811\\
     3&0.0271903146&0.2380590127\\
     4&0.0154373897&0.2431852264\\
     5&0.0099228217&0.2456089819\\
    \hline
  \end{tabular}\quad
  \begin{tabular}{|r|c|c|}
    \hline
    $|n|$&$\alpha\indup{cond}$&$\tilde{E}\indup{cond}/4\pi^2$\\
    \hline
     6&0.0069071371&0.2469394280\\
     7&0.0050818745&0.2477464054\\
     8&0.0038944180&0.2482720935\\
     9&0.0030790284&0.2486333844\\
    10&0.0024951483&0.2488922545\\
    \hline
  \end{tabular}\\
  \caption{Maximal fillings $\alpha_{\mathrm{cond}}$ and the
  corresponding energies $\tilde{E}_{\mathrm{cond}}$ for one-cut
  solutions for the first few mode numbers $n$. Above these
  fillings, solutions require a condensate for stability.}
\label{tab:maxfill}
\end{table}

\subsection{Fluctuations}
\label{sec:onecut.fluct}

Fluctuations are very small cuts which
can exist in a background the one long cut \cite{Beisert:2003xu}.
For the one-cut solution they were discussed in
\cite{Beisert:2003xu,Minahan:2004ds,Beisert:2005bv}.
Their position $x_\ast$ is determined by the long cut,
\[
p(x_\ast)=\pi n_\ast,
\]
but they are not strong enough to
back-react on the position of the long cut substantially.
We shall again assume that $n>0$.
The solution to the above equation reads
\[
\frac{1}{x_\ast}=2\pi n(1-2\alpha)+2\pi(n_\ast-n)\sqrt{1-\frac{4n^2\alpha(1-\alpha)}{(n_\ast-n)^2}}\,.
\]
The momentum and energy of a fluctuation mode quantum
with $\delta\alpha=1/L$
are given by \cite{Minahan:2004ds,Beisert:2005bv}
\[\label{eq:MomEngFluct}
\delta P =
\frac{2\pi n_\ast}{L}\,,
\qquad
\delta\tilde{E} =
\frac{4\pi^2}{L}  \lrbrk{
n(2n_\ast-n)(1-2\alpha)
+(n_\ast-n)^2\sqrt{1-\frac{4n^2\alpha(1-\alpha)}{(n_\ast-n)^2}}
}.
\]

If the long cut is rather short,
the fluctuations will reside close
to their vacuum positions $x_\ast\approx 1/2\pi n_\ast$.
For fluctuations of the same mode number as the cut,
$n_\ast=n$, the condition
$p(x_\ast)=\pi n_\ast$ is solved by the branch points,
which means that this particular fluctuation will increase the size of the long cut.
The longer the long cut gets the more will it attract fluctuations
with nearby mode numbers $n_\ast\approx n$ towards it \cite{Beisert:2005bv}.
At some value of $\alpha$, the fluctuation with $n_\ast=n+1$
collides with the long cut (from the left).
This happens precisely when the density reaches unity
at the real axis, cf.\ \eqref{eq:densityonaxis},
and so the above discussion applies.
Note that at this point the fluctuation with $n_\ast=n-1$
is still at some distance from the cut.

In conclusion, we can infer that for all stable one-cut solutions
the fluctuations are well-separated from the long cut.
When the fluctuation collides with the cut
we can actually argue independently for an instability:
Now the long cut can be filled not only from both ends, but also
from the middle. In practice we expect that a condensate cut
(with unit density) will form in the middle of the long cut.

\subsection{Condensate Formation}
\label{sec:onecut_cond}

\begin{figure}
  \centering
  \subfloat[$\alpha<\alpha\indup{cond}$]{\centering\includegraphics[width=0.3\textwidth]{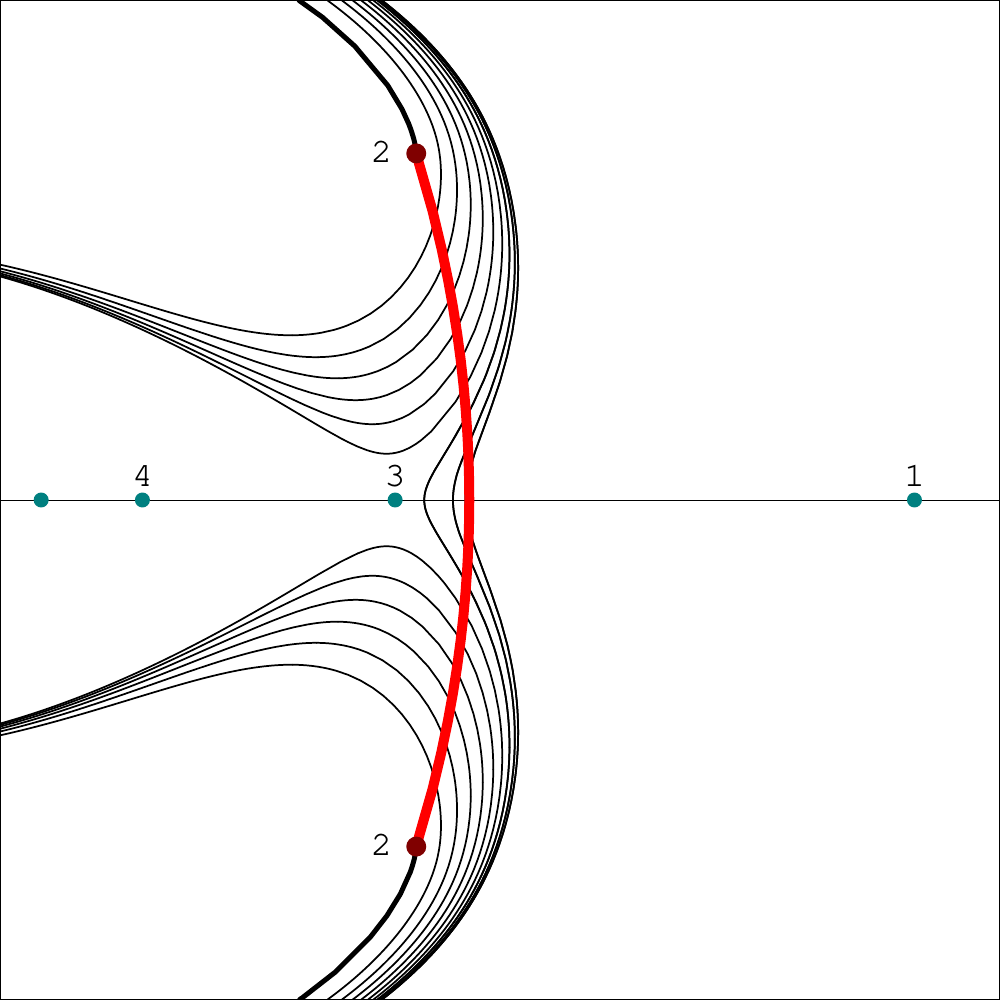}\label{fig:onebubblecutsa}}\hfill
  \subfloat[$\alpha\indup{cond}<\alpha<\alpha\indup{crit}$]{\centering\includegraphics[width=0.3\textwidth]{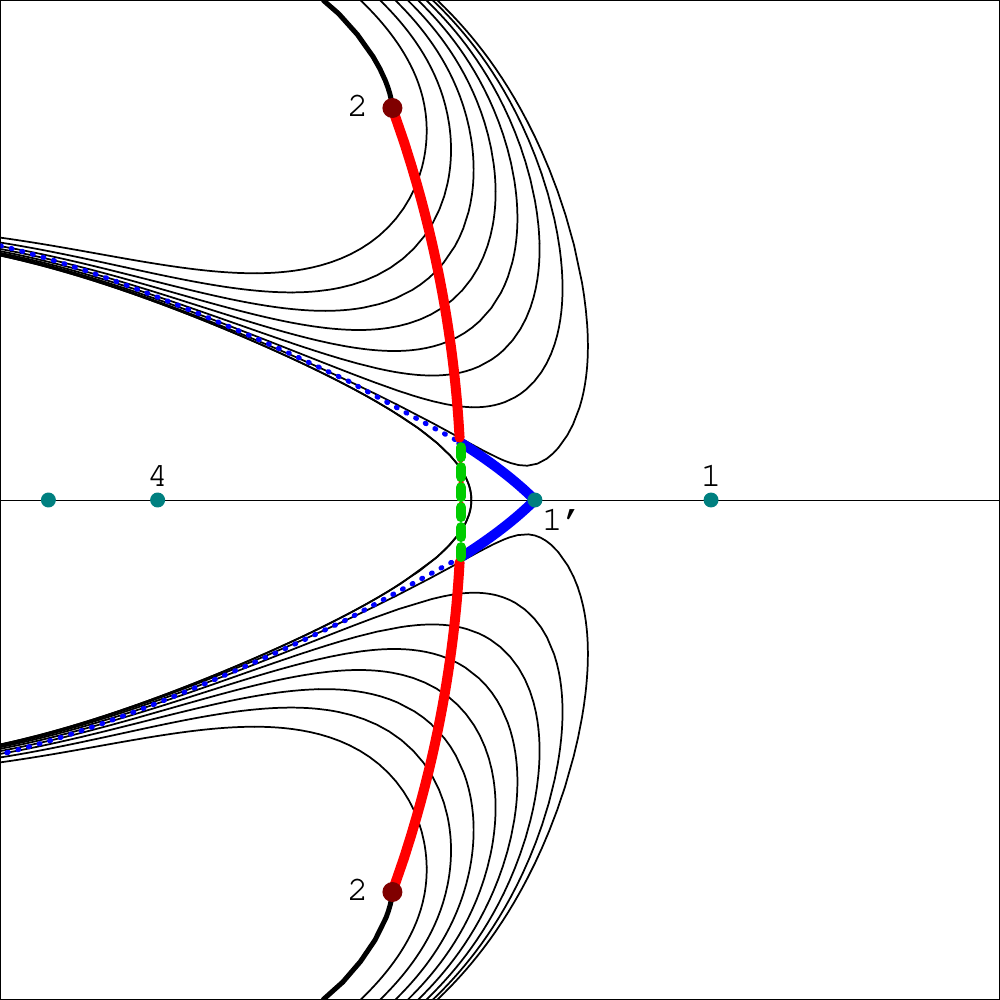}\label{fig:onebubblecutsb}}\hfill
  \subfloat[$\alpha\indup{crit}<\alpha$]{\centering\includegraphics[width=0.3\textwidth]{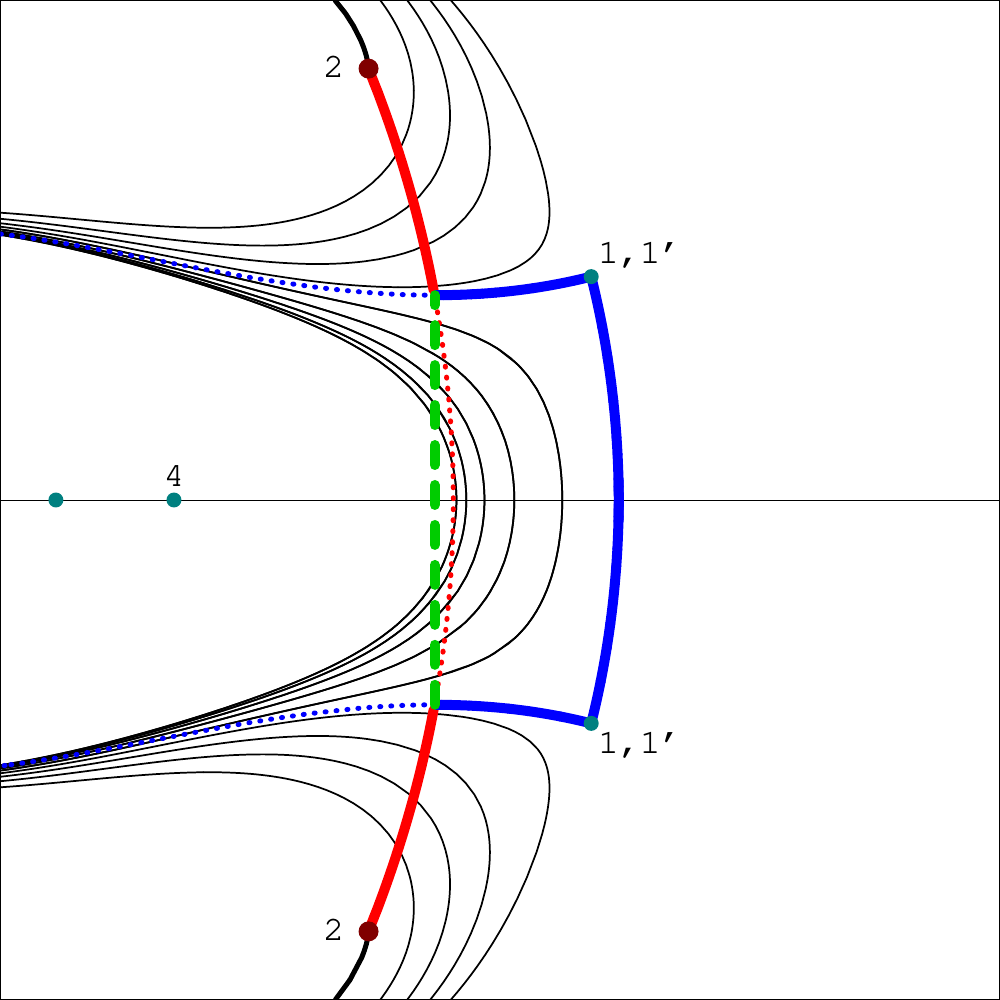}\label{fig:onebubblecutsc}}
  \caption{Potential deformations (thin/black) of a branch cut
  (solid/red) after inserting a condensate (dashed/green) for $n=2$
  and \protect\subref{fig:onebubblecutsa} $\alpha=0.03$,
  \protect\subref{fig:onebubblecutsb} $\alpha=0.065$,
  \protect\subref{fig:onebubblecutsc} $\alpha=0.8$.
  The physical deformation (solid/blue) exists only for
  $\alpha>\alpha\indup{crit}$. The contours originating from near
  the ends of the branch cut close up after encircling the origin. }
\label{fig:onebubblecuts}
\end{figure}

\begin{figure}
\centering
\includegraphics{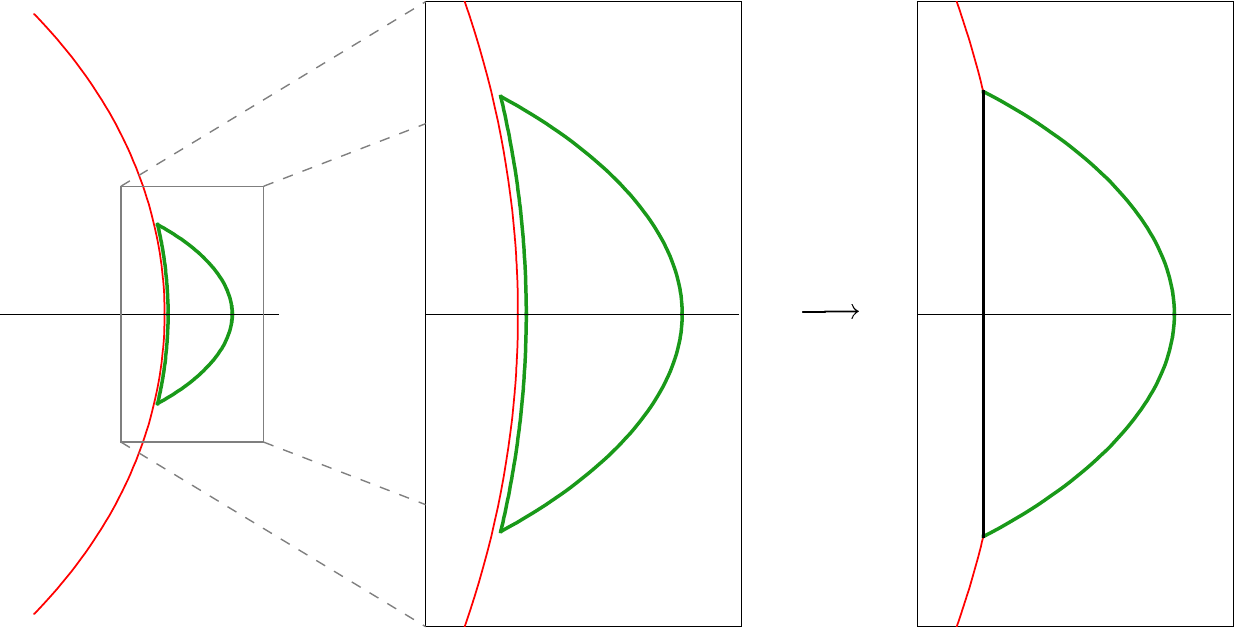}
\caption{Possible deformation of a branch cut: For large enough
filling $\alpha>\alpha\indup{cond}$, a closed loop cut (thick/green)
can be added to the branch cut (thin/red). In the region enclosed by
the loop cut, the quasi-momentum $p(x)$
is changed to $-p(x)+2\pi n\indup{L}$,
where $n\indup{L}$ is the mode number of the loop cut. As shown
in \protect\figref{fig:condensate_cut}, the density $\rho\indup{C}$ on the
common part of the two contours equals $i(n\indup{L}-n\indup{B})$, where
$n\indup{B}$ is the mode number of the branch cut. Hence the resulting
configuration has a condensate core, as shown in the right Figure.}
\label{fig:deformation}
\end{figure}

Let us add a vertical condensate cut ending on the existing branch
cut, see \figref{fig:onebubblecuts}. This is achieved by shifting the
quasi-momentum $p(x)$ by $-2\pi$ in the region enclosed by the two
contours, cf.\ \figref{fig:deformation}. Clearly the new curve
satisfies all conditions for the classical spectral curve. In effect,
the shift decreases the mode number on the inner part of the original
branch cut by one unit according to \eqref{eq:sqrtcut}. Furthermore,
the density function is changed according to \eqref{eq:density,eq:rho}
and thus the inner part of the branch cut has to be moved to obtain a
real density
\[ \rho(x) = \frac{1}{\pi i} \bigbrk{\pi (n-1)-p(x+\varepsilon) }. \]
The condensate cut can end on any point of the branch cut, and we show
the various potential configurations in \figref{fig:onebubblecuts}.
Any of these configurations with a condensate appear to be possible
from the point of view of spectral curves. However, the Bethe
equations will single out one particular configuration as the
distribution of Bethe roots in the thermodynamic limit. To understand
the physical distribution we have to distinguish three qualitatively
different cases, see \figref{fig:onebubblecuts}.

In the first case the filling is below
the threshold for condensate formation, $\alpha<\alpha\indup{cond}$,
as discussed in the previous subsections.
Here all potential deformations leave the branch cut almost vertically
and then circle around the origin.
Due to the residue at the origin the filling
of the configuration is altered and becomes larger than $\half$.
Inserting a condensate cut therefore leads to an
unphysical configuration when $\alpha<\alpha\indup{cond}$.

At $\alpha=\alpha\indup{cond}$ the fluctuation point
with mode number $n_\ast=n+1$ crosses the branch cut
and effectively acquires the mode number $n_\ast=n-1$.
We now have two fluctuation points with the same
mode number $n-1$. For definiteness, let us call the
new point $n_\ast=(n-1)'$.
When increasing $\alpha$ further,
it will eventually collide with the other fluctuation point.
This happens at $\alpha=\alpha\indup{crit}$ with
\[
  \alpha\indup{crit}=\frac{1}{2}-\frac{1}{2}\sqrt{1-\frac{1}{n^2}}\,.
\label{eq:onecut.alphacrit}
\]

\begin{figure}[t]
\centering
  \subfloat[]{\includegraphics{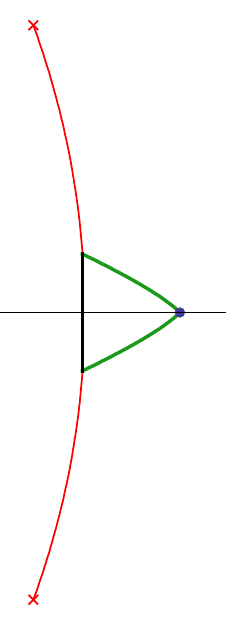}\label{fig:onecut2}}\quad
  \subfloat[]{\includegraphics{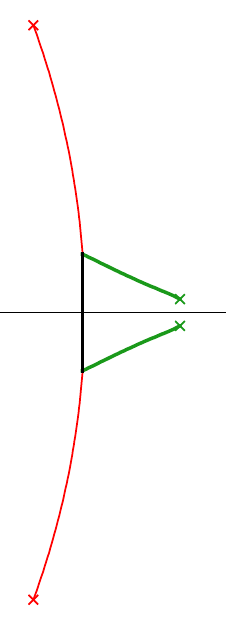}\label{fig:onecut2reg}}\quad\qquad
  \subfloat[]{\includegraphics{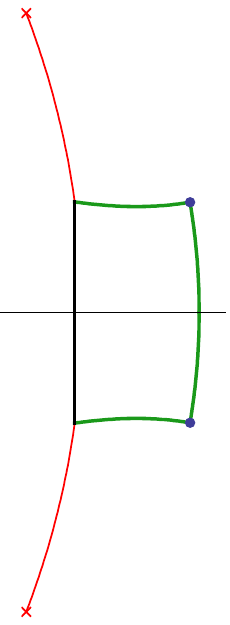}\label{fig:onecut3}}\quad
  \subfloat[]{\includegraphics{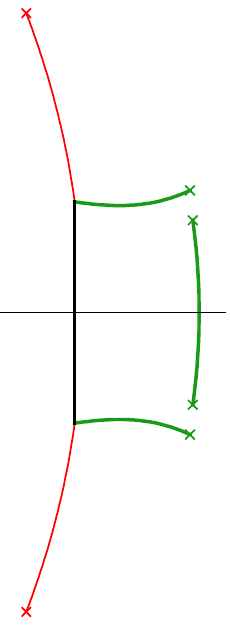}\label{fig:onecut3reg}}\quad\quad
  \subfloat[]{\includegraphics{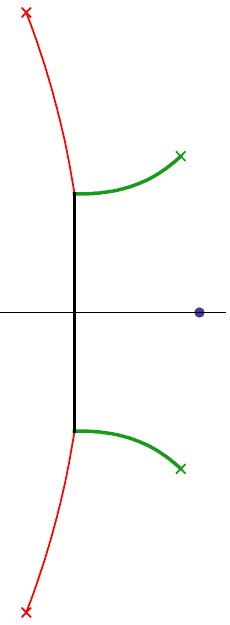}\label{fig:onecut32}}
  \caption{For large fillings $\alpha$, one-cut solutions are degenerate
  cases of two- and three-cut solutions. When a macroscopic number of
  Bethe roots gets added to the fluctuation point (blue dot) in \protect\subref{fig:onecut2},
  two square-root branch points form, the
  closed loop cut splits at its cusp and yields a genuine two-cut
  solution \protect\subref{fig:onecut2reg}. This singles out the
  deformation shown in \protect\subref{fig:onecut2} as the
  physical one for the case
  $\alpha\indup{cond}<\alpha<\alpha\indup{crit}$. At even higher
  fillings $\alpha>\alpha\indup{crit}$ \protect\subref{fig:onecut3} an
  excitation of the fluctuation points results in a three-cut solution
  \protect\subref{fig:onecut3reg}. A local minimum of the energy is reached when
  all roots from the third cut are moved to the second, leaving a
  two-cut configuration and a bare
  fluctuation point as in \protect\subref{fig:onecut32}.}
\label{fig:onecutreg}
\end{figure}

Consider now the case $\alpha\indup{cond}<\alpha<\alpha\indup{crit}$.
The deformations originating from the branch cut can now have two
qualitatively different shapes. The contours originating from near the
ends of the branch cut behave like in the first case above and are
thus unphysical. The contours originating from near the centre however
are just small deformations of the original cut and consequently they
lead to the same filling. All of these configurations are in principle
okay, however, only one can be physical. Indeed one of the
configurations is special: the limiting shape, i.e.\ the largest
possible small deformation. It is distinguished by a cusp at the
fluctuation point for $n_\ast=(n-1)'$, see
\figref{fig:onebubblecutsb}. We can argue that this is the physical
configuration: When we add macroscopically many Bethe roots to the
fluctuation point, it will split up and form two square root
singularities. The deformed cut will split at the cusp and form a
condensate with four tails as in \figref{fig:onecut2reg}. This
configuration is a genuine two-cut solution as we shall see in the
next section. When we take the new Bethe roots away we should return
to the one-cut solution. This is possible only if the deformed contour
meets the fluctuation point with mode number $n_\ast=(n-1)'$, as in
\figref{fig:onebubblecutsb,fig:onecut2}. In conclusion the one-cut
solution for $\alpha\indup{cond}<\alpha<\alpha\indup{crit}$ is a
degenerate case of a two-cut solution.

The final case is $\alpha\indup{crit}<\alpha$. Here the two
fluctuation points with mode numbers $n_\ast=(n-1)'$ and
$n_\ast=(n-1)$ have joined and branched off into the complex plane.
The configurations are similar to the above case of
$\alpha\indup{cond}<\alpha<\alpha\indup{crit}$, but now the limiting
shape meets both fluctuation points and thus has two cusps joined by
an approximately vertical contour. In this case we cannot add Bethe
roots to any of the two fluctuation points individually because it
would violate the reality condition (reflection symmetry of the
configuration about the real axis). They can only be excited in pairs
to create four new square root singularities and a three-cut solution
as in \figref{fig:onecut3reg}. In other words, the configuration is a
degenerate case of a three-cut solution. The third cut is the line
segment joining the two fluctuation points. By removing Bethe roots
from this cut one lowers the energy and we might therefore call this
solution unstable. Nevertheless we expect it to be a perfectly
well-behaved solution of the Bethe equations in the thermodynamic
limit albeit with non-minimal energy. A local minimum is obtained by
shifting all Bethe roots from the third cut to the second, as in
\figref{fig:onecut32}. This is a true two-cut solution which will be
discussed in generality in the next section.

Finally we remark that for $n=1$ the third case does not exist because
$\alpha\indup{crit}=\half$. This is because the fluctuation point with
$n_\ast=0$ is always at $x_\ast=\infty$ and cannot join with the one
for $n_\ast=0'$. Therefore the solution with $n=1$ represents a local
minimum of the energy for all $0<\alpha<1/2$.

\section{The General Two-Cut Solution}
\label{sec:twocut}

As described in \secref{sec:spectral_curves}, a solution to the
Bethe equations \eqref{eq:xxxbethe} in the thermodynamic limit is
given by a quasi-momentum $p$, whose domain is a multi-sheeted cover
of $\bar{\Complex}$. The corresponding spectral curve $\ud p$ has
B-periods in $2\pi\Integers$ and vanishing A-periods, while the
quasi-momentum $p(x)$ has a simple pole of residue $1/2$ at $x=0$. In
this section, the general spectral curve with two cuts will be
constructed and its properties will be investigated. Spectral curves
with two cuts have a Riemann surface of genus one and thus can be
described in terms of complete elliptic integrals of the first, second
and third kind. These are denoted by $\ellK$, $\ellE$ and $\ellPi$ and
are defined as
\begin{align}
  \ellK(q) & := \int _0 ^1 \frac{1}{\sqrt{(1 - t^2)(1 - q t^2)}} \, \ud t
  \,, \nn\\
  \ellE(q) & := \int _0 ^1 \frac{\sqrt{1 - q t^2}}{\sqrt{1 - t^2}} \,
  \ud t \,, \nn\\
  \ellPi(z\mathpunct{|}q) & := \int _{0} ^{1} \frac{1}{(1 - zt^2) \sqrt{(1 -
  t^2)(1 - qt^2)}} \, \ud t \,. \label{eq:ellpi}
\end{align}

The two-cut solution can be obtained by direct integration of the
general ansatz \eqref{eq:dp_ansatz}. Here a different approach will
be followed which makes use of the result obtained for the symmetric
case in \cite{Beisert:2004hm}.

\subsection{The Symmetric Two-Cut Solution}
\label{sec:twocut_sym}

We will obtain the general two-cut solution by generalising the
symmetric solution \cite{Arutyunov:2003rg,Beisert:2004hm}
\[
  p_0 (z) := - \frac{\oldDelta n \, a_0}{z} \sqrt{\frac{a_0^2 (b_0^2 - z^2)}{b_0^2 (a_0^2 - z^2)}}\,
\ellPi \Bigl( \frac{q z^2}{z^2 - a_0^2}\mathpunct{\Big|}q \Bigr),
\label{eq:p0}
\]
where $q = 1 - a_0^2/b_0^2$. As a function
of $z$, the elliptic integral of the third kind $\ellPi(z\mathpunct{|}q)$ has
branch points at $z=1$ and $z=\infty$. Its discontinuity across the
branch cut between these two points
(from the lower to the upper half plane)
equals $2 \pi i$ times the residue of the integrand in
\eqref{eq:ellpi} at $t = 1/\sqrt{z}$:
\<
  \ellPi(z + i \varepsilon\mathpunct{|}q) - \ellPi(z - i \varepsilon\mathpunct{|}q)
  \eq \oint \limits _{1/\sqrt{z}} \frac{1}{(1 - zt^2) \sqrt{(1 - t^2)(1 -  qt^2)}} \, \ud t
\nln\eq
  - \pi i \sqrt{\frac{z}{(z-1)(z-q)}} \,.
\label{eq:ellpi_discont}
\>
The function $\elltPi(z) := \ellPi \bigbrk{q z^2/ (z^2 - a_0^2)\mathpunct{|}q}$
used in \eqref{eq:p0} is chosen such that it has branch points
$\left\{ a_0, b_0, -a_0, -b_0 \right\}$, hence $\elltPi(z)$ has two
branch cuts that lie symmetrically around $z=0$. In principle, $a_0$
and $b_0$ can be arbitrary complex numbers. However, physical branch
cuts must be symmetric about the real axis, therefore only the case
$b_0 = \bar{a}_0$ is physical. The discontinuity across the cuts can
be inferred from \eqref{eq:ellpi_discont} and equals
\[
  \elltPi(z+\epsilon) - \elltPi(z-\epsilon)
  = \pm \frac{\pi z}{a_0} \sqrt{\frac{b_0^2 (a_0^2 - z^2)}{a_0^2
  (b_0^2 - z^2)}} \,.
\label{eq:pitilde_discont}
\]
In \eqref{eq:p0}, the factor in front of $\elltPi(z)$
makes the discontinuity constant, but also switches sign across the
cut. As a result, the sum of the limiting values of $p_0$ on either
side of a cut is
\[
 p_0 (z+\epsilon) + p_0 (z-\epsilon)
= \pm \pi \oldDelta n \,.
\]
This means that for $\oldDelta n \in 2\Integers$, the Bethe equations
\eqref{eq:xxxbethe} are satisfied and the cuts $(a_0,b_0)$ and
$(-a_0,-b_0)$ have mode numbers $\oldDelta n/2$ and $-\oldDelta n/2$
respectively.%
\footnote{The sign in \eqref{eq:pitilde_discont} depends on the choice
of $a_0$, $b_0$ and $\varepsilon$. For $\Re(a_0)>0$, $\Im(a_0)$
sufficiently small and $\varepsilon>0$, the positive sign holds on the
left, the negative sign on the right cut.}
Since $\elltPi (0) = \ellK (q)$, $p_0 (z)$ has a simple pole
at $z=0$ with residue
\[
  p_0 (z) = - \oldDelta n \, a_0 \ellK(q)\,\frac{1}{z}+\ldots
\label{eq:b_res0}
\]
For $b_0 = \bar a_0$, $q=1-\exp(4i\arg a_0)$ lies on
a unit circle centred at $1$
\footnote{Conventionally, the elliptic integrals have
a branch cut on the real axis for $q\geq 1$.
The unit circle crosses the cut at $q=2$ 
and for $\Im q<0$ 
one has to analytically continue the elliptic integrals suitably.
Alternatively one can apply the Landen transformation used in 
\protect\cite{Beisert:2003ea} which maps 
the unit circle to the interval $[0,1]$ where no
branch cut is encountered.}
and hence depends only on the argument of $a_0$,
not on its modulus.
As a result, one real degree of freedom remains after
imposing the condition $\res_0 p_0 (z) = 1/2$, which corresponds to
the filling of the two symmetric cuts.

\subsection{Construction of the General Two-Cut Solution}

The solution \eqref{eq:p0} can be generalised to the non-symmetric
case by forming the composition $p_0 \circ \mu$, where $\mu$ is a
M\"obius transformation
\[
  \mu : \bar{\Complex} \rightarrow \bar{\Complex}, \quad x \mapsto
  z = \mu (x) = \frac{t x + u}{r x + s} \,.
\label{eq:moebius}
\]
It turns out that transforming the symmetric solution this way
preserves enough of its structure while providing sufficient freedom
for constructing arbitrary two-cut solutions. By suitably choosing the
M\"obius transformation $\mu$ and the modulus $q$, one can map any
four points $\left\{ a,b,c,d \right\} \in \bar{\Complex}$ onto the
branch points $\left\{ a_0, b_0, -a_0, -b_0 \right\}$, i.e.\ one can
construct a function with any four branch points $\left\{ a,b,c,d
\right\}$ by solving the system of equations
\[
  \mu(a) = a_0 \,, \quad
  \mu(b) = b_0 \,, \quad
  \mu(c) = -a_0 \,, \quad
  \mu(d) = -b_0
\label{eq:bps}
\]
for $a_0$, $b_0$ and the parameters of the transformation $\mu$. In
fact, the equations \eqref{eq:bps} do not completely fix $a_0$, $b_0$
and $\mu$. However, only solutions whose cuts are symmetric under
reflection about the real axis are physical. Hence it is
useful to set $b_0 = \bar{a}_0$ and restrict oneself to real M\"obius
transformations (i.e.\ $s,t,u,r \in \Reals$), which map complex
conjugate pairs to complex conjugate pairs.

Applying the transformation to $p_0$ directly would move the pole from
$z = 0$ to $x = \mu^{-1}(0) = u/t$. In order to have the pole at $x=0$
in the transformed function, it has to be moved to $z = \mu(0) = u/s$
in the original function $p_0$. This can be achieved by adding a term
to $p_0$
\[
\tilde p_0(z):=p_0(z)
  -\sqrt{\frac{a_0^2 (b_0^2 - z^2)}{b_0^2 (a_0^2 - z^2)}}\,
  \frac{a_0^2 - z^2}{a_0^2}\, \frac{\oldDelta n \, a_0 \ellK(q) u}
        {z(zs-u)} \,.
\]
The prefactor is necessary for retaining the structure of
cuts of the original function without introducing additional poles at
$z = \pm a_0$. The function to be transformed now reads
\<
  \tilde{p}_0 (z) \eq - \frac{\oldDelta n}{a_0 z (zs-u)}
  \sqrt{\frac{a_0^2 (b_0^2 - z^2)}{b_0^2 (a_0^2 - z^2)}}
\nl
 \left( u (a_0^2 - z^2) \ellK(q) + a_0^2
  (zs-u) \ellPi \Bigl( \frac{q z^2}{z^2 - a_0^2}\mathpunct{\Big|}q \Bigr) \right)
  \,,
\>
and the candidate for the generalised solution is
\[
  p := \tilde{p}_0 \circ \mu + C\,.
\label{eq:p_def}
\]
The constant $C$ must be chosen such that $p(\infty) = 0$,
therefore it has to be defined as
\[
  C := - \tilde{p}_0 \circ \mu (\infty) = - \tilde{p}_0 \left( t/r
  \right) \,.
\]
The function $p$ has branch points $\left\{ a,b,c,d \right\} =
\mu^{-1}(\left\{ a_0, b_0, -a_0, -b_0 \right\})$ and branch cuts
$\contour{C}_1 := (a,b)$, $\contour{C}_2 := (c,d)$. As can be verified
directly by computing the derivative of \eqref{eq:p_def}, $\ud p$ is
of the form \eqref{eq:dp_ansatz}. The following sections will show
that the function $p$ indeed meets all requirements for being a valid
quasi-momentum while providing maximal freedom in the choice of
physical parameters.

\subsection{A-periods and Mode Numbers}

For being a valid quasi-momentum, $p$ must have vanishing A-periods
and integral mode numbers (B-periods), as was established in
\secref{sec:spectral_curve}. Because the function \eqref{eq:p_def} is
single-valued by construction, the integrals of $\ud p$ over A-cycles
vanish, as long as one chooses the cuts $\contour{C}_1$ and
$\contour{C}_2$ of $p$ between the branch points $(a,b) =
\mu^{-1}(\left\{ a_0, b_0 \right\})$ and $(c,d) = \mu^{-1}(\left\{
-a_0, -b_0 \right\})$ in a way that they do not cross each other.
Since $p$ vanishes at infinity, its mode numbers $n_1$ and $n_2$ are
(cf.\ \eqref{eq:Bj})
\<
  2 \pi n_1 \earel{:=} p_0 (x+\epsilon) + p_0 (x-\epsilon)
\nln\eq
    \tilde p_0 (\mu(x+\epsilon)) + \tilde p_0 (\mu(x-\epsilon))  + 2 C
\nln\eq 2 C + \pi \oldDelta n , \qquad x \in \contour{C}_1 ,
\label{eq:n1}
\>
and similarly
\[
  2 \pi n_2  :=p_0 (x+\epsilon) + p_0 (x-\epsilon) = 2 C - \pi \oldDelta n,
\qquad x \in \contour{C}_2 .
\label{eq:n2}
\]
For obtaining physical solutions, the parameters of the function $p$
must be chosen such that $n_1$ and $n_2$ are integer numbers.

\subsection{Energy and Momentum}

The residue of the pole as well as the total energy and the total
momentum of a solution $p$ are encoded in the expansion of $p(x)$ at
$x=0$. Performing this expansion, comparing with \eqref{eq:p_exp_0}
and demanding that the residue of the pole be $1/2$ yields the
equations
\begin{align}
  \frac{1}{2} & = \frac{\oldDelta n \,s^2 W(u/s)}{b_0 (st - ru)} \ellK(q)
  \,, \label{eq:solres0}\\
  P & = \frac{\oldDelta n\, a_0 s^2 W(u/s)}{b_0 (a_0^2 s^2 - u^2)} \Biggl( \frac{a_0 s}{u} \elltPi(u/s) \nn\\
  & \mspace{60mu} - \frac{u^3 \left( b_0^2 r s - t u \right) + a_0^2 s \left( r u^3 + b_0^2 s^2 ( s t - 2 r u ) \right)}{a_0 u \left( s t - r u \right) \left( b_0^2 s^2 - u^2 \right)} \ellK (q) \Biggr) - C , \nn\\
  \tilde{E} & = \frac{\oldDelta n (st-ru) s^2 W(u/s)}{b_0
  (a_0^2 s^2 - u^2) (b_0^2 s^2 - u^2)} \cdot \nn\\
  & \mspace{60mu} \cdot \Biggl( b_0^2 \ellE(q) - \frac{a_0^4 b_0^2
  s^4 + b_0^2 u^4 + a_0^2 (b_0^4 s^4 - 4 b_0^2 s^2 u^2 + u^4)}{2
  (b_0^2 s^2 - u^2) (a_0^2 s^2 - u^2)} \ellK(q) \Biggr) \,,
\end{align}
where
\[
  W (z) := \frac{b_0 (a_0^2 - z^2)}{a_0} \sqrt{\frac{a_0^2 (b_0^2 -
  z^2)}{b_0^2 (a_0^2 - z^2)}} = \pm \sqrt{(a_0^2
  - z^2) (b_0^2 - z^2)} \,.
\label{eq:rootform}
\]

\subsection{Fillings}

The partial filling fractions $\alpha_1, \alpha_2$ can be obtained by
finding the total filling $\alpha$ with the help of
\eqref{eq:p_exp_inf},
\[
  \alpha = \frac{1}{2} - \frac{\oldDelta n}{r^2 W(t/r)} \left(
  \frac{(a_0^2 r s - t u) (b_0^2 r s - t u)}{b_0 (st - ru)} \ellK
  (q) - b_0 (st -ru) \ellE (q) \right) \,,
\]
and using $\alpha = \alpha_1 + \alpha_2$ together with $n_1 \alpha_1 +
n_2 \alpha_2 = P/2\pi$ \eqref{eq:sum_nl_al}.

Alternatively, $\alpha_1$ and $\alpha_2$ can be calculated directly by
performing the contour integration \eqref{eq:fillingcurve} over $p(x)$
along A-cycles around the cuts. This is done explicitly in \appref{app:int_alphas},
the result is
\[
\begin{split}
  \alpha_{1,2} = \frac{\oldDelta n}{b_0 \pi} \Biggl( &
  \frac{b_0^2 \pi (st-ru)}{2 r^2 W (t/r)} \ellE(q)
  + \frac{\pi s^2}{2 (st-ru)} \biggl( W (u/s) - \frac{(a_0^2 -
  \frac{tu}{rs}) (b_0^2 - \frac{tu}{rs})}{W (t/r)} \biggr) \ellK(q)
  \\
  & \pm \frac{b_0 u}{t} \ellE(q) \ellK(q)
  \mp \frac{a_0^2 b_0 s^2}{u t} \ellK^2(q)
  \pm \frac{a_0^2 (b_0^2 s^2 - u^2) s}{b_0 u (st - ru)} \ellK(q)
  \elltPi (u/s) \\
  & \pm \frac{b_0 r (st-ru)}{r^2 t - t^3} \ellE(q)
  \elltPi (t/r)
  \pm \frac{a_0^2 r (a_0^2 r s - tu) (b_0^2 r s - tu)}{b_0 t
  (st-ru) (t^2 - a_0^2 r^2)} \ellK(q) \elltPi (t/r) \Biggr)
\end{split}
\label{eq:alpha12}
\]
Provided that $\res_0 p(x) = 1/2$ holds (which can be implemented by
imposing \eqref{eq:solres0}), one can verify that indeed
$\alpha_1 + \alpha_2 = \alpha$ and $n_1 \alpha_1 + n_2 \alpha_2 =
P/2\pi$, as required by \eqref{eq:p_exp_inf2} and
\eqref{eq:sum_nl_al}.

\subsection{Solving for Physical Parameters}

The parameters of the solution \eqref{eq:p_def} are $a_0$,
$b_0$, $\oldDelta n$ and the three parameters of the M\"obius
transformation $\mu$. The solution $p$ is only physical if these
parameters are adjusted such that all physicality conditions are
satisfied. Namely, the mode numbers \eqref{eq:n1} and \eqref{eq:n2}
must be integral, the filling fractions \eqref{eq:alpha12} must be
real and \eqref{eq:solres0} must hold. The correct behaviour at
infinity \eqref{eq:p_exp_inf} is provided by construction.

Setting $b_0 = \bar{a}_0$ and restricting to real M\"obius
transformations results in complex conjugate branch points, which
guarantees that the filling fractions are real. Further, equation
\eqref{eq:solres0} can be solved for $t$ (or $r$, alternatively):
\[
  t = \frac{r u}{s} - \frac{2 \oldDelta n s}{b_0}\, W(u/s) \ellK(q) \,.
\]
Imposing these constraints leaves the complex $a_0$, the real
$\oldDelta n$ and two real parameters of $\mu$ as free parameters.
Taking into account that the curve $p$ is invariant under the
rescaling $\left\{ a_0, t, u \right\} \rightarrow \lambda \left\{ a_0,
t, u \right\}$, $\lambda \in \Reals$, only four real parameters remain.
They correspond to the freedom in the choice of the mode numbers
$n_1$, $n_2$ and the filling fractions $\alpha_1$, $\alpha_2$. This is
consistent with the result of \cite{Kazakov:2004qf}, according to
which the algebraic curve with $k$ cuts has $2k$ parameters: $k$
discrete mode numbers and $k$ continuous fillings.

The expressions for the mode numbers \eqref{eq:n1} and \eqref{eq:n2}
can be easily solved for $\oldDelta n$, yielding
\[
  \oldDelta n = n_1 - n_2 \,.
\]
The physicality constraints cannot be solved analytically for the
remaining parameters, hence one has to resort to numerical methods for
finding explicit solutions $p$ for given mode numbers and filling
fractions (or other sets of physical quantities, such as energy
$\tilde{E}$ and total filling $\alpha$).

\subsection{Finding the Cut Contours and the Density}
\label{sec:condensates}

For a given solution $p$, the physical contours of the two branch cuts
are determined by the condition that $\rho(x) \ud x$ be real. Together
with the relation \eqref{eq:rho} between the density and the
quasi-momentum, this condition yields a first-order differential
equation for the cut contours. The physical cut contours can be
obtained by numerically integrating this equation. Alternatively, one
can obtain an explicit expression for the integrated density, similar
to \eqref{eq:onecut_Lambda} in the one-cut case. The integral of the
density reads
\[
  \int _{x_*} ^{x} \rho (y) \ud y
  = \frac{1}{\pi i} \int _{x_*} ^x \bigl( p(y) - \pi n_j \bigr) \ud y
  = \frac{1}{\pi i} \bigl( \Lambda(x) - \Lambda(x_*) \bigr) + i n_j
  (x - x_*) \,,
\label{eq:rho_int}
\]
where $\Lambda (x)$ is the integral of the two-cut quasi-momentum
$p(x)$ and $n_j$ is the mode number of the respective cut. The
integral $\Lambda$ can be calculated analytically and is given in
\appref{app:int_density}. On the physical contour, the
expression \eqref{eq:rho_int} must be real when $x_*$ is one of the
branch points.

Note that the function $\Lambda(x)$ can also be used for obtaining
initial positions of Bethe roots for the computation of discrete Bethe
root distributions. Namely, when one wants to approximate a given
two-cut spectral curve $p$ with a solution to the discrete Bethe
equations on a chain of length $L$, then this solution has to consist
of $M=\alpha L$ roots. $M_j=\alpha_j L$ of these roots have mode number
$n_j$ and must lie on or close to cut $\contour{C}_j$, $j=1,2$. Taking
the solutions $x_k$ to the equations
\[
  \frac{1}{\pi i} \bigl( \Lambda(x_k) - \Lambda(x_{*,j}) \bigr) + i
  n_j (x_k - x_{*,j}) = \frac{k-1/2}{L} \,, \quad k=1,\dots,M_j \,,
  \quad j=1,2\label{eq:discretd}
\]
as initial positions for the Bethe roots reproduces the density
function on the respective classical cut well and thus gives a good
chance of numerically finding an approximating solution of the Bethe
equations.
Of course, the same method can be employed for the
one-cut solution, using the respective integral
\eqref{eq:onecut_Lambda}. This way,
the solutions shown in \figref{fig:shortstring,fig:longstring} were
obtained.

In \secref{sec:nummethod}, we describe another simple way for obtaining
discrete distributions of Bethe roots which become exact in the
limit of large length $L$ and fixed number of roots $M$. They can be easily
constructed for any number of cuts and we use them to obtain numerical
solutions of the Bethe equations in \secref{sec:numsols}. However
\eqref{eq:discretd} gives a better approximation when $L$ is not sufficiently large.

\section{Moduli Space for Consecutive Mode Numbers and Stability}
\label{sec:consecutive}

In the last section, the general two-cut spectral curve has been
constructed. For given mode numbers $n_1$ and $n_2$, the moduli space
of this solution is two-dimensional. It can be parametrised by the
two fillings $\alpha_1$ and $\alpha_2$, by the total filling $\alpha$
and the energy $\tilde{E}$ or by any other pair of physical
quantities. In this section, the moduli space of configurations with
two \emph{consecutive} mode numbers is investigated. It turns out that
this space has a very rich structure and even is globally connected,
that is the spaces of different pairs of mode numbers are
interconnected. Some features of the moduli space for non-consecutive
mode numbers are presented in the following section.

\subsection{General Features of the Moduli Space}

As discussed in \secref{sec:onecut}, a single cut with mode
number $n\indup{c}$ and a small filling is centred around $1/(2\pi
n\indup{c})$. When the filling of the cut increases, the neighbouring
excitation points with mode numbers $n\indup{c} \pm 1$ get attracted
by the cut, until the excitation $n\indup{c} +1$ (for positive
$n\indup{c}$) collides with the cut when the absolute density at its
centre reaches unity. The same happens when the excitation $n\indup{c}
+ 1$ has a small but finite filling, i.e.\ when it is replaced by a
small cut. As will be shown in more detail below, larger cuts in
general attract smaller cuts when their filling is increased.

\begin{figure}[p]
\centering
  \includegraphics{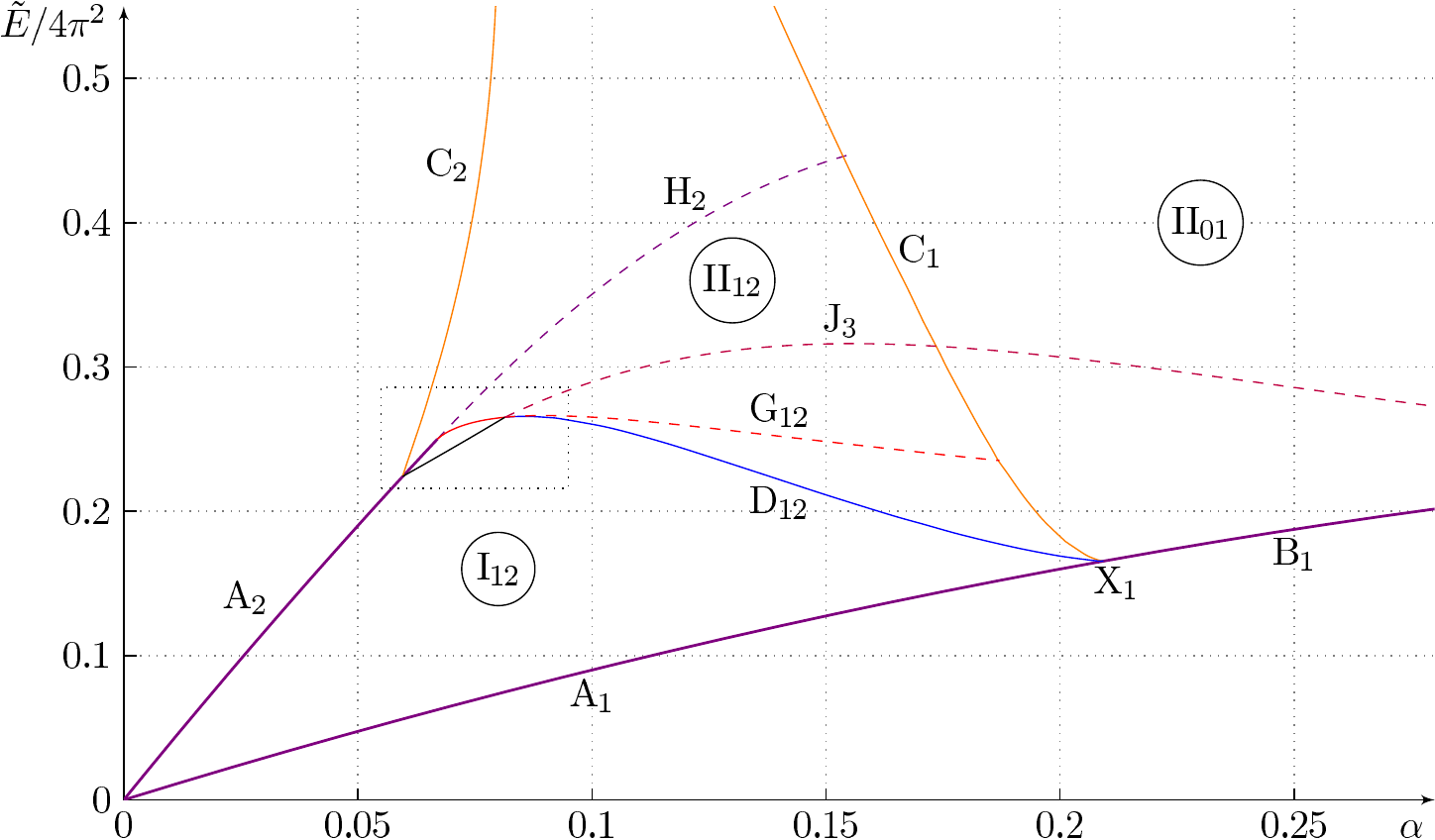}
  \caption[Moduli space for mode numbers $n_1 = 1$, $n_2 = 2$ in the
  $(\alpha,\tilde{E})$ plane]{Moduli space of configurations with
  mode numbers $n_1 = 1$, $n_2 = 2$, shown in the
  $(\alpha,\tilde{E})$ plane. Regions of the space are named by
  roman numbers $\mathrm{I}$-$\mathrm{III}$, while lines are
  assigned Latin letters $\mathrm{A}$-$\mathrm{J}$. Special points
  have labels $\mathrm{X}$-$\mathrm{Z}$. The dotted area is
  magnified in \figref{fig:12_eva_magnif}. The moduli spaces of
  higher mode numbers look very similar,
  cf.\ \protect\figref{fig:23_eva}.}
\label{fig:12_eva}
\end{figure}

\begin{figure}[p]
\centering
  \includegraphics{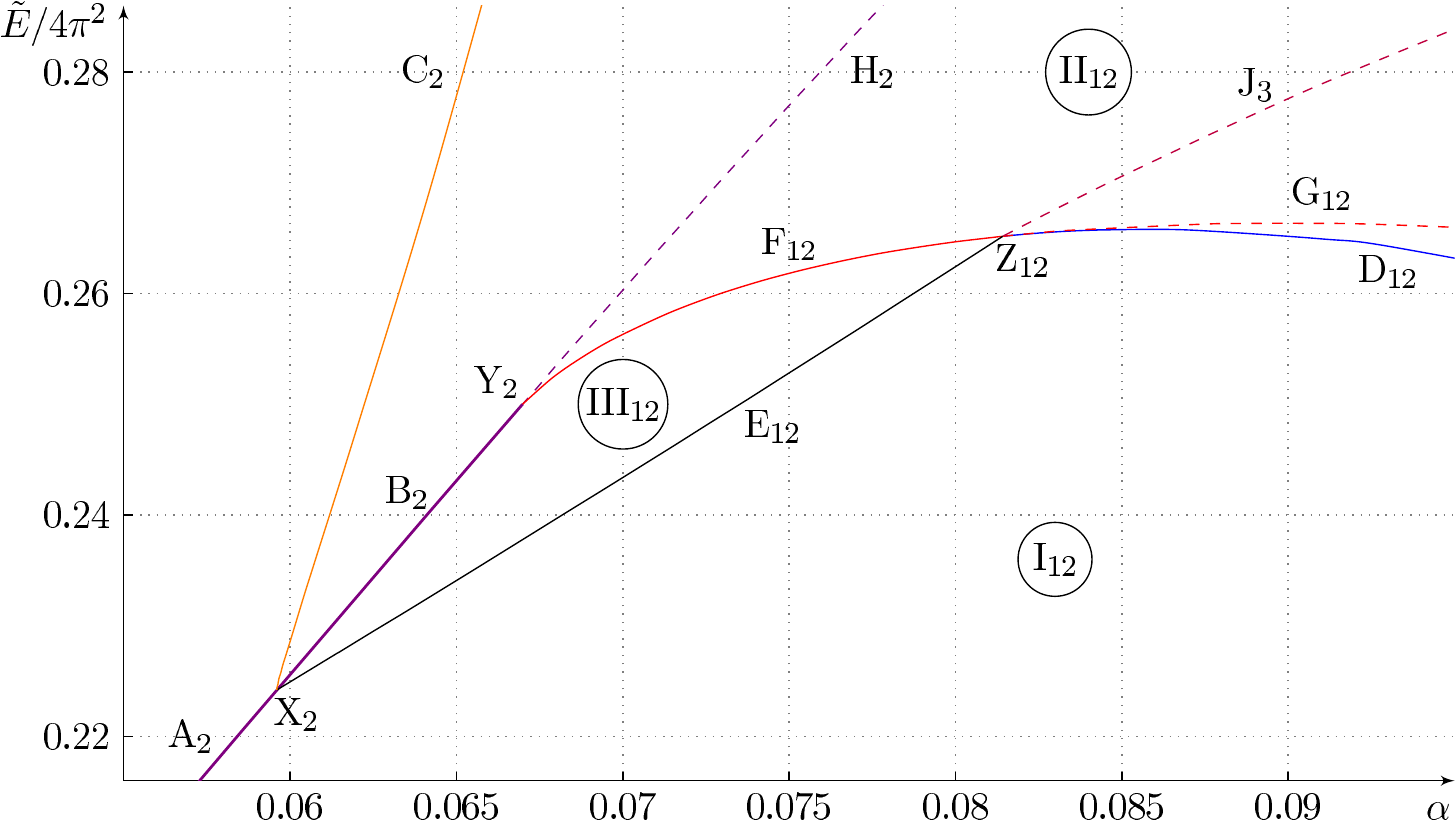}
  \caption{Magnification of the dotted area in \protect\figref{fig:12_eva}}
\label{fig:12_eva_magnif}
\end{figure}

\begin{figure}[tbp]
\centering
  \includegraphics{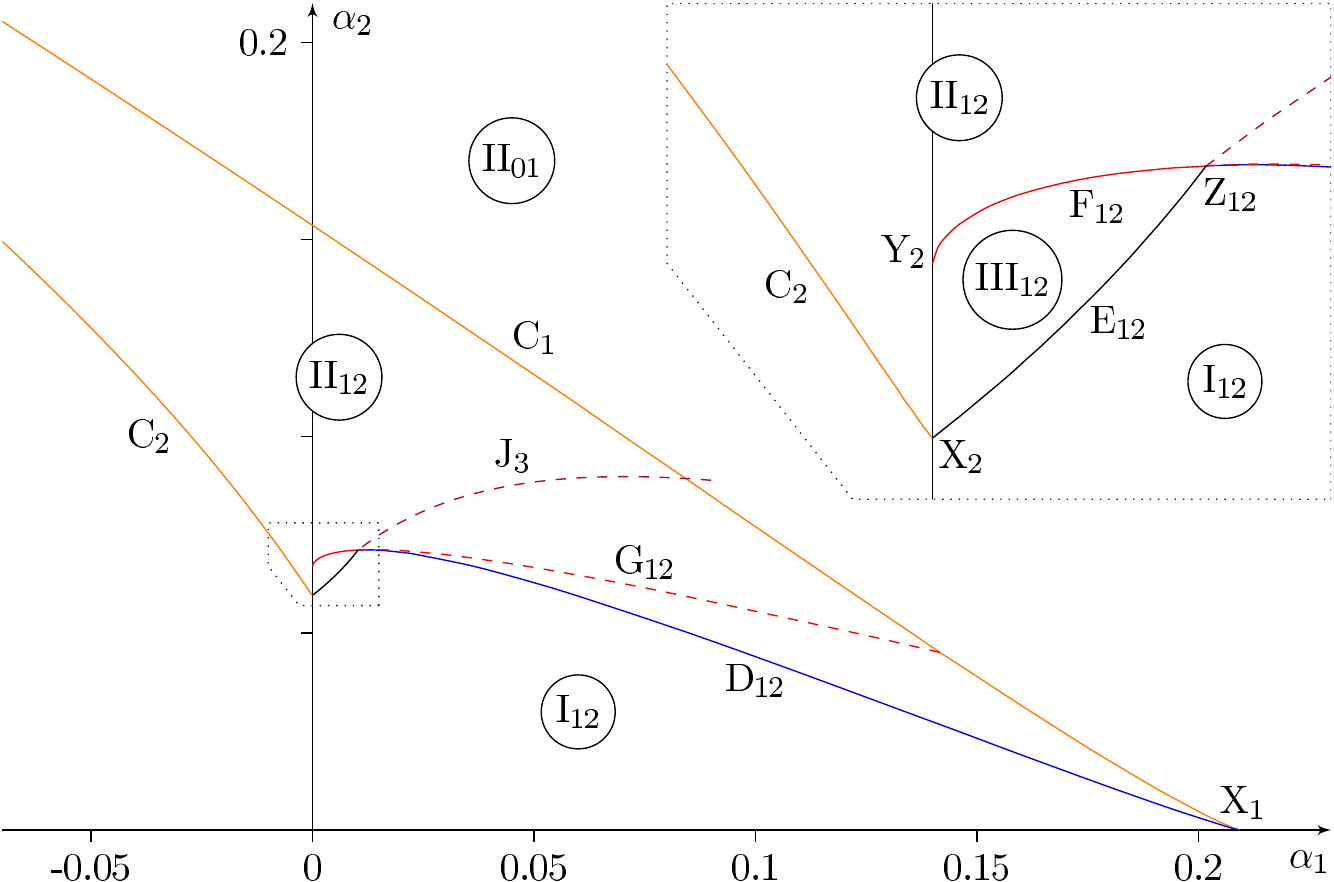}\\
  \caption[Moduli space for mode numbers $n_1 = 1$, $n_2 = 2$ in the
  $(\alpha_1,\alpha_2)$ plane]{The same moduli space as in \protect\figref{fig:12_eva},
shown in the $(\alpha_1, \alpha_2)$ plane. The
  lines $\mathrm{A}_1$, $\mathrm{B}_1$, $\mathrm{A}_2$ and
  $\mathrm{H}_2$ lie on the axes.}
\label{fig:12_aa}
\end{figure}

\begin{figure}[p]
\centering
  \includegraphics{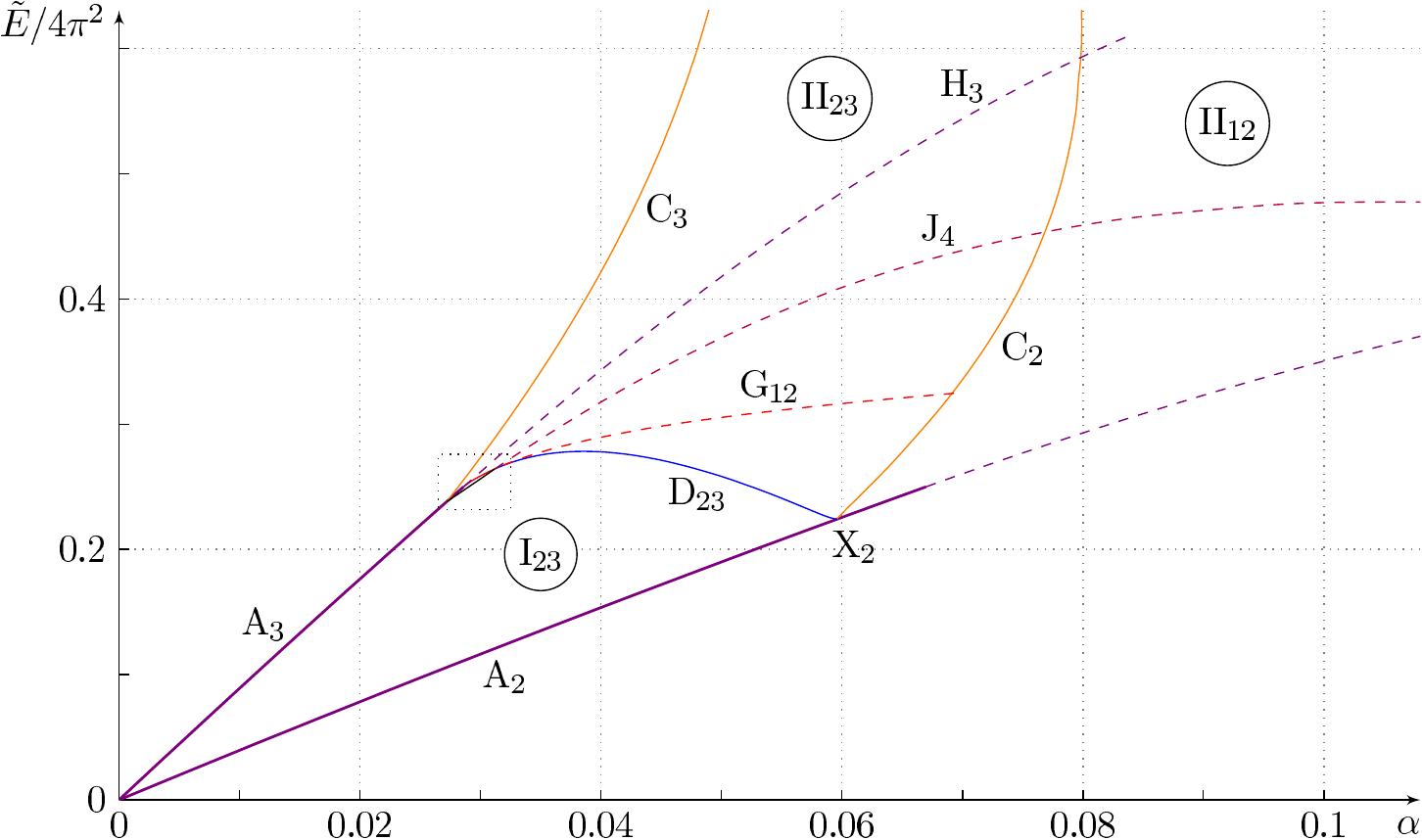}
  \caption[Moduli space for mode numbers $n_1 = 2$, $n_2 = 3$ in the
  $(\alpha,\tilde{E})$ plane]{Moduli space of configurations with
  mode numbers $n_1 = 2$, $n_2 = 3$, shown in the
  $(\alpha,\tilde{E})$ plane. The dotted area is magnified in
  \protect\figref{fig:23_eva_magnif}. Structurally, the moduli spaces of higher
  pairs of mode numbers like this one are very similar to the case
  $n_1=1$, $n_2=2$ shown in \protect\figref{fig:12_eva}. One can already
  see that the part of the space shown here is connected to the part
  shown in \protect\figref{fig:12_eva}.}
\label{fig:23_eva}
\end{figure}

\begin{figure}[p]
\centering
  \includegraphics{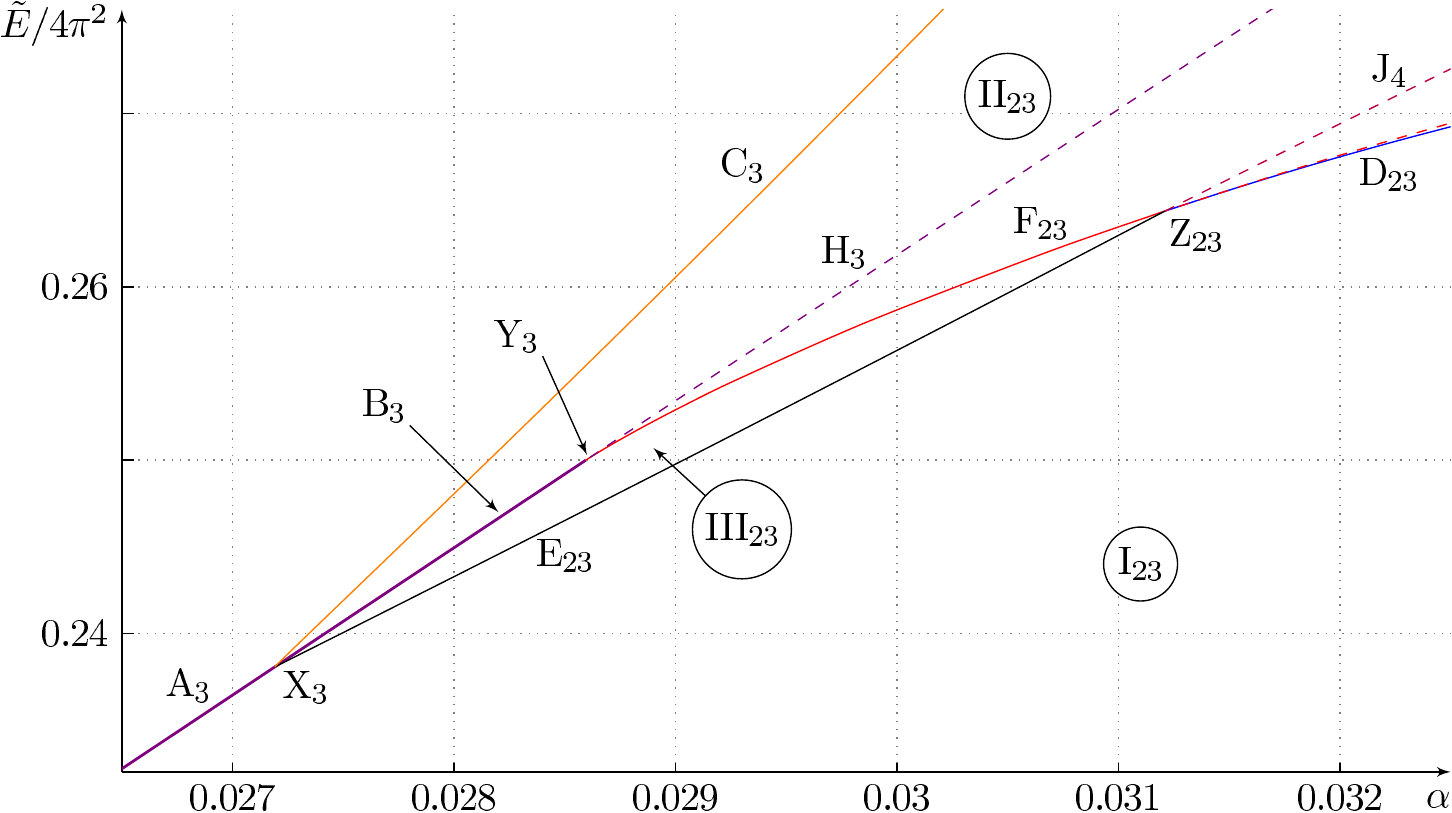}
  \caption{Magnification of the dotted area in \protect\figref{fig:23_eva}}
\label{fig:23_eva_magnif}
\end{figure}

To begin with, the moduli spaces of configurations with two
consecutive mode numbers can be divided into several regions, as is
shown in \figref{fig:12_eva,fig:12_eva_magnif,fig:12_aa,fig:23_eva,fig:23_eva_magnif}
 for the cases of $n_1 = 1$, $n_2 = 2$ and $n_1 = 2$, $n_2 = 3$. Solutions in each
region and on each line separating the regions share certain
characteristics. In the following sections, the different regions,
lines and points of the moduli space will be discussed in detail and
examples will be given. For simplicity, only non-negative mode numbers
$n_i \geq 0$ are considered, which correspond to cuts that lie to the
right of the imaginary line. The discussion for negative mode numbers
is completely analogous. The two cuts are named ``cut one'' and ``cut
two''. Unless otherwise stated, the mode number $n_1$ of cut one is
smaller than the one of cut two, $n_2 = n_1 + 1$.

In all figures that show cut contours or densities, cut one has blue
colour (thick lines), while cut two has red colour (thin lines). Plots
of cut contours in the complex plane always have an aspect ratio of
$1:1$. In all density plots, the density is shown versus the distance
from the cut's centre, measured along the cut's contour.

\subsection{Regions}
\label{sec:regions}

The moduli space of configurations with two given consecutive mode
numbers can be divided into three regions. Configurations in each
region share certain characteristics: In region $\mathrm{I}$, the two
cuts are disjoint; in region $\mathrm{II}$, the cuts join with each
other and form a condensate with four tails; in region $\mathrm{III}$,
the two cuts are disjoint but require a third, closed cut and a
condensate for stability. In the following paragraphs, the
characteristics of the different regions are described in more detail
and example solutions are shown.

\paragraph{Region $\mathrm{I}_{n,n+1}$: Two Disjoint Cuts.}

\begin{figure}[tbp]
\centering
  \subfloat[Cut contours]{\includegraphics{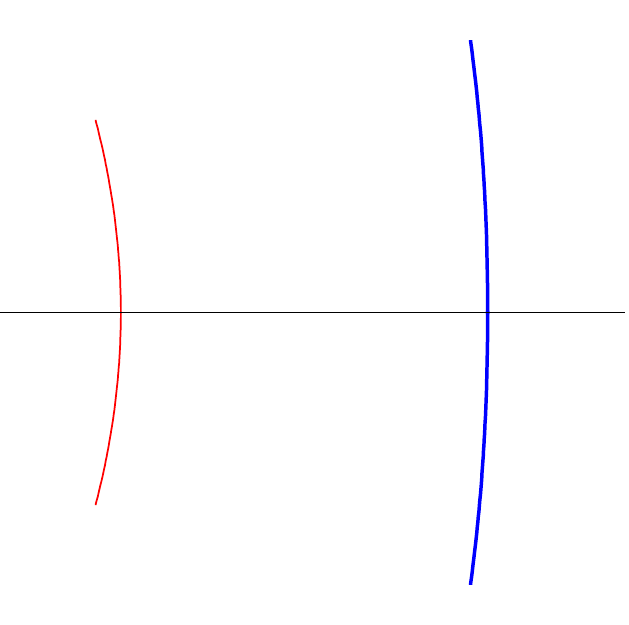}}\qquad
  \subfloat[Absolute densities along the cuts vs.\ distance from the cuts' centres (measured along the contour)]{\includegraphics{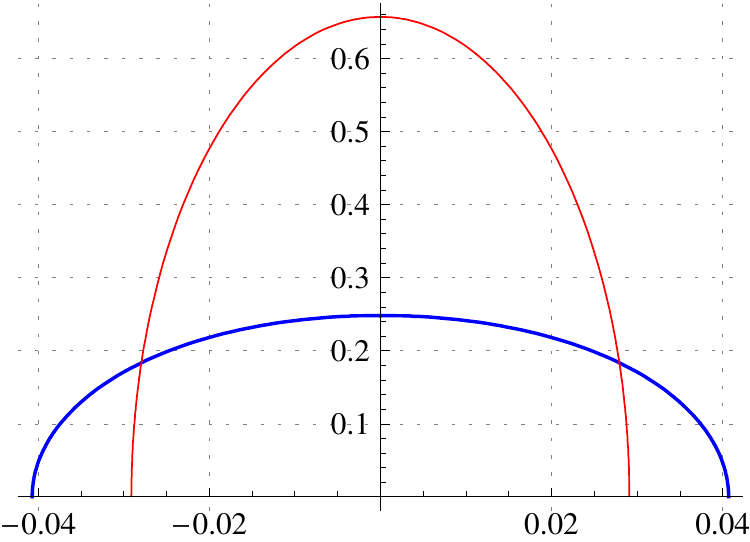}}
  \caption[Example solution in region $\mathrm{I}_{12}$.]{Cut
  contours and absolute densities of an example solution in region
  $\mathrm{I}_{12}$ (see \protect\figref{fig:12_eva}). The fillings are
  $\alpha_1 = 0.016$ (blue/thick, $n_1 = 1$) and $\alpha_2 = 0.03$
  (red/thin, $n_2 = 2$).}
\label{fig:12_region1}
\end{figure}

\begin{figure}[tbp]
\centering
  \subfloat[Cut contours]{\includegraphics{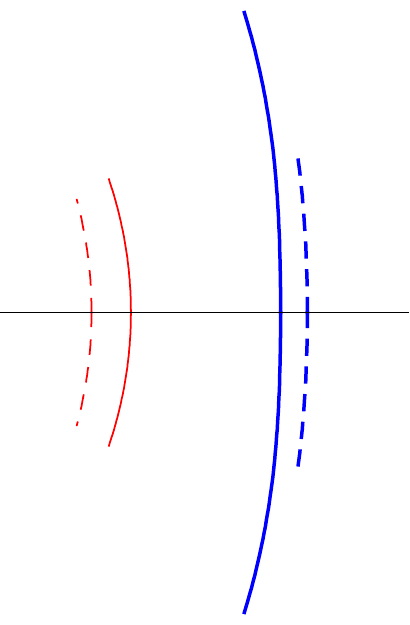}}\qquad
  \subfloat[Absolute densities along the cuts vs.\ distance from the cuts' centres (measured along the contour)]{\includegraphics{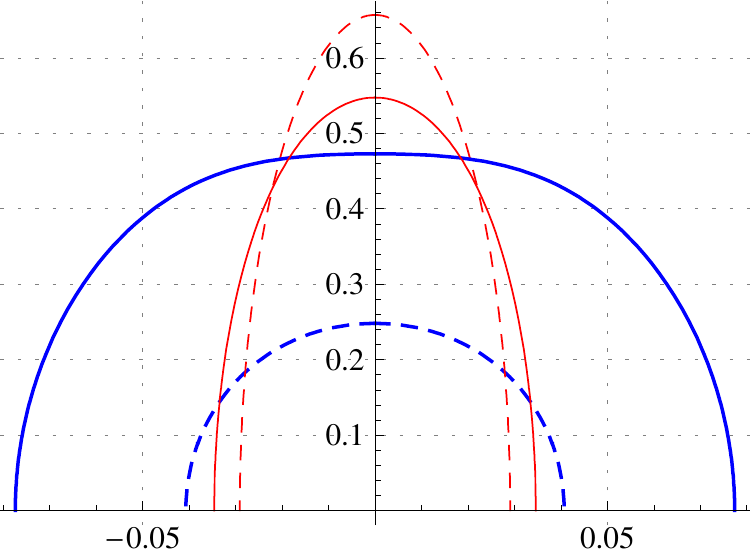}}
  \caption[Example solution in region $\mathrm{I}_{12}$, cut one has
  larger filling than in \protect\figref{fig:12_region1}]{Another
  configuration in region $\mathrm{I}_{12}$. The cut contours and
  densities of \protect\figref{fig:12_region1} are shown with dashed
  lines. Cut one has a larger filling than in
  \protect\figref{fig:12_region1} and attracts cut two ($\alpha_1 = 0.06$,
  $\alpha_2 = 0.03$).}
\label{fig:12_region1_cutonelong}
\end{figure}

\begin{figure}[tbp]
\centering
  \subfloat[Cut contours]{\includegraphics{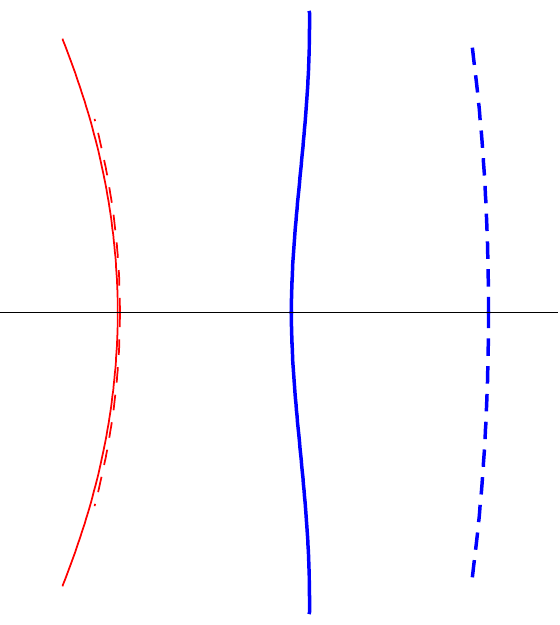}}\qquad
  \subfloat[Absolute densities along the cuts vs.\ distance from the cuts' centres (measured along the contour)]{\includegraphics{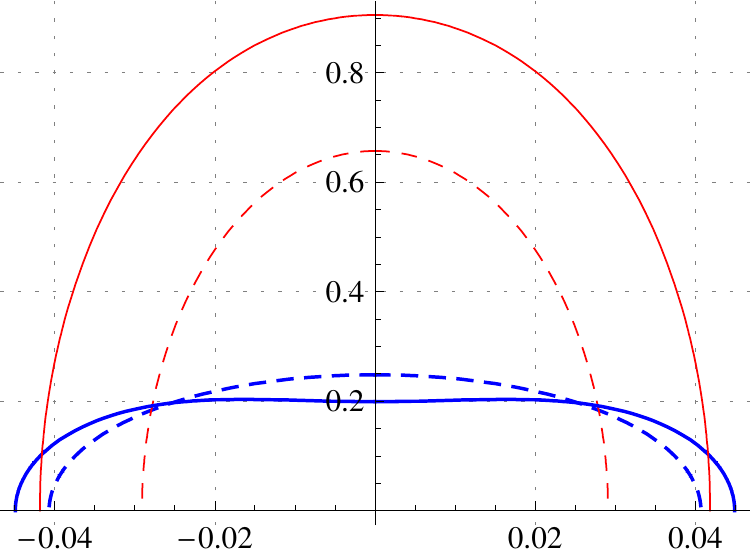}}
  \caption[Example solution in region $\mathrm{I}_{12}$, cut two has
  larger filling than in \protect\figref{fig:12_region1}]{A third
  configuration in region $\mathrm{I}_{12}$. The cut contours and
  densities of \protect\figref{fig:12_region1} are shown with dashed
  lines. Cut two has a larger filling than in
\protect\figref{fig:12_region1} and attracts cut one ($\alpha_1 = 0.016$,
  $\alpha_2 = 0.06$).}
\label{fig:12_region1_cuttwolong}
\end{figure}

In this region, the contours of the two cuts with mode numbers $n$ and
$n+1$ are disjoint, and the cut with mode number $n+1$ lies further
left (closer to the origin) than the one with mode number $n$. The
absolute density is bounded above by $1$ everywhere on both cuts. An
example is shown in \figref{fig:12_region1}. When the filling of
one of the cut increases, this cut attracts the other one, as can be
seen in \figref{fig:12_region1_cutonelong,fig:12_region1_cuttwolong}.

\paragraph{Region $\mathrm{II}_{n,n+1}$: Condensate with Four Tails.}

\begin{figure}[htpb]
\centering
  \subfloat[Cut contours]{\quad\includegraphics{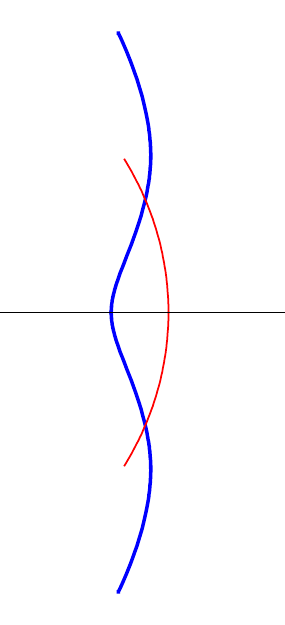}\quad}\qquad
  \subfloat[Absolute densities along the cuts  vs.\ distance from the cuts' centres (measured along the contour)]{\includegraphics{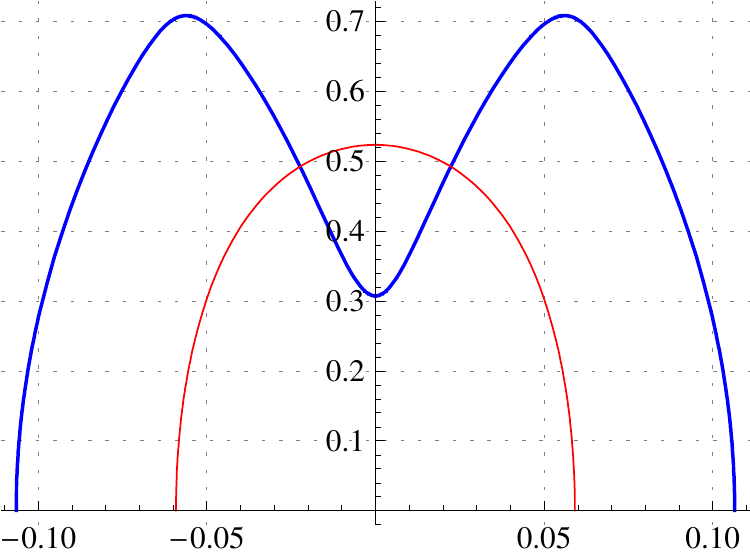}
    \label{fig:12_region3_densities}}
  \caption[Example configuration in region $\mathrm{II}_{12}$,
  without condensate cut]{Standard cut contours and absolute
  densities of an example configuration in region $\mathrm{II}_{12}$
  (see \protect\figref{fig:12_eva}). The fillings fractions are
  $\alpha_1 = 0.11$ (blue/thick, $n_1 = 1$) and $\alpha_2 = 0.05$
  (red/thin, $n_2 = 2$). This particular configuration lies below
  line $\mathrm{G}_{12}$ in the $(\alpha,\tilde{E})$ plane. The physical
  version of the configuration has a condensate and is shown in
  \protect\figref{fig:12_region3_cond}.}
\label{fig:12_region3}
\end{figure}

\begin{figure}[htpb]
\centering
  \subfloat[Cut contours]{\quad\includegraphics{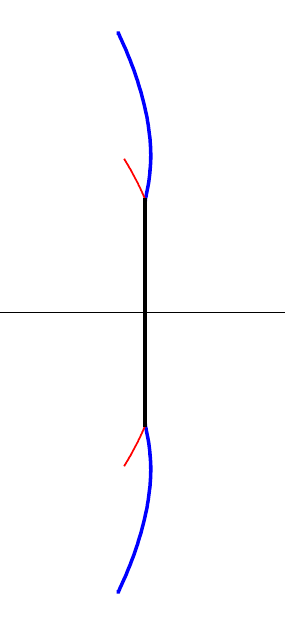}\quad}\qquad
  \subfloat[Absolute densities along the cuts  vs.\ distance from the cuts' centres (measured along the condensate and the cuts' contour)]{\includegraphics{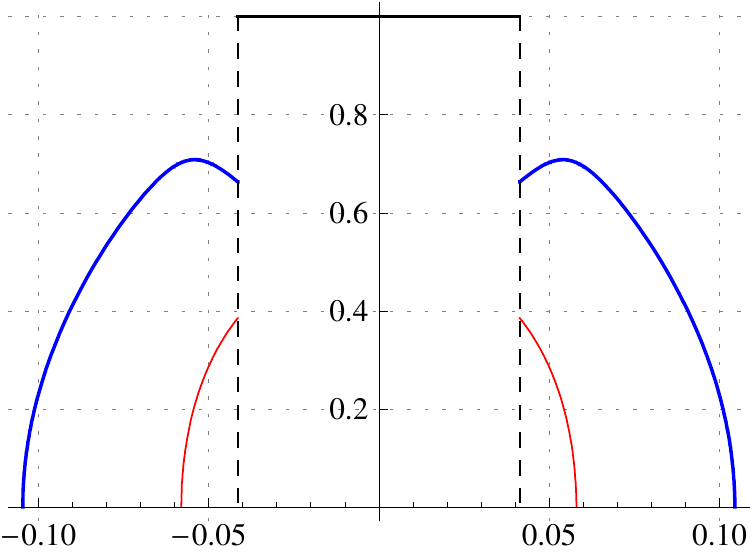}}
  \caption[Example configuration in region $\mathrm{II}_{12}$, with
  condensate cut]{Physical version of the configuration shown in
  \protect\figref{fig:12_region3}. The two common points of the standard
  contours are connected by a condensate.}
\label{fig:12_region3_cond}
\end{figure}

In this region, the standard contours%
\footnote{Here and in the following, the cut contours that are
determined through the requirement that $\rho(x)\,\ud x\in\Reals$, and
which do not involve condensate cuts will be referred to as ``standard
contours''.}
of the two cuts bend towards and cross each other. An example is shown
in \figref{fig:12_region3}.
As explained below \eqref{eq:logcut}, the sum of the densities at the
two common points of the standard contours equals $-i$ (since the
difference of the mode numbers is $n_1-n_2=-1$) and solutions in this
region allow for straight condensate cuts with constant density
$\rho\indup{c} (x) = -i$ between the two common points of the cuts.
The condensate-cut version of the configuration in
\figref{fig:12_region3} is shown in \figref{fig:12_region3_cond}.
Beyond the two common points, the remaining parts of the standard cut
contours form four tails that are attached to the ends of the
condensate. The tails of cut two (with mode number $n+1$) lie further
left than the ones of cut one (with mode number $n$).

Whenever a particular solution $p$ allows for a condensate cut, it
will be assumed that the configuration with the condensate cut is the
physical version of the solution. This assumption may not seem very
motivated at the moment, but will become more justified during the
subsequent discussion. A first sign of its verity is the fact that
there is no well-defined direction on standard cut contours that cross
each other. This can be seen as follows: According to \eqref{eq:rho},
the density $\rho$ on a standard cut contour is proportional to the
difference between the limiting values of $p(x)$ on either side of the
cut contour: $\rho(x) \sim p(x+\epsilon) -p(x-\epsilon)$.
When a standard cut contour crosses
another standard cut contour, the value of $p(x\pm \epsilon)$
on either side of the contour changes to $-p(x\pm \epsilon)+2\pi n$,
where $n$ is the mode number of the cut that is traversed;
this follows from the Bethe equations \eqref{eq:xxxbethe}.
Hence the density $\rho$ on the contour switches
its sign, and in order that the corresponding root density
$\rho(x)\,\ud x$ stays positive, the direction on the contour must
change as well. However a unique direction on each contour was assumed in
the derivation of the spectral curve and quasi-momentum
in \secref{sec:branch_cuts}. This argument makes it seem very unlikely that
configurations with crossing standard cut contours can be realised as
limits of Bethe root distributions.

\begin{figure}[tbp]
\centering
  \subfloat[Cut contours]{\includegraphics{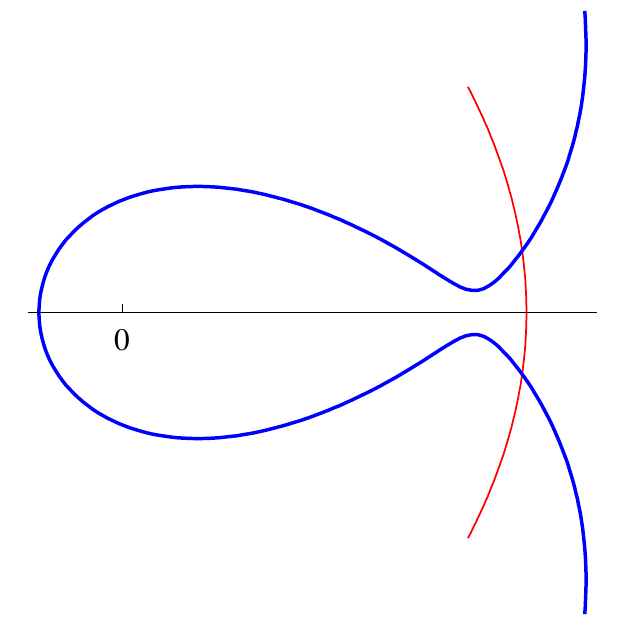}}\qquad
  \subfloat[Absolute densities along the cuts  vs.\ distance from the cuts' centres (measured along the contour)]{\includegraphics{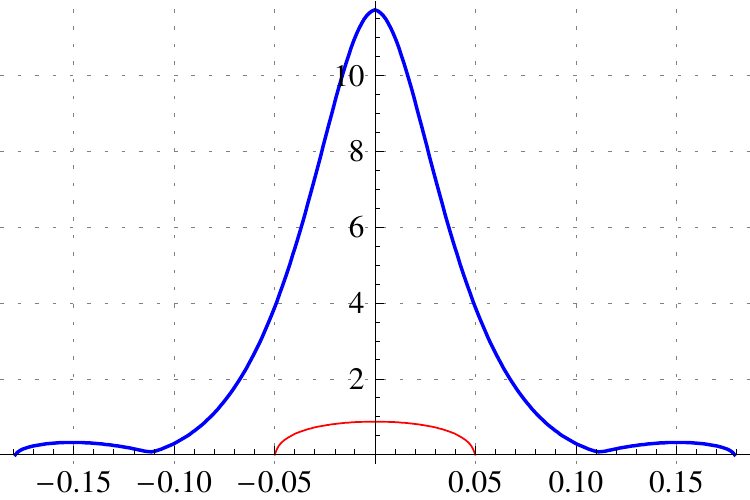}}
  \caption[Example configuration beyond line $\mathrm{G}_{12}$,
  without condensate cut]{An example configuration in region
  $\mathrm{II}_{12}$ beyond line $\mathrm{G}_{12}$ (see
  \protect\figref{fig:12_eva}). The filling fractions are $\alpha_1 = 0.032$
  (blue/thick, $n_1 = 1$) and $\alpha_2 = 0.07$ (red/thin, $n_2 =
  2$). The physical version of this solution has a condensate cut
  and is shown in \protect\figref{fig:12_region4_cond}.}
\label{fig:12_region4}
\end{figure}

\begin{figure}[tbp]
\centering
  \subfloat[Cut contours]{\includegraphics{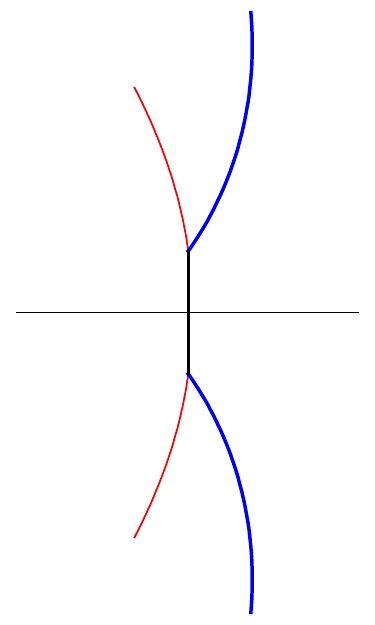}}\qquad
  \subfloat[Absolute densities along the cuts  vs.\ distance from the cuts' centres (measured along the condensate and the cuts' contours)]{\includegraphics{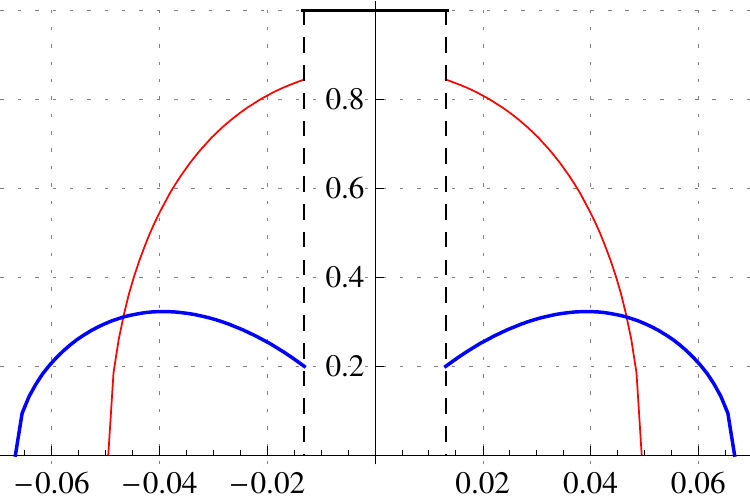}}
  \caption[Example configuration beyond line $\mathrm{G}_{12}$, with
  condensate cut]{Physical version of the solution shown in
  \protect\figref{fig:12_region4}. The part of cut one that encircles the
  origin is hidden in the condensate cut.}
\label{fig:12_region4_cond}
\end{figure}

As indicated by the dashed lines in
\figref{fig:12_eva,fig:12_eva_magnif,fig:12_aa,fig:23_eva,fig:23_eva_magnif},
region $\mathrm{II}_{n,n+1}$ can be further subdivided. Between the
lines $\mathrm{D}_{n,n+1}$, $\mathrm{G}_{n,n+1}$ and $\mathrm{C}_n$,
the standard cut contours have the usual form, as in
\figref{fig:12_region3}: Neither do they encircle the origin, nor do
they extend to infinity. In contrast, beyond line $\mathrm{G}_{n,n+1}$
the standard contour of cut one encircles the origin, as in the
example in \figref{fig:12_region4}.
Moving a contour with filling $\alpha_1$ past the pole at the origin
changes its filling to $\alpha_1' = \alpha_1 + 1$; this follows from
the expression of the filling fraction as a contour integral
\eqref{eq:fillingcurve}, the residue theorem and the fact that the
residue at $x=0$ switches its sign as the contour moves past it. As a
result, the absolute density on the standard contour typically exceeds
unity at its centre, as in \figref{fig:12_region4}. This and the
fact that the pole at the origin has the wrong sign in this
configuration are strong indications that the physical realisation of
such a solution is the one with a condensate cut, as in
\figref{fig:12_region4_cond}.

Physically, there is no qualitative difference between configurations
on either side of line $\mathrm{G}_{n,n+1}$, as their physical
versions always contain a condensate cut and thus do not see the
central part of the standard cut contours. Note that below line
$\mathrm{G}_{n,n+1}$, as in the example in
\figref{fig:12_region3}, the absolute density does not exceed unity on
the standard cut contours. Hence the stability criterion
\eqref{eq:stabcrit} is obeyed by the standard contours and no
condensate is necessary to fulfil \eqref{eq:stability}. Yet a
condensate version of the configuration is possible. Beyond line
$\mathrm{G}_{n,n+1}$ a condensate is necessary for stability, and for
continuity reasons one would assume that also below line
$\mathrm{G}_{n,n+1}$ the physical version is the one with a condensate
cut. This is another sign in favour of the assumption that if a
condensate cut is possible, it is also realised physically.

The other dashed lines $\mathrm{H}_{n,n+1}$ and $\mathrm{J}_{n,n+1}$
that further subdivide region $\mathrm{II}_{n,n+1}$ are discussed in
\secref{sec:lines} below.

\paragraph{Region $\mathrm{III}_{n,n+1}$: Two Disjoint Cuts, One with a
Condensate.}

\begin{figure}[tbp]
\centering
  \subfloat[Cut contours] {\includegraphics{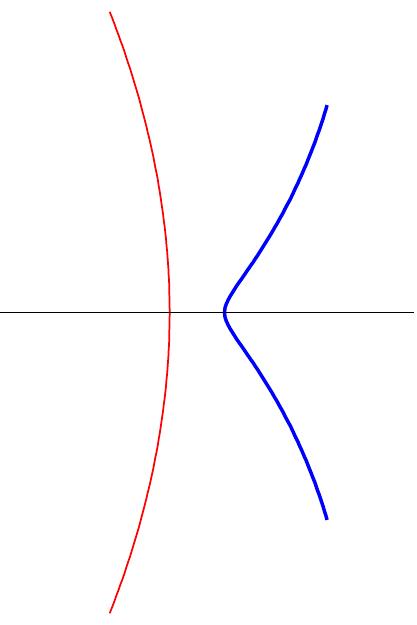}}\qquad
  \subfloat[Absolute densities along the cuts  vs.\ distance from the cuts' centres (measured along the contour)]{\includegraphics{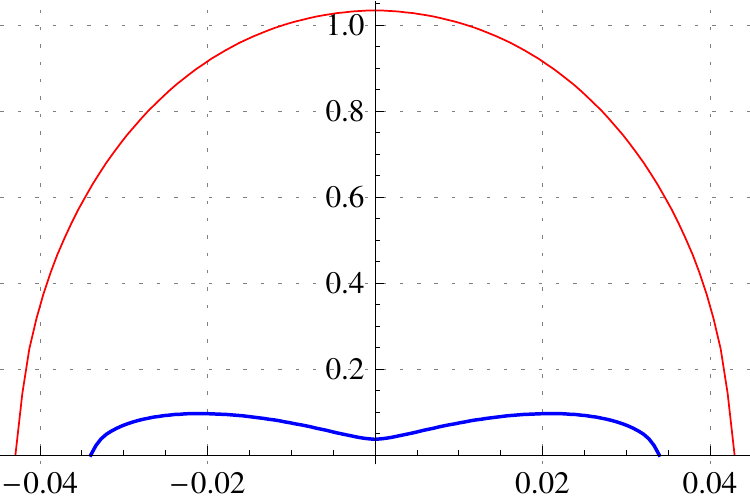}}
  \caption[Example solution in region $\mathrm{III}_{12}$, without
  condensate cut]{An example solution in region $\mathrm{III}_{12}$
  (see \protect\figref{fig:12_eva}). The fillings are $\alpha_1 =
  0.0038$ (blue/thick, $n_1 = 1$) and $\alpha_2 = 0.07$ (red/thin,
  $n_2 = 2$). The absolute density exceeds unity at the centre of
  cut two, hence the stability criterion \eqref{eq:stabcrit} is
  violated. The physical version of the configuration involves a
  condensate cut and is shown in \protect\figref{fig:12_region2_phys}.}
\label{fig:12_region2}
\end{figure}

\begin{figure}[tbp]
\centering
  \subfloat[Central part of the cut contours]{\includegraphics{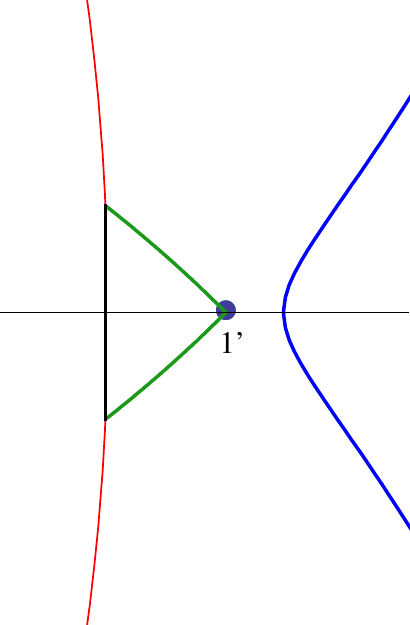}}\qquad
  \subfloat[Absolute density along cut two, the closed cut and the condensate vs.\ distance from the condensate's centre (measured along the contour)]{\includegraphics{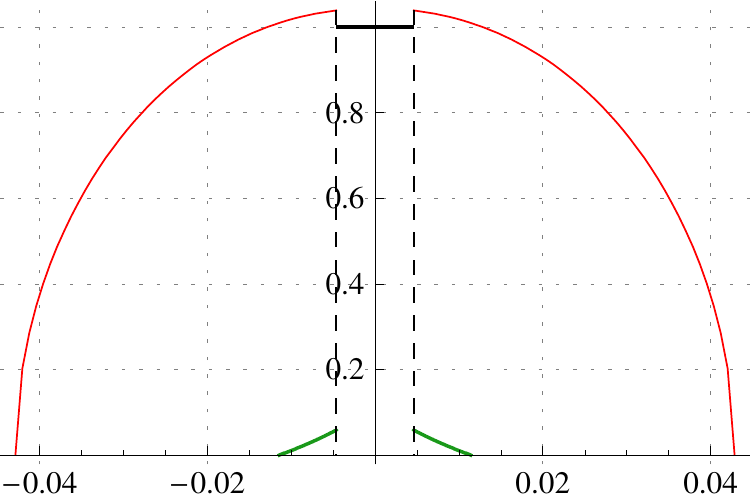}}
  \caption[Example solution in region $\mathrm{III}_{12}$, with
  condensate cut]{Physical version of the configuration shown in
  \protect\figref{fig:12_region2}. A third, closed cut forms a
  condensate with cut one, as in the one-cut case discussed in
  \protect\secref{sec:onecut_cond}. The closed cut starts and ends at
  the excitation point $1'$, which passed through cut two as the
  density at the centre of cut two reached unity.}
\label{fig:12_region2_phys}
\end{figure}

In this region, the contours of the two cuts are disjoint as in region
$\mathrm{I}_{n,n+1}$, but the absolute density exceeds unity at the
centre of cut two, i.e.\ the stability condition \eqref{eq:stabcrit} is
violated. An example configuration is shown in
\figref{fig:12_region2}.
Configurations in this region are still physical, because the same
discussion as in \secref{sec:onecut_cond} is applicable: A third,
closed cut can be added in order to form a condensate at the centre of
cut two. Since the absolute density at the centre of cut two exceeds
unity, the excitation point with mode number $n+2$ has already passed
through the contour of cut two and changed its mode number to $n$. The
closed cut originates and ends at this excitation point. The resulting
physical version of such a configuration is shown in
\figref{fig:12_region2_phys}.

\subsection{Lines}
\label{sec:lines}

In the following paragraphs, the features of configurations on the
various lines shown in \figref{fig:12_eva,fig:12_eva_magnif,fig:12_aa,fig:23_eva,fig:23_eva_magnif}
are presented and appropriate examples are
given.

\paragraph{Line $\mathrm{A}_n$: Only One Cut.}

This line represents configurations where only the cut with mode
number $n$ is present and has a filling $\alpha <
\alpha_{\mathrm{cond}}$ (cf.\ \tabref{tab:maxfill}), such that the
absolute density at the cut's centre always stays below unity. Among
all configurations with mode numbers $n$ and $n+1$ and given total
filling, the state on this line is the one with the lowest energy.
Adding a second cut with higher mode number while keeping the total
filling constant obviously increases the energy. Conversely, adding a
second cut with lower mode number while keeping the total filling
constant decreases the total energy.

\paragraph{Line $\mathrm{B}_n$: Only One Cut with a Condensate.}

This line is the continuation of line $\mathrm{A}_n$. Solutions on
this line have only one cut with mode number $n$ which has a filling
$\alpha_\mathrm{cond} < \alpha < \alpha_{\mathrm{crit}}$ and were
discussed in \secref{sec:onecut}. The absolute density at the
centre of the cut's standard contour exceeds unity; hence a second,
closed cut that starts and ends at the excitation $(n-1)'$ and forms a
condensate with the central part of the cut is required for stability,
as in the second example in \figref{fig:onebubblecuts}.

\paragraph{Line $\mathrm{C}_n$: Condensate with Two Tails.}

Line $\mathrm{C}_n$ separates region $\mathrm{II}_{n,n+1}$ from region
$\mathrm{II}_{n-1,n}$, which both consist of solutions with a
condensate cut. Solutions on line $\mathrm{C}_n$ consist of a larger
cut (cut one) with mode number $n$ and a smaller cut (cut two), whose
end points lie exactly on the standard contour of the larger cut.
Therefore, the physical version of such a configuration consists of a
condensate with one tail at each of its ends; the smaller cut is
completely hidden in the condensate. Since the density of cut two at
the cut's end points vanishes, the density of cut one at these points
must equal $-i$, as explained below \eqref{eq:logcut}. Consequently,
also the direction of cut one at these points is purely imaginary,
i.e.\ vertical. Hence, the cut contour of the condensate version of
this configuration and the density along it are smooth. Solutions on
this line were investigated by Sutherland in \cite{Sutherland:1995aa},
refer to Figure 2 in that reference for a picture of the absolute
density on the condensate and the tails.

\begin{figure}
\centering
  \subfloat[0.12]{\quad\includegraphics{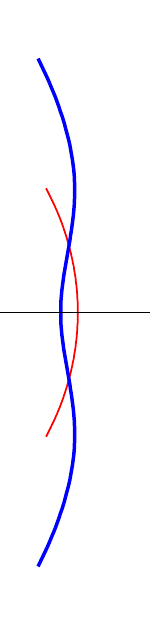}\label{fig:12_lineD_series_a}}
  \subfloat[0.13]{\quad\includegraphics{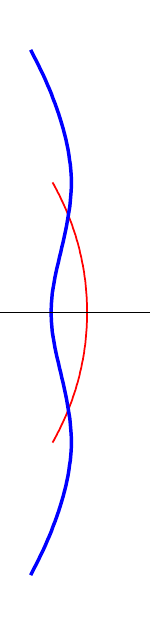}\label{fig:12_lineD_series_b}}
  \subfloat[0.14]{\quad\includegraphics{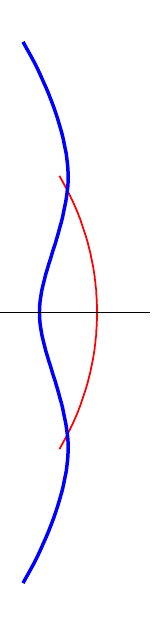}\label{fig:12_lineD_series_c}}
  \subfloat[0.1487]{\quad\includegraphics{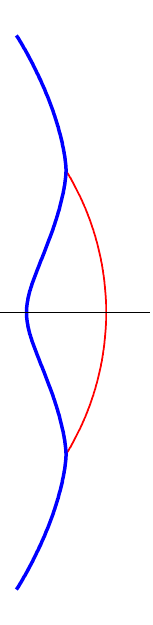}\label{fig:12_lineD_series_d}}
  \subfloat[0.1487]{\quad\includegraphics{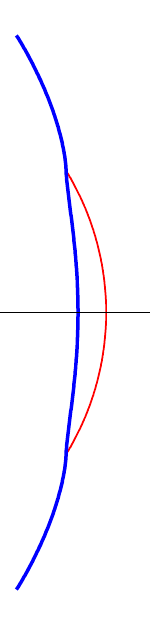}\label{fig:12_lineD_series_e}}
  \subfloat[0.155]{\quad\includegraphics{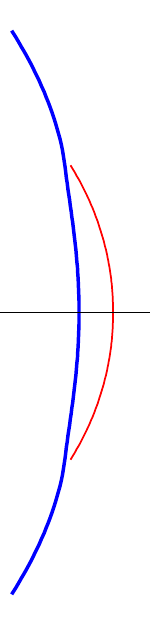}\label{fig:12_lineD_series_f}}
  \subfloat[0.165]{\quad\includegraphics{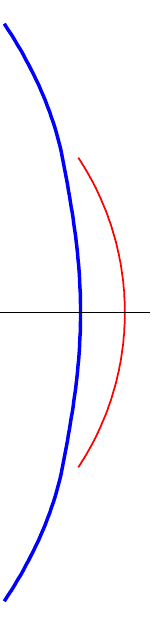}\label{fig:12_lineD_series_g}}
  \caption[Series of configurations that passes through line
  $\mathrm{C}_1$, unphysical version]{A Series of configurations
  that passes through line $\mathrm{C}_1$ in
  \protect\figref{fig:12_eva}. At all points of the series, the filling
  fraction of the small cut equals $\alpha_2 = 0.04$. The filling
  fraction $\alpha_1$ of the large cut is indicated below each plot.
  Configurations (d) and (e) are the same in terms of mode numbers,
  fillings and branch points, but, as explained in the main text,
  this solution allows for two different contours for cut one. All
  solutions have a condensate in their physical version as displayed
  in \protect\figref{fig:12_lineD_series_cond}.}
\label{fig:12_lineD_series}
\end{figure}

A series of configurations in which the branch points of cut two pass
the contour of cut one is shown in \figref{fig:12_lineD_series}.
The sequence passes line $\mathrm{C}_1$ from left to right in \figref{fig:12_eva}.
During the series, the filling fraction of cut two
stays constant, while the filling fraction of cut one increases. At
the beginning, the configuration is in region $\mathrm{II}_{12}$: The
branch points of cut two lie to the left of cut one, and the mode
number of cut two is $n_2 = 2$, while the mode number of cut one is
$n_1 = 1$ (\figref{fig:12_lineD_series_a,fig:12_lineD_series_b,fig:12_lineD_series_c}).
While the filling fraction of cut one
increases, cut two gets attracted by it and moves further right.
During this process, the cuts increasingly bend towards each other and
both become longer. At a certain value of the large cut's filling, the
end points of the small cut lie exactly on the large cut's contour,
the configuration has arrived at line $\mathrm{C}_1$ (\figref{fig:12_lineD_series_d}). One can see that the contour of cut one
is indeed vertical at the end points of cut two. The filling of cut
two stays constant during the series, but its length increases, hence
the absolute density along its contour decreases.

When the filling of cut one increases further, the small cut's branch
points move on to its other side and the configuration reaches region
$\mathrm{II}_{n-1,n}$ (\figref{fig:12_lineD_series_f,fig:12_lineD_series_g}).
While the end points of cut two pass the
large cut, multiple quantities change: According to \eqref{eq:n1}, the
mode number of a cut is $n_i = \pi p(x_*)$, where $x_*$ is a branch
point of the cut. Across a cut with mode number $n_1$, $p$ switches
sign and shifts by $2 \pi n_1$ \eqref{eq:sqrtcut}. Therefore, a cut
with mode number $n_2$ whose branch points pass another cut with mode
number $n_1$ changes its mode number to $n_2' = -n_2 + 2 n_1$.
Specifically, if $n_2=n_1+1$, as in the series in \figref{fig:12_lineD_series},
then $n_2'=n_1-1$. Further, while cut two
passes through the contour of cut one, the sign of the density
\eqref{eq:rho} along cut two changes, as shown above in the
discussion of region $\mathrm{II}_{n,n+1}$. This does not affect the
density of roots along the cut, but it switches the sign of the
filling: $\alpha_2 \rightarrow \alpha_2' = - \alpha_2$. Thirdly, the
contour of cut one and its filling change by a finite amount when cut
two passes. The filling changes from $\alpha_1$ to $\alpha_1' =
\alpha_1 + 2 \alpha_2$, which is consistent with the fact that the
total filling $\alpha_1 + \alpha_2$, as a coefficient of the expansion
of $p$ at infinity \eqref{eq:p_exp_inf}, must vary continuously during
the series. Also the total momentum $m = n_1 \alpha_1 + n_2 \alpha_2$
varies continuously during the transition from region
$\mathrm{II}_{n,n+1}$ to region $\mathrm{II}_{n-1,n}$.

\begin{figure}[tbp]
\centering
  \subfloat[Cut contours]{\includegraphics{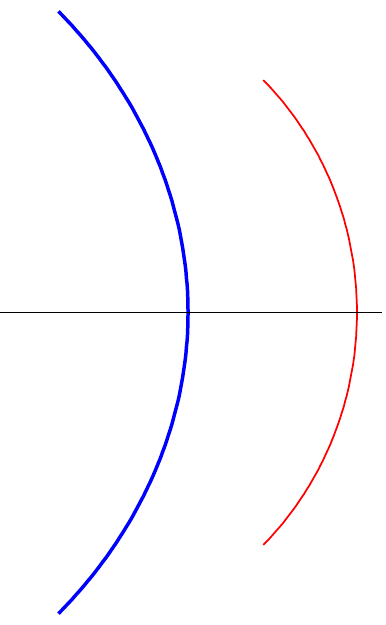}}\qquad
  \subfloat[Absolute densities along the cuts  vs.\ distance from the cuts' centres (measured along the contour)]{\includegraphics{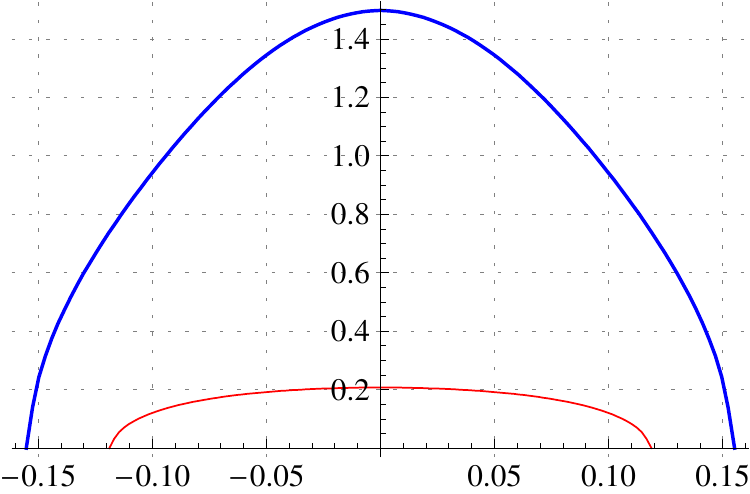}\label{fig:12_region5_densities}}
  \caption[Example solution in region $\mathrm{II}_{n-1,n}$ beyond
  line $\mathrm{C}_1$, unphysical version]{Standard cut contours and
  absolute densities of an example solution in region
  $\mathrm{II}_{n-1,n}$, beyond line $\mathrm{C}_1$ (see
  \protect\figref{fig:12_eva}). The fillings are $\alpha_1 = 0.25$ (blue/thick,
  $n_1 = 1$) and $\alpha_2 = 0.04$ (red/thin, $n_2 = 0$). The
  absolute density exceeds unity at the centre of cut one, hence
  this configuration is unstable. The physical, stable form of this
  solution is shown in \protect\figref{fig:12_region5_cond}.}
\label{fig:12_region5}
\end{figure}

\begin{figure}[tbp]
\centering
  \subfloat[Cut contours]{\includegraphics{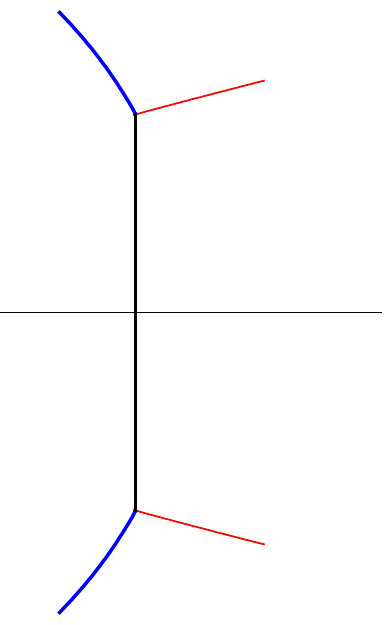}}\qquad
  \subfloat[Absolute densities along the cuts  vs.\ distance from the cuts' centres (measured along the condensate and the cuts' contours)]{\includegraphics{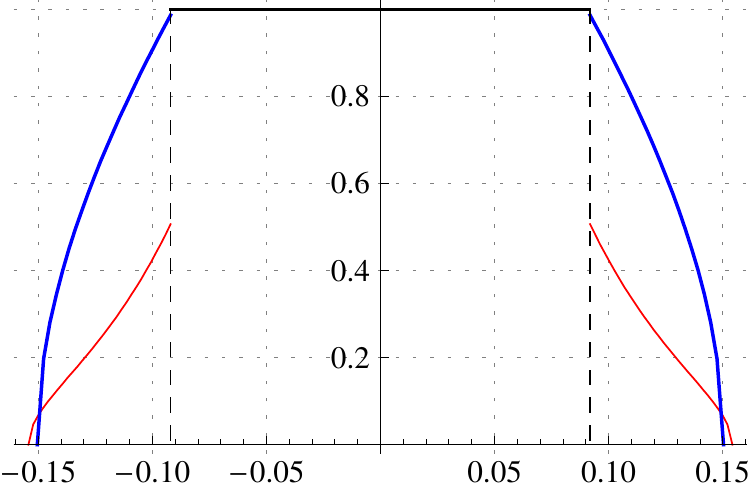}}
  \caption[Example solution in region $\mathrm{II}_{n-1,n}$ beyond
  line $\mathrm{C}_1$, physical version]{Physical version of the
  solution shown in \figref{fig:12_region5}. Cut two (red/thin)
  has a different starting direction than in
  \protect\figref{fig:12_region5} and hence a different contour, which allows
  for the formation of a condensate cut that hides the central part
  of cut one (blue/thick) and renders the configuration stable.}
\label{fig:12_region5_cond}
\end{figure}

\begin{figure}
\centering
  \subfloat[0.12]{\quad\includegraphics{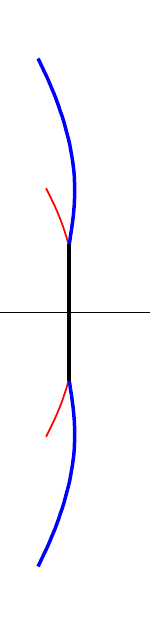}\label{fig:12_lineD_series_cond_a}}
  \subfloat[0.13]{\quad\includegraphics{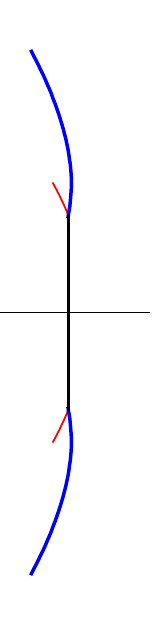}\label{fig:12_lineD_series_cond_b}}
  \subfloat[0.14]{\quad\includegraphics{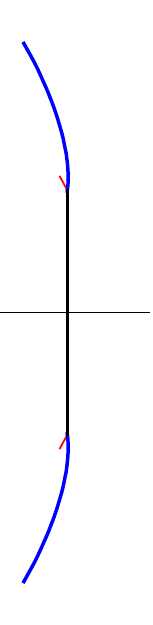}\label{fig:12_lineD_series_cond_c}}
  \subfloat[0.1487]{\quad\includegraphics{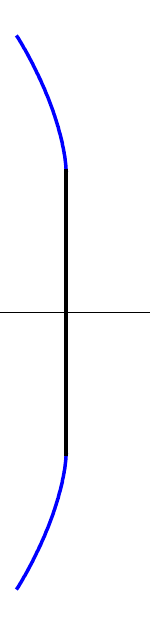}\label{fig:12_lineD_series_cond_d}}
  \subfloat[0.1487]{\quad\includegraphics{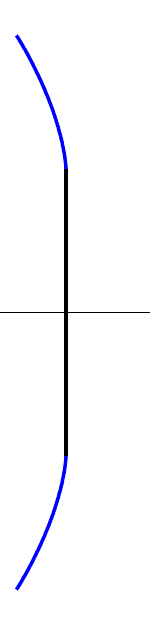}\label{fig:12_lineD_series_cond_e}}
  \subfloat[0.155]{\quad\includegraphics{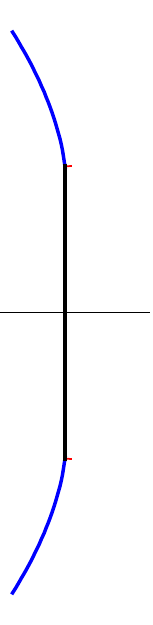}\label{fig:12_lineD_series_cond_f}}
  \subfloat[0.165]{\quad\includegraphics{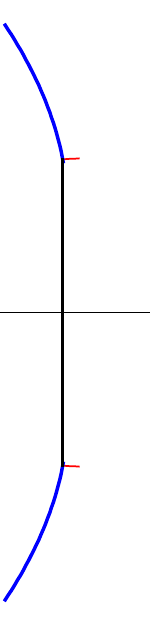}\label{fig:12_lineD_series_cond_g}}
  \caption[Series of configurations that passes through line
  $\mathrm{C}_1$, physical version]{Physical version of the series
  of solutions in \protect\figref{fig:12_lineD_series}. The
  configurations \protect\subref{fig:12_lineD_series_cond_f} and
  \protect\subref{fig:12_lineD_series_cond_g} show the physical,
  smooth continuation of the series beyond line $\mathrm{C}_1$.}
\label{fig:12_lineD_series_cond}
\end{figure}

For a qualitative picture of the densities on the cuts in the series,
refer to \figref{fig:12_region3_densities,fig:12_region5_densities}.
As a consequence of the change in the
contour and the filling of cut one, in all solutions beyond line
$\mathrm{C}_n$ (right of $\mathrm{C}_n$ in the $(\alpha,\tilde{E})$
plane) the absolute density exceeds unity at the centre of cut one's
standard contour. The standard contour of cut two is disjoint from cut
one, as in \figref{fig:12_region5}.
In this form, the configuration does not allow for a condensate cut
and is therefore unstable. However, there still is a stable version of
such solutions: As discussed below \eqref{eq:sqrtsing}, there are
three possible starting directions for a cut at each branch point.
Choosing a different direction for cut two, this cut's contour can be
made to cross the contour of the first cut, and the unphysical density
at the first cut's centre becomes hidden in a condensate. For the
solution in \figref{fig:12_region5}, his type of configuration
is shown in \figref{fig:12_region5_cond}.
These condensate-cut versions of the solutions beyond line
$\mathrm{C}_n$ not only cure the instability at the centre of cut one,
they also allow for a smooth transition from configurations on line
$\mathrm{C}_n$ to solutions beyond line $\mathrm{C}_n$. This can be
seen in \figref{fig:12_lineD_series_cond}, where the physical
version of the series of \figref{fig:12_lineD_series} is
displayed.

The changes that occur when the branch points of cut two pass the
contour of cut one imply that configurations on line $\mathrm{C}_n$
have an ambiguity in the mode number of cut two and in the standard
contour of cut one. This can be seen in the example in
\figref{fig:12_lineD_series_d,fig:12_lineD_series_e}. In the
limit where cut two approaches cut one from the left, the mode number
of cut two is $n_2 = n_1 + 1$, and the contour of cut one is bent to
the left. In the other limit, where cut two approaches cut one from
the right, the mode number of cut two is $n_2 = n_1 - 1$, and cut one
is bent to the right. The fillings of the cuts behave accordingly, as
described in the discussion of the series above. On line
$\mathrm{C}_n$, the contour of cut one is unique between its end
points and the branch points of cut two. The fact that the contour is
ambiguous between the two branch points of cut two is consistent with
the fact that the expansion of $p$ has a square root term at the
branch points $x_*$ (cf.\ \eqref{eq:sqrtsing}):
\[
  p (x_* + \varepsilon) = p (x_*) + A \sqrt{\varepsilon} +
  \order{\varepsilon} \,,
\]
because this results in the expansion of the density (as follows from
\eqref{eq:rho})
\[
  \rho_1 (x_* + \varepsilon)
  = \frac{1}{\pi i} \left( p(x_* + \varepsilon) - \pi n_1 \right)
  = i \left( n_1 - n_2 - \pi A \sqrt{\varepsilon} \right) +
  \order{\varepsilon} \,,
\]
which implies that the solution of the differential equation $\rho_1
(x) \ud x \in \Reals$ has an ambiguous second derivative at the branch
points $x_\pm$ and hence there are two choices for the continuation of
the contour beyond the branch points $x_*$. However, this ambiguity
has no physical significance, as the physical configuration with the
condensate is non-ambiguous.

\paragraph{Line $\mathrm{D}_{n,n+1}$: Two Tangential Cuts.}

\begin{figure}[tbp]
\centering
  \subfloat[Cut contours]{\quad\includegraphics{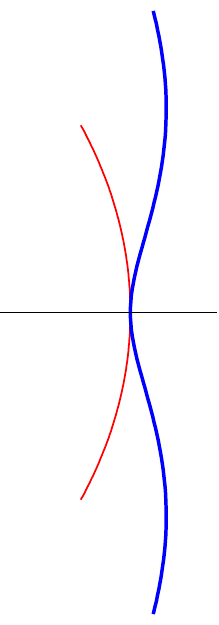}\quad}\qquad
  \subfloat[Absolute densities along the cuts  vs.\ distance from the cuts' centres (measured along the contour)]
    {\includegraphics{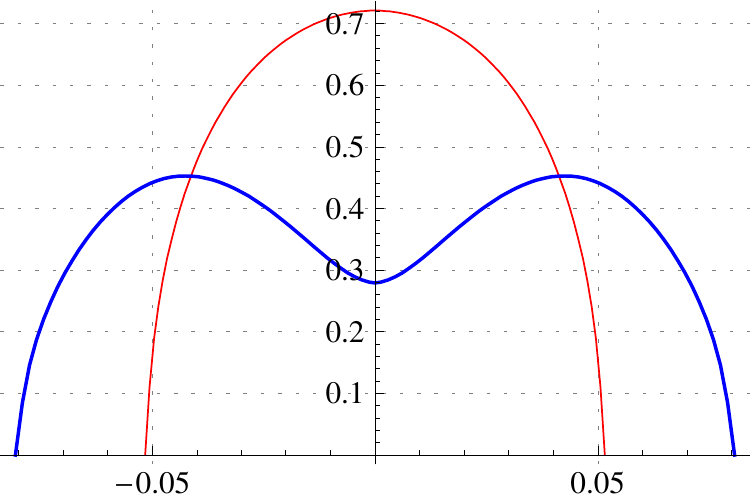}}
  \caption[Example configuration on line $\mathrm{D}_{12}$]{An
  example configuration on line $\mathrm{D}_{12}$. The fillings are
  $\alpha_1 = 0.05833$ (blue/thick, $n_1 = 1$) and $\alpha_2 =
  0.06$ (red/thin, $n_2 = 2$). The cuts touch at one point on the
  real axis, where the sum of their densities equals $-i$.}
\label{fig:12_lineC}
\end{figure}

This line separates region $\mathrm{I}_{n,n+1}$ from region
$\mathrm{II}_{n,n+1}$. The two cut contours of configurations on this
line bend towards each other and touch each other in one point on the
real axis. At the common point of the contours, the sum of the two
densities equals $-i$. An example is shown in
\figref{fig:12_lineC}. Beyond this line, the formation of condensate
cuts begins.

\paragraph{Line $\mathrm{E}_{n,n+1}$: Cut Two Violating Stability.}

This line can be reached from region $\mathrm{I}_{n,n+1}$ by
increasing the filling of cut two while keeping the filling of cut one
small, such that the two cuts stay disjoint. Line $\mathrm{E}_{n,n+1}$
is reached when the absolute density at the centre of cut one reaches
unity. At this point, the excitation with mode number $n+2$ collides
with the contour of cut two from the left. Further increasing the
filling of cut two makes the excitation pass to the right of the
contour and change its mode number to $n$. The absolute density on the
standard contour of cut two now exceeds unity and a third, closed cut
is required for stability; the configuration has reached region
$\mathrm{III}_{n,n+1}$ (cf.\ \figref{fig:12_region2_phys}).

\paragraph{Line $\mathrm{F}_{n,n+1}$: Two Disjoint Cuts and a Cusp on
Cut One.}
\label{sec:lineE}

\begin{figure}[tbp]
\centering
  \subfloat[0.0703]{\includegraphics{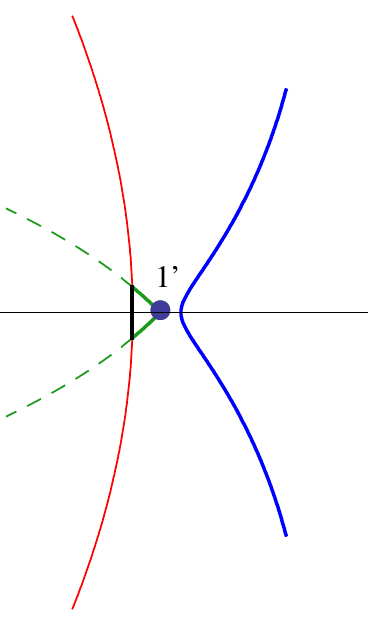}\qquad\label{fig:12_lineE_series_a}}
  \subfloat[0.07042]{\includegraphics{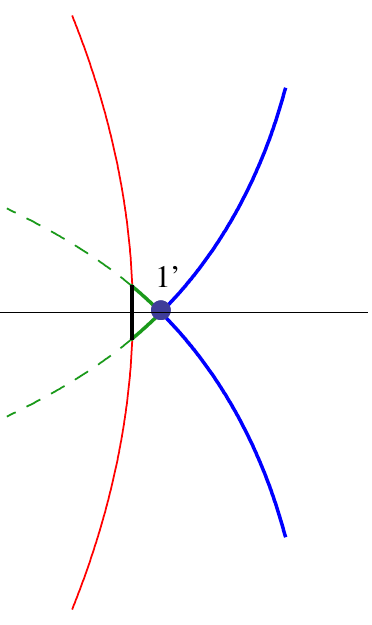}\qquad\label{fig:12_lineE_series_b}}
  \subfloat[0.0705]{\includegraphics{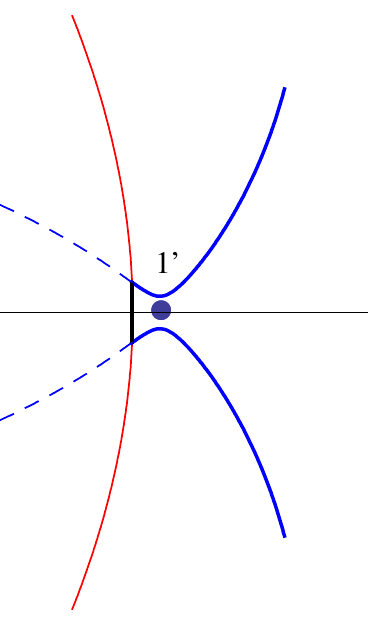}\label{fig:12_lineE_series_c}}
  \caption[Sequence of configurations that passes line
  $\mathrm{F}_{12}$]{A sequence of configurations that passes line
  $\mathrm{F}_{12}$. Cut one (blue/thick) has mode number $n_1=1$
  and constant filling $\alpha_1 = 0.005$, cut two (red/thin) has
  mode number $n_2=2$ and its filling $\alpha_2$ is indicated below
  each subfigure. The excitation with mode number $1$ is shown and
  carries a closed cut (green/thick) which forms a condensate cut at
  the centre of cut two. Configuration (a) lies in region
  $\mathrm{III}_{12}$, configuration (b) lies on line
  $\mathrm{F}_{12}$ and configuration (c) lies in region
  $\mathrm{II}_{12}$. The dashed contours all close around the
  origin. The energy increases during the sequence.}
\label{fig:12_lineE_series}
\end{figure}

This line separates region $\mathrm{III}_{n,n+1}$, where the standard
cut contours of the two cuts are disjoint, from solutions in region
$\mathrm{II}_{n,n+1}$, in which cut one encircles the origin and
consequently crosses the contour of cut two. A sequence of
configurations that passes line $\mathrm{F}_{12}$ from region
$\mathrm{III}_{12}$ to region $\mathrm{II}_{12}$ is shown in \figref{fig:12_lineE_series}.
The transition happens when the excitation $n'$, which carries a
closed loop that forms a condensate with cut two, reaches the contour
of cut one. At this point, the contour of the closed loop merges with
cut one, whose standard contour encircles the origin afterwards. The
transition point is marked by line $\mathrm{F}_{n,n+1}$. On this line,
the density \eqref{eq:density} on cut one decreases to zero at the
merging point $x_0$, as $p(x_0) = \pi n$. Around this point, the
density expands to $\rho_1 (x_0 + \varepsilon) \ud \varepsilon \sim
\varepsilon^2$, which implies that the contour has a cusp with opening
angle $90^\circ$.

\paragraph{Line $\mathrm{G}_{n,n+1}$: Condensate with Tails and a
Virtual Cusp.}

\begin{figure}[tbp]
\centering
  \subfloat[Cut contours]{\qquad\includegraphics{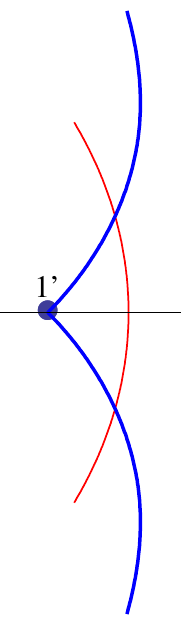}\qquad}\quad
  \subfloat[Absolute densities along the cuts  vs.\ distance from the cuts' centres]{\includegraphics{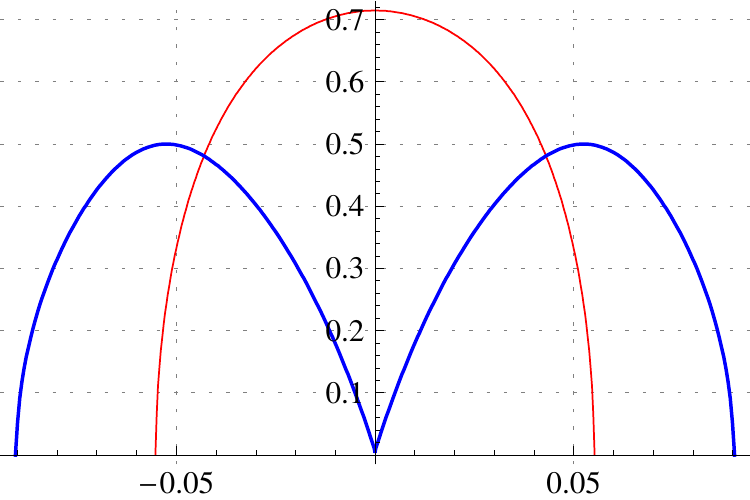}}
  \caption[Example configuration on line $\mathrm{G}_{12}$]{An
  example configuration on line $\mathrm{G}_{12}$. The fillings are
  $\alpha_1 = 0.0635$ (blue/thick, $n_1 = 1$) and $\alpha_2 = 0.064$
  (red/thin, $n_2 = 2$). The excitation with mode number $1$ is
  shown and lies on the standard contour of cut one, which implies
  that the density \eqref{eq:rho} decays to zero at that point.}
\label{fig:12_lineF}
\end{figure}

This line in region $\mathrm{II}_{n,n+1}$ separates solutions in which
the standard contour of cut one closes on the right side of the origin
from solutions in which it encircles the origin (cf.\
\figref{fig:12_region3,fig:12_region4}). The qualitative change
that happens to cut one's contour when crossing this line is the same
as on line $\mathrm{F}_{n,n+1}$: The excitation point $n'$ collides
with the standard contour, making the density decay to zero at its
centre. Unlike on line $\mathrm{F}_{n,n+1}$, here the process is less
significant because the central part of cut one is hidden in the
condensate. An example configuration that lies on line
$\mathrm{G}_{12}$ is displayed in \figref{fig:12_lineF}.

\paragraph{Line $\mathrm{H}_n$: One Cut with Vanishing Virtual Filling.}

\begin{figure}[tbp]
\centering
  \subfloat[Physical version of the cut contours. Dashed lines show the hidden part of the standard contours.] {\includegraphics{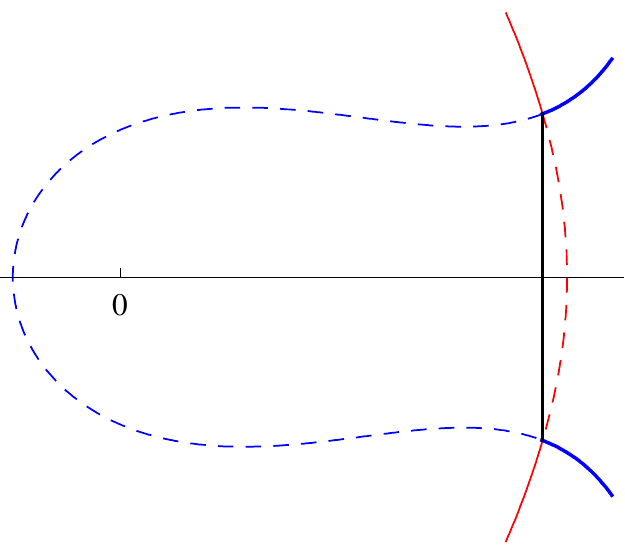}}\qquad
  \subfloat[Absolute densities along the condensate and the tails vs.\ distance from the condensate's centre (measured along the contour).]{\includegraphics{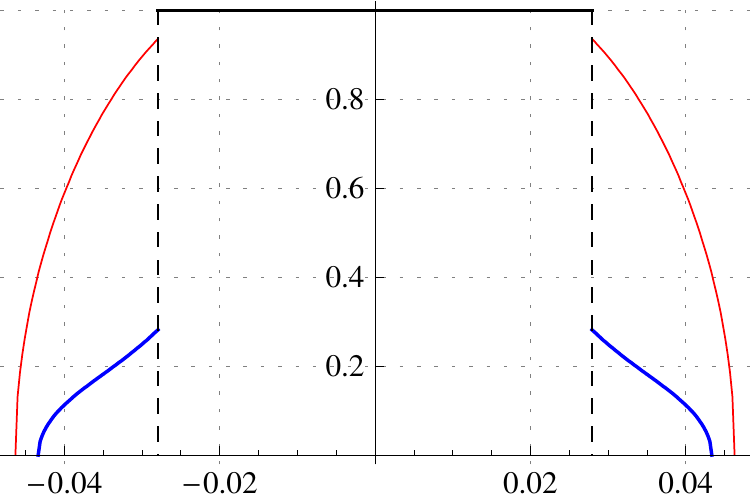}}
  \caption[Example configuration on line $\mathrm{H}_2$]{An example
  configuration on line $\mathrm{H}_2$. The fillings are $\alpha_1 =
  0$ (blue/thick, $n_1 = 1$) and $\alpha_2 = 0.085$ (red/thin, $n_2
  = 2$). The parts of the standard contours that are hidden in the
  condensate cut are shown as dashed lines. Since the standard
  contour of cut one encircles the origin, the integral of the
  absolute density along it is not zero, but unity.}
\label{fig:12_lineJ}
\end{figure}

On this line, the filling $\alpha_1$ of cut one (with mode number
$n-1$) vanishes. However, as shown above in the discussion of region
$\mathrm{II}_{n,n+1}$, the standard contour of cut one encircles the
origin and the filling on this contour equals one. Since most of cut
one is hidden in the condensate cut, the vanishing filling of the
physical cut contour does not have any further implications for the
physical configuration. An example is given in
\figref{fig:12_lineJ}. Because the filling of cut one vanishes, there is
a one-cut solution with the same mode numbers and fillings for each
solution on this line. These one-cut solutions are of the type shown
in \figref{fig:onebubblecutsc} and have a higher total energy than
the corresponding solutions on line $\mathrm{H}_n$. Refer to the
discussion of point $\mathrm{Y}_n$ below for more details on the
relation between these two types of solutions.

\paragraph{Line $\mathrm{J}_n$: Excitation $n$ Collides with the
Condensate.}

These lines mark all configurations where the excitation point with
mode number $n$ is located on the condensate cut. When line
$\mathrm{J}_n$ gets traversed, the excitation point with mode number
$n$ collides with the condensate from the left and passes to the right
(for positive mode numbers), thereby changing its mode number to
$n-2$. But unlike an excitation point passing a standard cut, this
process does not create an additional, closed cut (as on line
$\mathrm{E}_{n,n+1}$); in fact, the passage of the excitation point
has no effect on the configuration at all.

\subsection{Special Points}
\label{sec:cons_points}

As a result of the discussion of the various regions and lines in the
previous sections, three types of special points emerge, labelled by
$\mathrm{X}_n$, $\mathrm{Y}_n$ and $\mathrm{Z}_{n,n+1}$ in the
\figref{fig:12_eva,fig:12_eva_magnif,fig:12_aa,fig:23_eva,fig:23_eva_magnif}.
These points will be discussed in the following.

\paragraph{Point $\mathrm{X}_n$: Excitation Point Meets Single Cut.}

The points $\mathrm{X}_n$ mark the one-cut solutions with mode number
$n$ whose absolute density equals unity at the cut's centre, which
means that the excitation point with mode number $n+1$ is located on
the cut's contour. The filling fractions of these solutions for $n\leq
10$ can be found in \tabref{tab:maxfill}. The point $\mathrm{X}_n$
connects line $\mathrm{A}_n$, on which the absolute density on the
single cut's centre falls below unity, with line $\mathrm{B}_n$, on
which the absolute density at the centre of the standard contour
exceeds unity and a condensate begins to form with the help of a
second, closed cut. Also the lines $\mathrm{C}_n$,
$\mathrm{D}_{n,n+1}$ and $\mathrm{E}_{n-1,n}$ begin at the point
$\mathrm{X}_n$ (the last one only if $n>1$). Among all configurations
on line $\mathrm{A}_n$, the one at this point has the highest energy
(cf.\ \tabref{tab:maxfill}).

\paragraph{Point $\mathrm{Y}_n$: The Two Excitation Points with the
Same Mode Number Collide.}

Point $\mathrm{Y}_n$ marks the end of line $\mathrm{B}_n$ and thus
represents a one-cut solution with a condensate. On line
$\mathrm{B}_n$, the two excitation points $(n-1)'$ and $(n-1)$ both
lie on the real line, as in \figref{fig:onebubblecutsb}. At the
point $\mathrm{Y}_n$, these two excitation points collide.
Analytically continuing line $\mathrm{B}_n$ beyond point
$\mathrm{Y}_n$ lets the branch points diverge into the complex plane
and leads to one-cut solutions of the type shown in
\figref{fig:onebubblecutsc}. This analytic continuation of line
$\mathrm{B}_n$ shall be called line $\mathrm{B}_n'$. As discussed in
\secref{sec:onecut_cond}, solutions on this line are limiting
cases of three-cut solutions. Hence line $\mathrm{B}_n'$ is not part
of the moduli space of two-cut solutions (or of its boundary).

The only way to continue beyond point $\mathrm{Y}_n$ while staying in
the regime of consecutive mode numbers is to enter either region
$\mathrm{II}_{n-1,n}$ or region $\mathrm{III}_{n-1,n}$. In both cases,
the excitation points $(n-1)'$ and $(n-1)$ separate again. The former
stays on the real line, while the latter splits into two square root
branch points that diverge into the complex plane. Upon entering
region $\mathrm{III}_{n-1,n}$, the excitation point $(n-1)'$ continues
to carry a closed cut as on line $\mathrm{B}_n$ while the new branch
points form the ends of a second standard cut, as in
\figref{fig:12_lineE_series_a}. When the configuration enters region
$\mathrm{II}_{n-1,n}$, the afore closed cut opens up at its cusp and
becomes a standard cut that ends on the new pair of branch points, as
in \figref{fig:12_lineE_series_c}. Either way, a second standard
cut emerges, which can have a positive, negative or vanishing filling.
Configurations with vanishing filling belong to line $\mathrm{H}_n$
and lie in region $\mathrm{II}_{n-1,n}$. Since the closed cut in the
solutions on line $\mathrm{B}_n$ also has zero filling, line
$\mathrm{H}_n$ seems to be the most natural continuation of line
$\mathrm{B}_n$.

\begin{figure}[tbp]
\centering
  \includegraphics{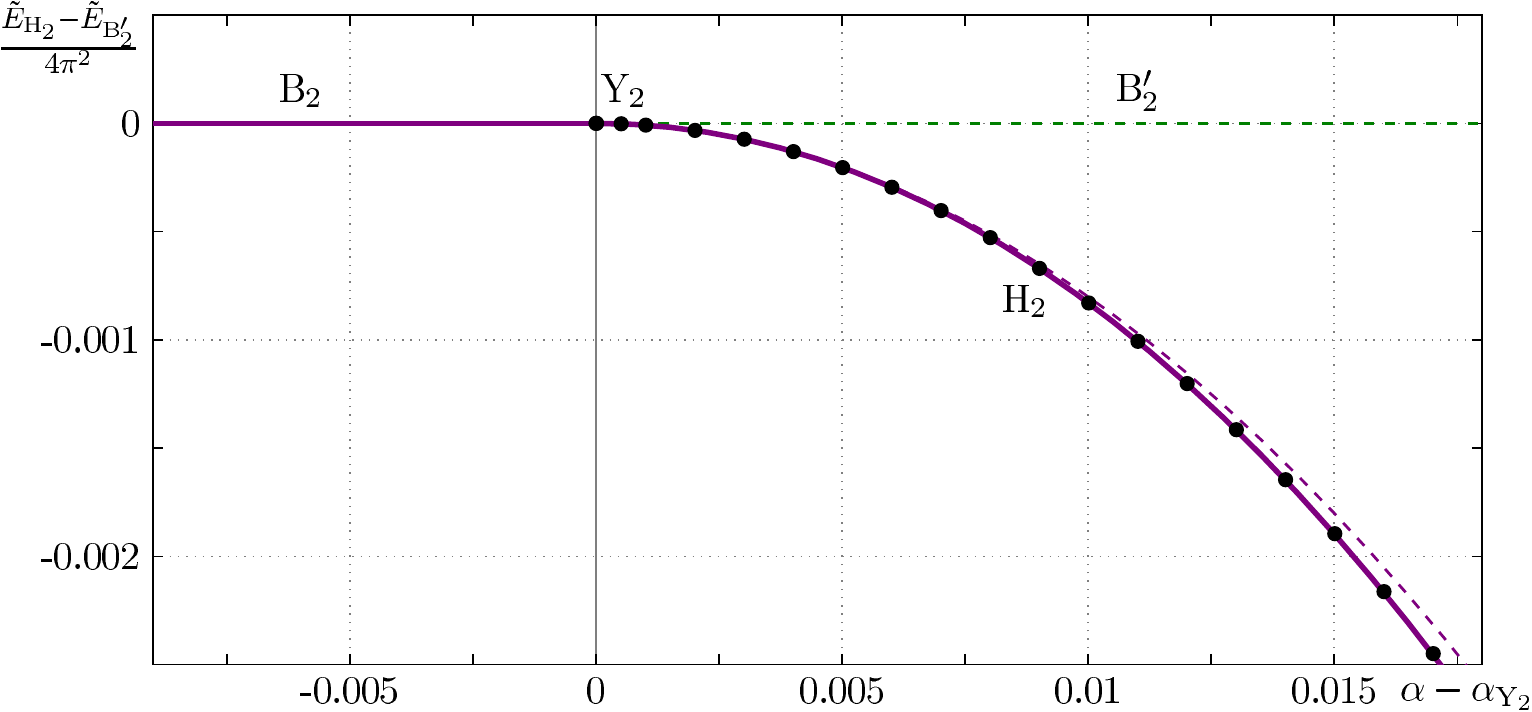}
  \caption[Energy deviation showing a second order phase transition
  at point $\mathrm{Y}_2$]{Deviation of the energy on line
  $\mathrm{H}_2$ from the energy on line
  $\mathrm{B}_2'$ (green/dashed), which is the analytic continuation
  of line $\mathrm{B}_2$ (purple/solid). Shown are the energies of
  numerically obtained solutions on line $\mathrm{H}_2$ (black dots),
  the corresponding function $\tilde{E}_{\mathrm{H}_n} -
  \tilde{E}_{\mathrm{B}_n'}$ (purple/solid) and the quadratic term of
  \eqref{eq:en_dev} (purple/dashed).
  The energy deviation is measured between solutions with the same total
  filling $\alpha$ and shows that a second order phase transition
  occurs upon the passage from line $\mathrm{B}_n$ to line
  $\mathrm{H}_n$.}
\label{fig:pointY_pt}
\end{figure}

As noted at the end of \secref{sec:onecut}, solutions on line
$\mathrm{B}_n'$ can be transformed into the corresponding solutions on
line $\mathrm{H_n}$ by separating the two segments of the closed cut
in \figref{fig:onebubblecutsc}. In doing so, the two excitation points
in the complex plane split into two square-root branch points each,
which results in a three cut solution. The filling of the third,
vertical cut can then be decreased to zero while the filling of the
second cut increases to zero, such that the total filling fraction
$\alpha$ stays constant. During this process, the energy decreases.
Comparing the energy of solutions on line $\mathrm{H}_n$ to the energy
of the corresponding one-cut solutions on line $\mathrm{B}_n'$
(\figref{fig:pointY_pt}) shows that a second order phase transition
occurs as a configuration follows line $\mathrm{B}_n$ and continues on
line $\mathrm{H}_n$. For example, the deviation of the energy on line
$\mathrm{H}_n$ from the energy on line $\mathrm{B}_n'$ has the
following expansion in the total filling at point $\mathrm{Y}_n$:
\[
  (\tilde{E}_{\mathrm{H}_n} - \tilde{E}_{\mathrm{B}_n'})(\alpha)
  = - 8\pi^2n^2(\alpha-\alpha_{\mathrm{Y}_n})^2 + \order{(\alpha-\alpha_{\mathrm{Y}_n})^3} \,.
\label{eq:en_dev}
\]
The prefactor of the quadratic deviation, including its dependency on
$n$, has been guessed on the basis of numerical data which, for the
first few values of $n$, matches this form with high accuracy. On a
technical level, this phase transition is related to the one discussed
by Douglas and Kazakov in \cite{Douglas:1993ii}: It involves a
transition between the same classes of algebraic curves and it is also
related to a density threshold, see \appref{sec:DKtrans} for a
technical review. The authors study the large-$N$ limit of the
partition function of continuous pure Yang-Mills theory on the
two-sphere and find a phase transition with respect to the area of the
sphere.%
\footnote{We thank V.\ Kazakov for remarking this point.}
%

\paragraph{Point $\mathrm{Z}_{n,n+1}$: Cusp on Cut One Meets Cut Two.}

At this point, the lines $\mathrm{F}_{n,n+1}$ and $\mathrm{G}_{n,n+1}$
meet: Cut one has a cusp at the centre of its contour whose tip lies
on the contour of cut two. Since the tip of the cusp marks the
excitation point with mode number $n+2$, this implies that the
absolute density of cut two at this point equals unity and that lines
$\mathrm{E}_{n,n+1}$ and $\mathrm{J}_{n+2}$ also end on point
$\mathrm{Z}_{n,n+1}$. The two cuts touch each other only in one point,
thus also line $\mathrm{D}_{n,n+1}$ ends here.

\subsection{Global Structure of the Moduli Space}
\label{sec:consecutive.global}

\begin{figure}[tbp]
\centering
  \includegraphics[angle=90]{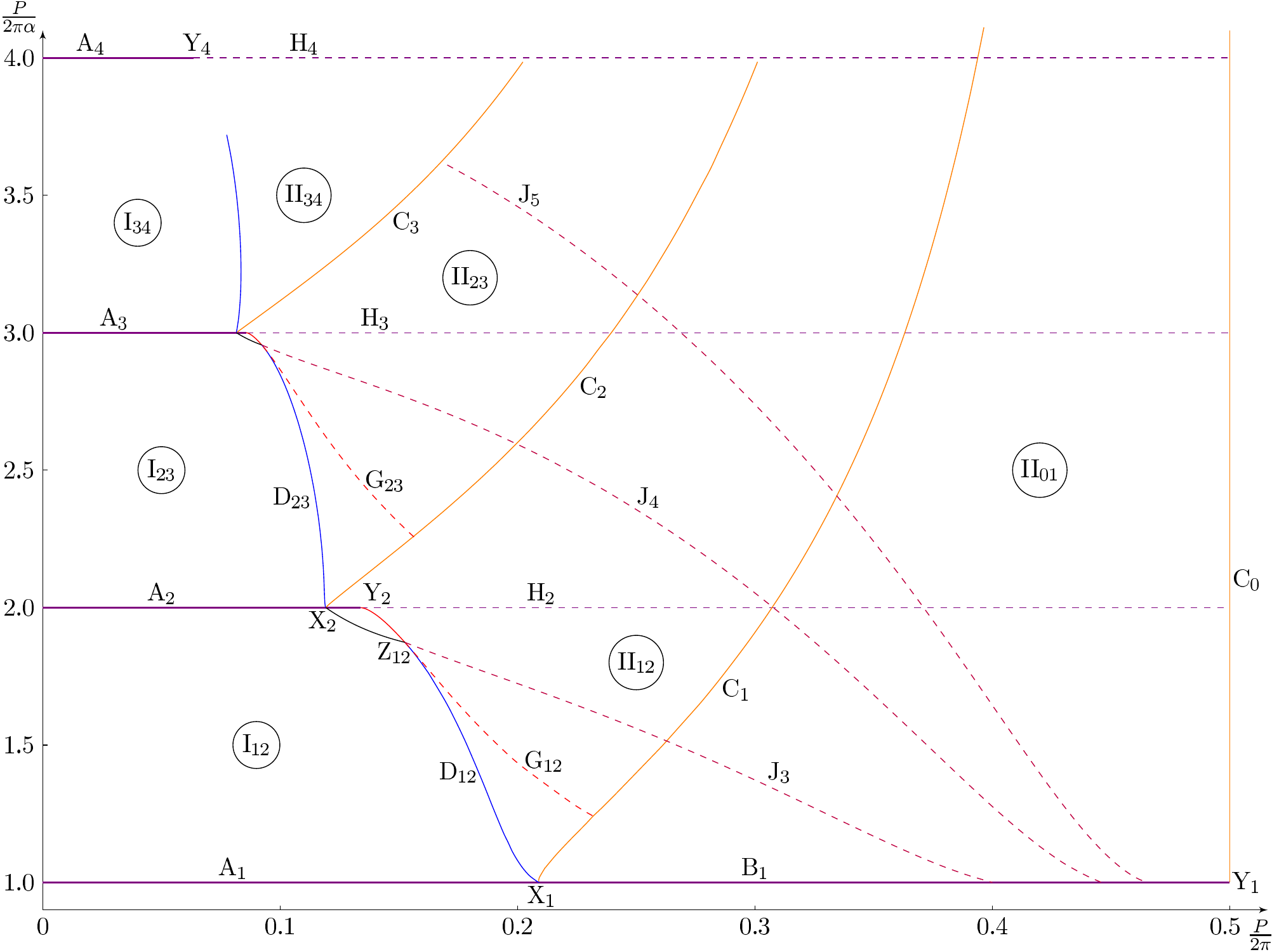}
  \caption[Global moduli space of consecutive mode numbers, $(P,
  P/\alpha)$ plane]{Global structure of the moduli space for
  consecutive mode numbers, shown in the $(P, P/\alpha)$ plane.}
\label{fig:poa_vs_p}
\end{figure}

\begin{figure}[tbp]
\centering
  \includegraphics[angle=90]{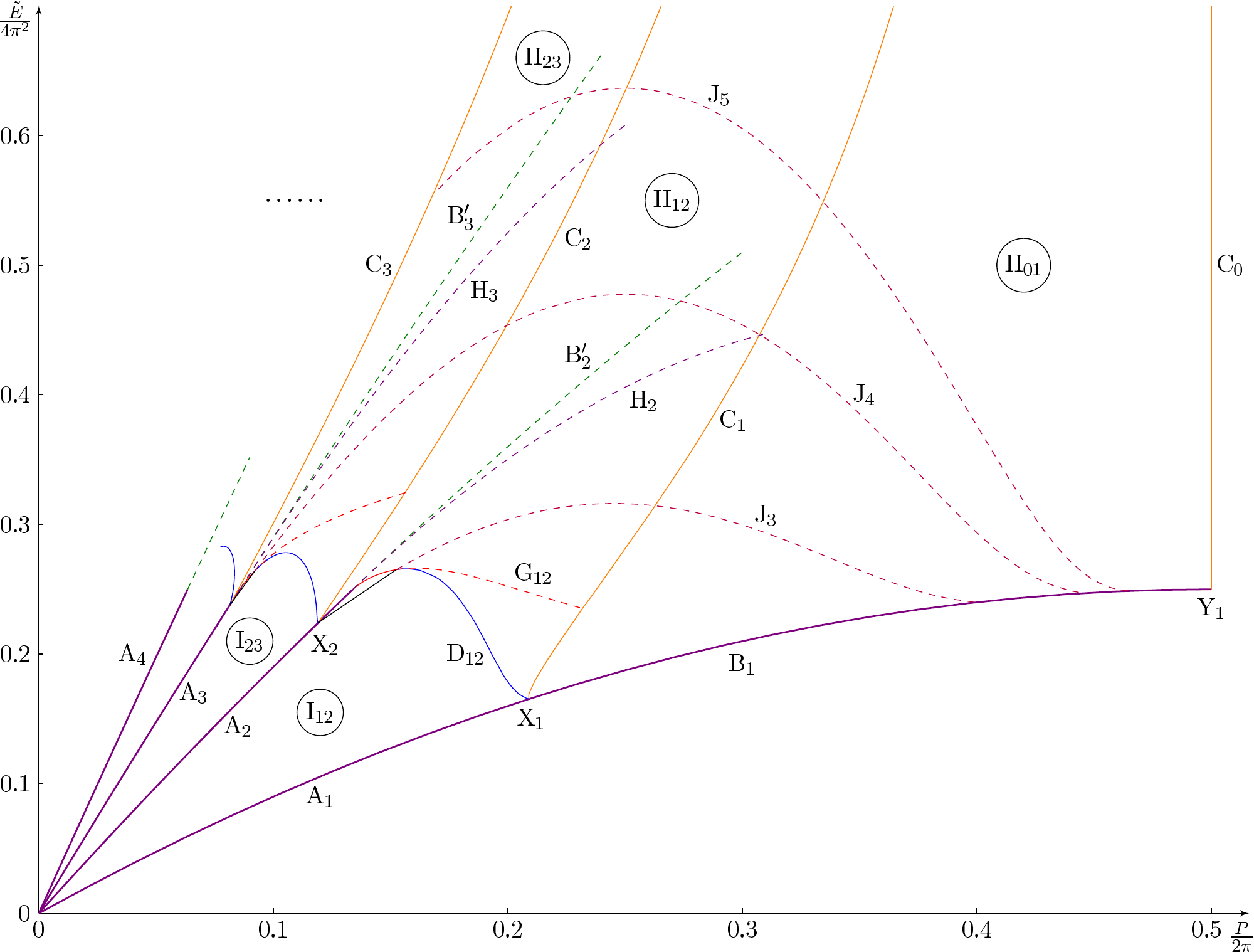}
  \caption[Global moduli space of consecutive mode numbers, $(P,
  \tilde{E})$
  plane]{Global structure of the moduli space for consecutive mode
  numbers, shown in the $(P, \tilde{E})$ plane.}
\label{fig:e_vs_p}
\end{figure}

As a consequence of the discussion about the lines $\mathrm{C}_n$ in
\secref{sec:lines}, all moduli spaces of two consecutive mode
numbers are analytically connected through the lines $\mathrm{C}_n$.
Pictures of the resulting global space are shown in
\figref{fig:poa_vs_p,fig:e_vs_p}. In the latter, also the lines
of unstable one-cut solutions $\mathrm{B}_n'$ are shown for
comparison, while in the former these lines are the same as the lines
$\mathrm{H}_n$. In particular, the global connectedness of the space
shows that mode numbers lose their intuitive meaning at large filling
fractions: A small change in the number of excitations (filling
fractions) can alter the mode numbers, hence they can no longer be
interpreted as the number of periods of a coherent state on the spin
chain.

Note that the case $n_1 = 0$ (cut one), $n_2 = 1$ (cut two)
qualitatively differs from the other cases with consecutive mode
numbers. The excitation point $n_* = 0$ always lies at $x = \infty$
and hence cannot be excited, only the excitation point $n_* = 0'$ can.
Consequently, line $\mathrm{B}_1$ cannot be crossed, and cut one
always has a negative filling and its standard contour winds around
the origin. That means that solutions with mode numbers $n_1 = 0$,
$n_2 = 1$ always have a condensate and regions $\mathrm{I}_{01}$ and
$\mathrm{III}_{01}$ do not exist.

The one-cut lines $\mathrm{A}_n$ and $\mathrm{B}_n$, $n>1$, form
branch cuts of the global moduli space; across them, the derivatives
of physical quantities such as the energy and the momentum are
discontinuous. The points $\mathrm{Y}_n$, $n>1$ are the corresponding
branch points.

\begin{figure}[htpb]
\centering
  \includegraphics{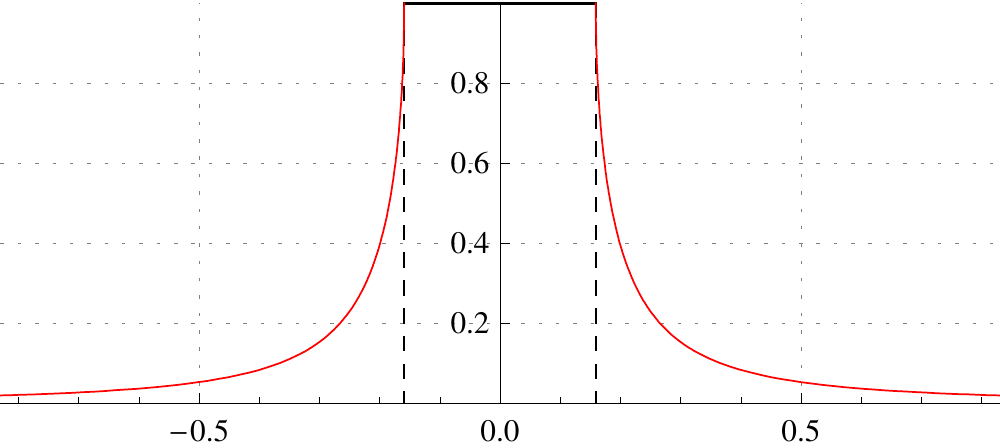}
  \caption[Density on the cuts at point $\mathrm{Y}_1$
  ($\alpha=1/2$)]{Absolute density along the condensate and the two
  tails of the solution at point $\mathrm{Y}_1$. Cut one ($n_1 = 0$,
  $\alpha_1=0$, red/thin) extends along the whole imaginary line,
  while the branch points of cut two ($n_2 = 1$, $\alpha_2=1/2$) are
  $x_\pm = \pm i/2\pi$. The absolute density on the tails falls
  off as $1/|x|$.}
\label{fig:halffilling}
\end{figure}

\begin{figure}[htpb]
\centering
  \subfloat[Cut contours]{\quad\includegraphics{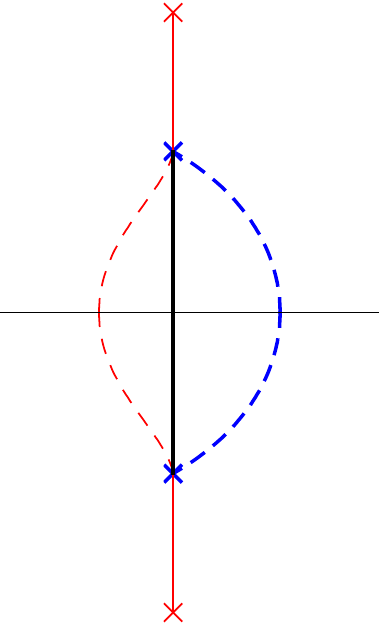}\quad}\quad
  \subfloat[Absolute densities along the condensate and the tails]{\includegraphics{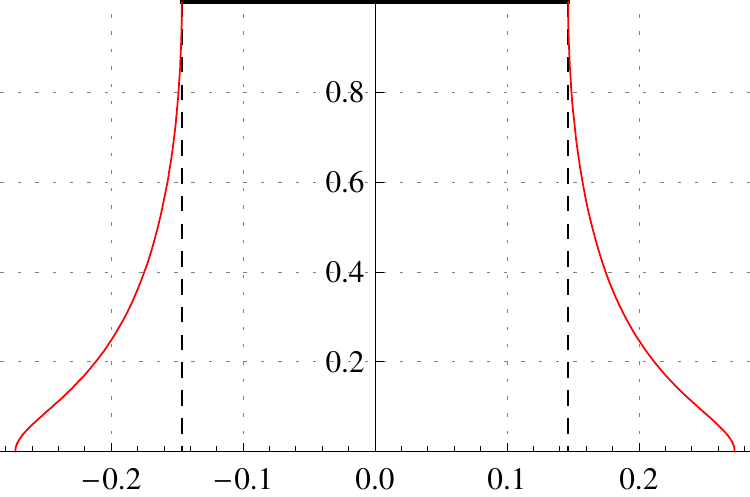}}
  \caption[Example solution on line $\mathrm{C}_0$]{An example
  solution on line $\mathrm{C}_0$. The condensate and the tails lie
  on the imaginary line. The standard contours are shown as dashed
  lines. The filling fractions are $\alpha_1 = -0.14$ (red/thin,
  $n_1 = 0$, $x_\pm = \pm 0.2726i$) and $\alpha_2 = 1/2$
  (blue/thick, $n_2 = 1$, $x_\pm = 0.146434i$). Cut one is
  completely hidden in the condensate.}
\label{fig:line_C0}
\end{figure}

As mentioned at the end of \secref{sec:onecut_cond}, at point
$\mathrm{Y}_1$ the excitation point $n_* = 0'$ reaches $x = \infty$.
At that point, the filling fraction of cut two equals $\alpha_2 =
1/2$ and cut one extends along the whole imaginary line while
the branch points of cut two are $x_\pm = \pm i/2\pi$. The
absolute density of this solution is shown in \figref{fig:halffilling}.
When the filling of cut one gets reduced to finite negative values
while $\alpha_2$ stays at $1/2$, the condensate and the tails of the
configuration remain on the imaginary line, but their length
decreases. Solutions of this type constitute line $\mathrm{C}_0$; an
example is shown in \figref{fig:line_C0}. All solutions on this
line have a momentum of $P = 2\pi (n_1 \alpha_1 + n_2 \alpha_2) =
\pi$.

\begin{figure}[tbp]
\centering
  \includegraphics{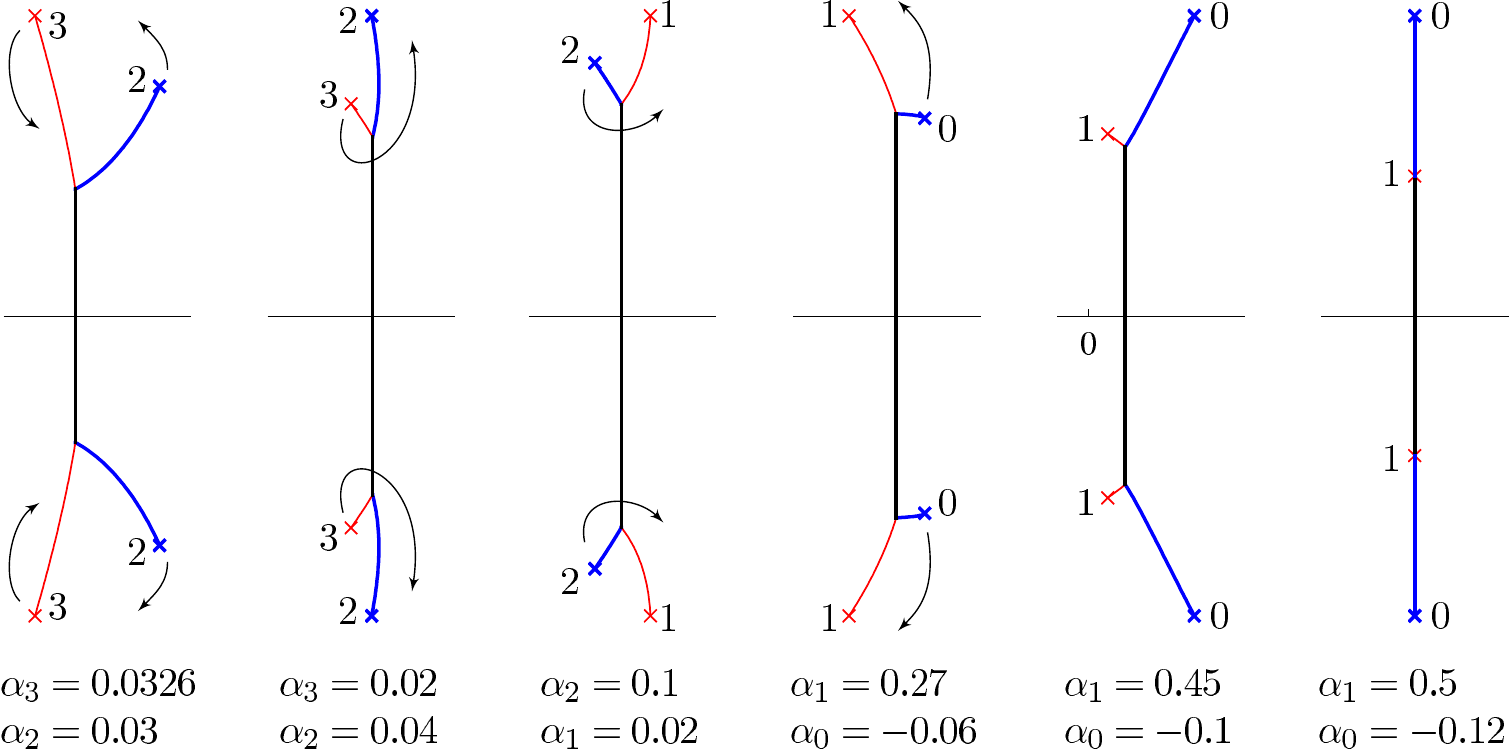}
  \caption[Sequence of solutions with decreasing mode numbers]{A
  sequence of configurations that evolves through the global moduli
  space of consecutive mode numbers. During the sequence, the branch
  cuts cycle around each other. The mode numbers are indicated next
  to each branch point. Each time a branch point passes a contour,
  its mode number decreases by two. The total filling fraction
  $\alpha$ increases during the sequence; the individual filling
  fractions are indicated below each subfigure. The sequence roughly
  follows a straight horizontal line in \protect\figref{fig:e_vs_p} at
  $\tilde{E}/4\pi^2 \approx 0.3$. The final configuration lies on line
  $\mathrm{C}_0$, i.e.\ all branch points lie on the imaginary line.
  The size of the cuts increases from left to right, the rightmost
  configuration is about 15 times larger than the leftmost.}
\label{fig:cycle}
\end{figure}

As mentioned, a consequence of the fact that the phase space of
consecutive mode numbers is connected is that the mode numbers can
change by a continuous variation of the fillings. In particular, if
one follows a straight horizontal line in \figref{fig:poa_vs_p}
from the left to the right, the branch points cycle around each other
while the mode numbers decrease until the configuration reaches line
$\mathrm{C}_0$ and both pairs of branch points and the contour lie on
the imaginary line. A schematic sketch of this process is shown in
\figref{fig:cycle}.
The sequence can even be continued further by following it in reverse,
but in the other (left) half of the complex plane: The mode numbers
become negative, but all other physical quantities are the same as
in the mirror solution with positive mode numbers. This means that
there exist two copies of the space in \figref{fig:poa_vs_p}, one
for positive and one for negative mode numbers, and the two parts are
connected through line $\mathrm{C}_0$.

Also the lines $\mathrm{B}_n'$, $n>0$ of unstable one-cut solutions
(as shown in \figref{fig:onebubblecutsc}) can be followed until
the branch points reach the imaginary axis, similar to the sequence of
two-cut solutions that approaches line $\mathrm{C}_0$. At this point,
also the fluctuation points $n_*=(n-1)$ and $n_*=(n-1)'$ lie on the
imaginary line and the filling of the branch cut reaches $\alpha=1/2$,
as follows from \eqref{eq:onecut_bp}. But, according to
\eqref{eq:onecut_EP}, this implies that $P=\pi n>\pi$. Since the
maximal physical momentum on a lattice is $P=\pi$, the physical
momentum is $P-2\pi$ for $P>\pi$. The maximal physical momentum
$P=\pi$ is reached by the configurations on line $\mathrm{B}_n'$ long
before the branch points arrive at the imaginary axis. Reaching the
regime of such high momenta is prevented by the phase transition
discussed at the end of \secref{sec:cons_points}, which makes the
one-cut solution transform into a two-cut solution at the critical
value of the filling $\alpha_{\mathrm{crit}}=\alpha_{\mathrm{Y}_n}$.
Physically, nothing peculiar would happen when the momentum would
increase to values larger than $\pi$, but it is interesting to see
that the two-cut solutions by themselves implement the restriction to
momenta $|P|\leq \pi$, as can be seen in the pictures of the phase
space in \figref{fig:poa_vs_p,fig:e_vs_p}.

\subsection{Stability and Gauge Theory States}
\label{sec:cons_stab}

The expositions of this section show that the moduli space of two-cut
solutions has a very rich structure, even though only solutions with
consecutive mode numbers have been considered. Most remarkably, there
seem to be no unstable two-cut solutions: For all spectral curves with
consecutive mode numbers there is a configuration of standard and
condensate cuts that generate the curve and that do not violate the
stability condition. Here, by stability condition is meant that the
absolute density never exceeds unity at any point on the cut contour
whose tangent is vertical:
\[
  |\rho(x_0)| \leq 1 \quad \text{for } \rho(x_0)\sim i
\label{eq:stab_refined}
\]
Such critical point are always hidden in a condensate cut. Thus it may
be conjectured that \emph{all} classically admissible two-cut
solutions are solutions of the Bethe equations in the thermodynamic
limit. In fact, numerical tests show strong evidence in favour of this
conjecture: For most of the cases discussed in the previous section,
numerically exact Bethe root distributions that approximate the cut
contours have been obtained, see \secref{sec:numerics}.

A remark about suitable gauge theory states is appropriate at this point.
Only spin chain states with a momentum $P\in 2\pi\Integers$ correspond
to gauge theory states. None of the solutions studied in this chapter
satisfies this condition: All two-cut configurations with consecutive
mode numbers have a non-zero momentum $0<P\leq\pi$. Nevertheless, all
qualitative results should generalise to valid gauge theory states.
Namely, configurations whose cut structure is point symmetric about
the origin manifestly have a vanishing momentum $P$.%
\footnote{Unless there is a condensate cut \emph{on} the imaginary
line that passes through the origin. In that case, the momentum is
maximal, $P=\pi$.}
It is reasonable to expect that such point symmetric solutions with
mode numbers $(n,n+1)$ and $(-n,-n-1)$ and with symmetric fillings
exhibit the same qualitative features as the solutions investigated
here (such as condensate formation and connectedness of the moduli
space). Some point-symmetric configurations of two-cut and degenerate
multi-cut solutions are studied in the next section.

\section{Moduli Space of Non-Consecutive Mode Numbers}
\label{sec:nonconsec}

All qualitative features of the general moduli space with arbitrary
mode numbers can already be seen in the case of consecutive mode
numbers, which was studied in the previous section. In general, cuts
with non-consecutive mode numbers lie further away from each other and
hence influence each other less. Some examples will be studied in this
section.

\subsection{Mode Numbers $n_1=1$, $n_2=3$}
\label{sec:13}

\begin{figure}[tbp]
\centering
  \includegraphics[angle=90]{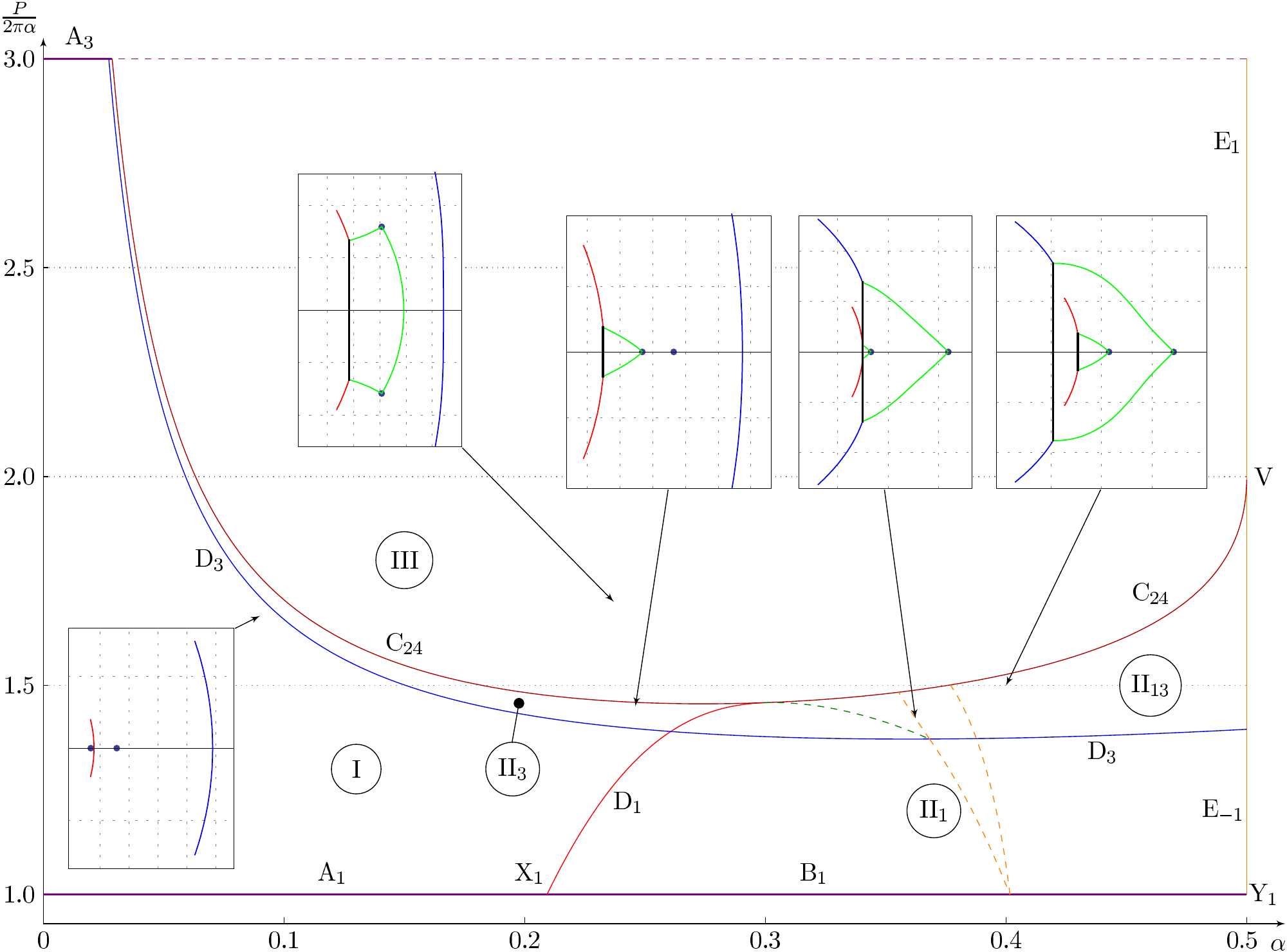}
  \caption{The space of solutions with mode numbers $n_1=1$,
  $n_2=3$, shown in the $(\alpha, P/2\pi\alpha)$ plane. Some
  example solutions in different regions are shown.}
\label{fig:13_poava}
\end{figure}

First, the comparatively simple space of solutions with mode numbers
$n_1=1$ (cut one), $n_2=3$ (cut two) shall be exposed. It is shown in
the $(\alpha, P/2\pi\alpha)$ plane in \figref{fig:13_poava}.
In region $\mathrm{I}$, the fillings of both cuts are small, their
contours are disjoint and their absolute densities are bound by unity,
hence there are no condensates. Line $\mathrm{D_1}$ marks the
configurations in which the density on cut one equals $-i$. Beyond
this line lies region $\mathrm{II_1}$, whose configurations have a
condensate on cut one and an associated closed loop-cut with mode
number $0=2'$. Similarly, configurations on line $\mathrm{D_2}$ have
an absolute density of $-i$ at the centre of cut two; beyond this line
lies region $\mathrm{II_3}$, where cut two has a condensate core and
an associated closed loop-cut with mode number $2=4'$. In region
$\mathrm{II_{13}}$, both cuts have a condensate core and associated
closed loop-cuts. Finally, line $\mathrm{C_{24}}$ marks configurations
in which the excitation points with mode numbers $n=2$ and $n=4$ have
collided. Configurations beyond this line are unstable in the sense
that they are degenerate cases of solutions with higher genus
(higher number of cuts).

Regions $\mathrm{II_1}$ and $\mathrm{II_{13}}$ can be further
subdivided, which is signified by the dashed and dotted lines in
\figref{fig:13_poava}. To the left of these lines (at low total
filling $\alpha$), the two cuts with their condensates and loop-cuts
are disjoint. On the green, dashed line, the excitation point with
mode number $2'=4$ (which is the cusp of the loop-cut of cut two)
collides with the condensate core of cut one. Upon further increasing
the total filling, the condensate core of cut two (or its standard
contour) collides with the condensate core of cut one; this event is
marked by the orange, dotted line. At even higher total filling, the
branch points of cut two pass through the condensate core of cut one
(orange, dashed line), such that the whole contour of cut two (with
its associated loop-cut) is encircled by the loop-cut of cut one.
Nothing special happens to the condensate core of cut one during the
passage of cut two -- this is sensible, as the condensate core is
invisible in the spectral curve $\ud p$, which contains all the
information about the solution. When cut two has completely passed
through the condensate core of cut one, its mode number has changed to
$n_2'=n_2-2=1$.

\begin{figure}
\centering
  \subfloat[]{\parbox{5.5cm}{\includegraphics{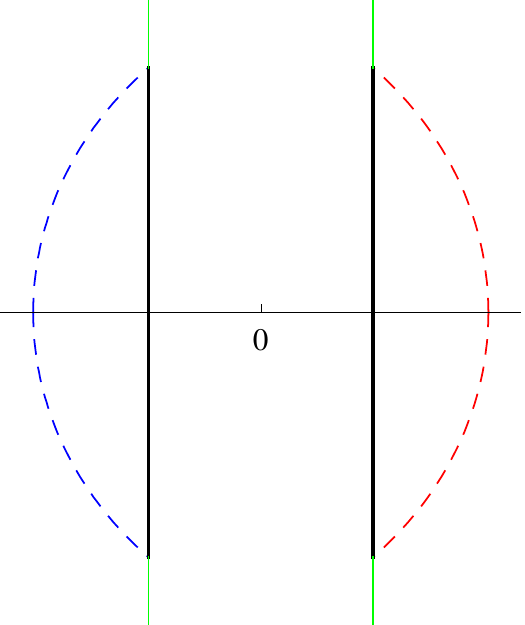}}\label{fig:point_V_1}}
  \quad $\longleftrightarrow$ \quad
  \subfloat[]{\parbox{4cm}{\includegraphics{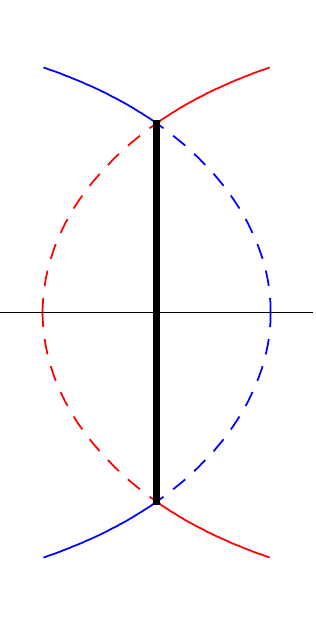}}\label{fig:point_V_2}}
  \caption{Configuration at point $\mathrm{V}$ in
  \protect\figref{fig:13_poava}. This solution is symmetric about the origin
  and hence represents a valid gauge-theory state (The momentum equals
  $P=2\pi$). Figure \protect\subref{fig:point_V_1} shows the configuration that
  is reached when coming from region $\mathrm{II_{13}}$ or from line
  $\mathrm{E_{-1}}$ in \protect\figref{fig:13_poava}. There is a dual
  configuration of cuts, shown in \protect\subref{fig:point_V_2}, that
  generates the same solution. It has a ``double-condensate'' with root density
  $|\rho|=2$ at it's centre.}
\label{fig:point_V}
\end{figure}

When the total filling gets increased up to $\alpha=1/2$ (line
$\mathrm{E_{-1}}$), the excitation point with mode number $2=0'$,
which marks the cusp of the closed loop-cut that is connected to cut
one, reaches $x=\infty$. This results in a configuration similar to
the one at the point $\mathrm{Y_1}$ (see \figref{fig:halffilling}):
The contour of the closed cut passes through
the branch points of cut one, such that cut one is completely hidden
in a condensate that has two vertical, infinitely long tails.
Consequently, the mode number of cut one is ambiguous: It is either
$n_1=1$ or $n_1=-1$. In addition, there is cut two, whose mode number
is $n_2'=1$. It has a standard contour (below line $\mathrm{D_3}$), or
a standard contour with a condensate core and a closed loop-cut (above
line $\mathrm{D_3}$). At the upper end of line $\mathrm{E_{-1}}$ lies
point $\mathrm{V}$, at which also the excitation point that marks the
cusp of cut two's closed loop-cut has reached $x=\infty$ and the
configuration is completely symmetric about the origin
(\figref{fig:point_V_1}).
Consequently, the momentum vanishes at this point, $P=2\pi$, and the
solution represents a valid gauge-theory state.
\figref{fig:point_V_2} shows another cut configuration that realises the
solution at point $\mathrm{V}$. It can be constructed by taking
another possible standard contour for each cut and replacing their
common centre with a \emph{double condensate}, that is a condensate
with root density $|\rho|=2$. Bethe root distributions that show this
type of pattern have been found before
\cite{Dhar:2000aa,Beisert:2003xu}.
Behind point $\mathrm{V}$ lies line $\mathrm{E_1}$. Configurations on
this line are ``mirror solutions'' of the ones on line
$\mathrm{E_{-1}}$: Cut three has mode number $1/-1$ and consists of a
condensate with two vertical tails that extend to infinity, while cut
one has mode number $-1$ and consists of a standard contour with or
without a condensate core and an associated closed loop-cut.

\subsection{Mode Numbers $n_1=1$, $n_2=-1$}

\begin{figure}[tbp]
\centering
  \includegraphics{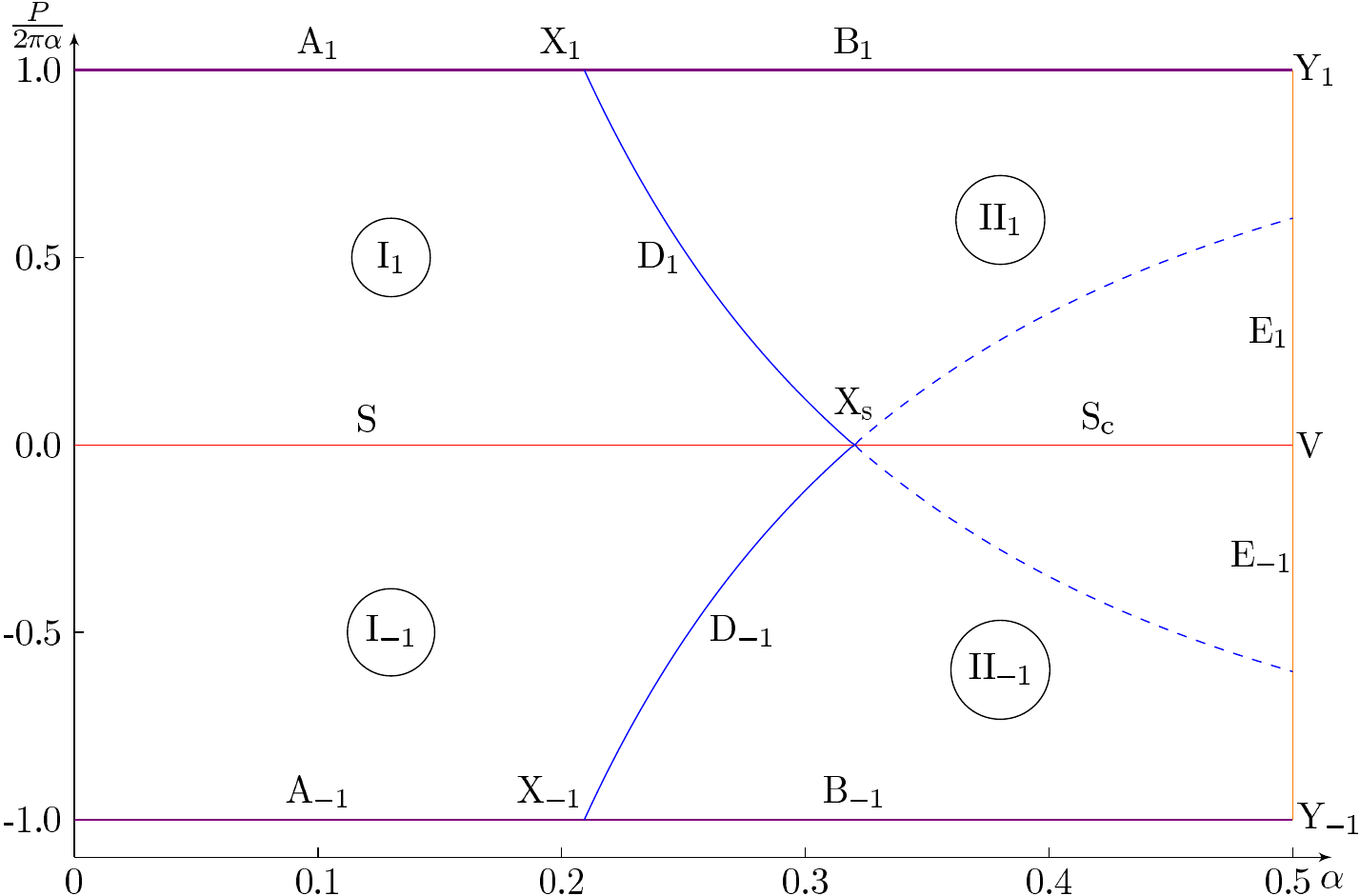}
  \caption{The space of configurations with mode numbers $n_1=1$,
  $n_2=-1$, shown in the $(\alpha, P/2\pi\alpha)$ plane.
  Configurations in regions $\mathrm{I}_{\pm 1}$ have plain standard
  contours, while configurations in regions $\mathrm{II}_{\pm 1}$ have
  condensates on either or both of the contours.}
\label{fig:1-1_poava}
\end{figure}

\begin{figure}[tbp]
\centering
  \includegraphics{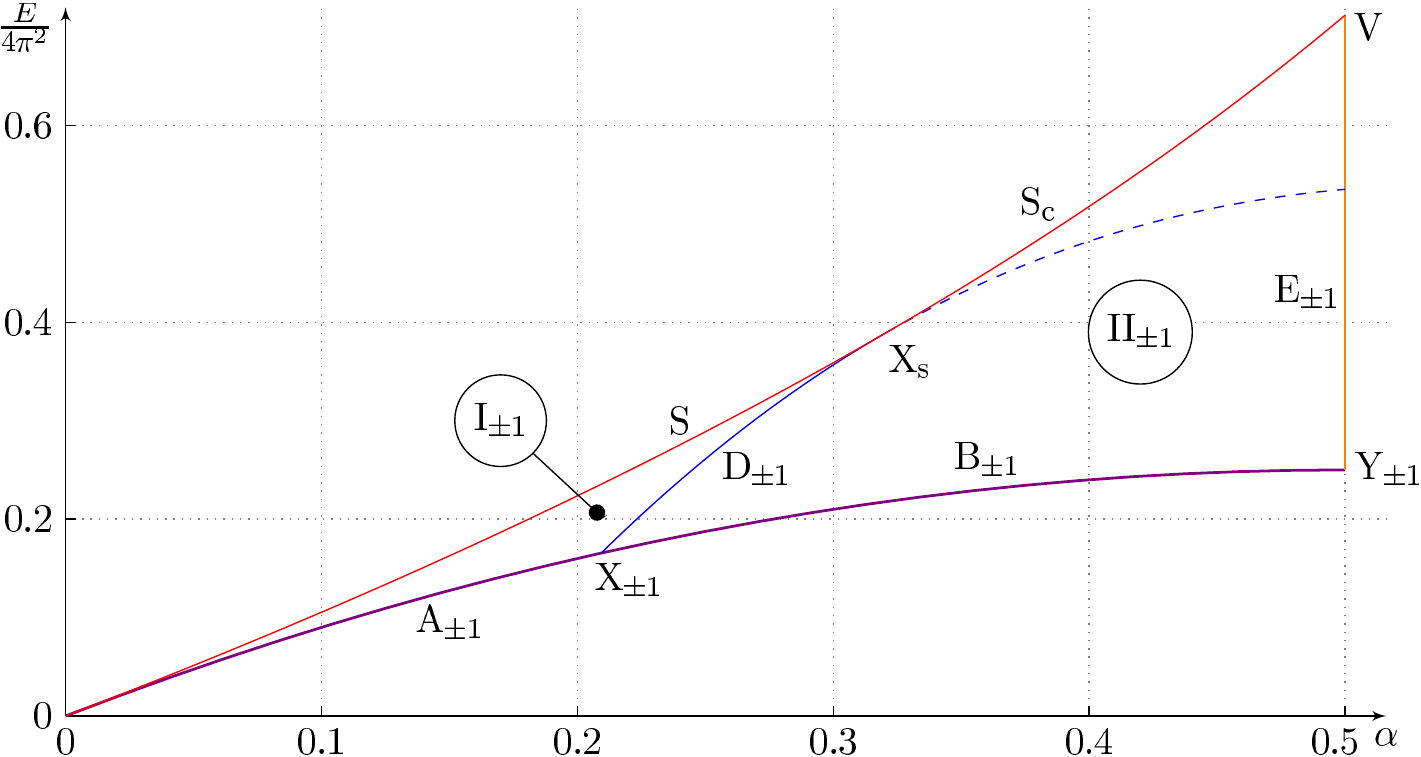}
  \caption{The space of configurations with mode numbers $n_1=1$,
  $n_2=-1$, shown in the $(\alpha, E)$ plane. Configurations that
  only differ by a reflection about the imaginary line have the same
  energy, hence the upper and the lower lines and regions of
  \protect\figref{fig:1-1_poava} lie on top of each other in this
  representation.}
\label{fig:1-1_eva}
\end{figure}

The moduli space of solutions with mode numbers $n_1=1$ (cut one),
$n_2=-1$ (cut two) is shown in \figref{fig:1-1_poava,fig:1-1_eva}
in the $(\alpha, P/2\pi\alpha)$ and in the $(\alpha,
E)$ plane.
Regions, lines and points with the same labels but subscripts $1$ and
$-1$ fall together in \figref{fig:1-1_eva}, since their
configurations only differ by a reflection about the imaginary line
and hence have the same energy.

As in previous examples, the lines $\mathrm{A}_{\pm 1}$ and
$\mathrm{B}_{\pm 1}$ represent solutions with a single cut with mode
number $n$, solutions on the lines $\mathrm{A}_{\pm 1}$ have a
standard contour without a condensate core, while solutions on the
lines $\mathrm{B}_{\pm 1}$ have a condensate core and a closed
loop-cut. In the regions $\mathrm{I}_{\pm 1}$ and $\mathrm{II}_{\pm
1}$, both cuts with mode numbers $n_1=1$ and $n_2=-1$ are present.
Line $\mathrm{D}_{\pm 1}$ separates solutions that only have a
standard contour (region $\mathrm{I}_{\pm 1}$) from solutions that
have a condensate core with a closed loop-cut (region
$\mathrm{II}_{\pm 1}$). The closed loop-cut has mode number $n=0$ for
both cuts. Hence on line $\mathrm{D}_{\pm 1}$, the absolute density at
the centre of the cut with mode number $\pm 1$ equals $|\rho|=1$.

In the regions $\mathrm{I}_1$ and $\mathrm{II}_1$, the filling
fraction of cut one is larger than the filling fraction of cut two,
$\alpha_1>\alpha_2$. In the regions $\mathrm{I}_{-1}$ and
$\mathrm{II}_{-1}$, it is the other way round, $\alpha_1<\alpha_2$.
These regions are separated by the lines $\mathrm{S}$ and
$\mathrm{S_c}$, which consist of symmetric solutions with
$\alpha_1=\alpha_2$. Such solutions have vanishing momentum $P$ and
hence represent valid gauge theory states, they were studied in
\cite{Beisert:2004hm}. Obviously, solutions on line $\mathrm{S}$ have
plain standard contours, while solutions on line $\mathrm{S_c}$ have
condensate cores in both contours. Point $\mathrm{X_s}$ marks the
symmetric solution with maximal filling that does not have
condensates. The lines $\mathrm{E}_{\pm 1}$ and point $\mathrm{V}$ are
the same as the ones discussed above in \secref{sec:13}, hence the
diagrams in \figref{fig:13_poava,fig:1-1_poava} are connected. The
points $\mathrm{Y}_{\pm 1}$ actually both represent the same solution,
namely the one shown in \figref{fig:halffilling} whose cut
configuration lies entirely on the imaginary line.

\section{Numerical Solution of the Bethe Equation}
\label{sec:numerics}

In this section we develop a method for obtaining numerically exact
distributions of Bethe roots for the Heisenberg spin chain. The main goal
of this section is to test our predictions from the analysis of the
one-cut and two-cut spectral curves in
\secref{sec:onecut,sec:consecutive,sec:nonconsec}. Thus we are especially interested in distributions
which can be compared with the spectral curve, i.e.\ solutions with
large numbers of roots that align on a set of contours.

We examine in detail some subtle points of the previous analysis, such as
stability (\secref{sec:stab}), condensate formation
(\secref{sec:onecut_cond}) and phase
transitions between spectral curves with different topologies
(\secref{sec:cons_points}).
We will see that condensate cores with roots
separated by $\oldDelta u\approx i$ indeed appear in all cases
considered. 
We will also confirm that the phase transition from one-cut 
to two-cut solutions indeed happens. 
Furthermore, in \secref{sec:numfinite} we will analyse small deviations from
the thermodynamic limit. 
In particular we will numerically analyse the
leading finite-size corrections and show that they do not exhibit any
non-analyticity due to the formation of the condensate.

The numerical configurations shown in this section are 
approximate solutions to the
discrete Bethe equations \eqref{eq:xxxbethe} with finite length $L$.
It is astonishing to observe how well the predictions from the thermodynamic limit 
describe the actual distributions of Bethe roots.

\subsection{Description of the Numerical Method}
\label{sec:nummethod}

Here we describe the numerical method used to solve the
Bethe equations \eqref{eq:xxxbethe} in detail.
In order to find roots of a set of equations numerically, it is important to
specify a good set of starting points that are sufficiently close to the
exact solution. We will focus on this problem first and then describe a
simple programme for finding numerical solutions.

\paragraph{Starting Points.}

In \secref{sec:branch_cuts} we introduced the mode number associated
to a branch cut in the thermodynamic limit. We will need to
generalise the notion of mode numbers to the quantum case. The mode
numbers naturally appear in the quantum theory when one takes the
logarithm of \eqref{eq:xxxbethe}:
\begin{equation}
iL \log\frac{u_k+\frac{i}{2}}{u_k-\frac{i}{2}}+2\pi n_k
=\sum_{j\neq k}^Mi\log\frac{u_k-u_j+i}{u_k-u_j-i}\,,\qquad k=1,\dots,M.
\label{eq:logbae}
\end{equation}
The integers $n_k$ represent the $2\pi i$-ambiguity of the logarithm,
in the thermodynamic limit they exactly become the mode numbers
introduced in equation \eqref{eq:sqrtcut}. This can be seen by taking
the large-$L$ limit of \eqref{eq:logbae}, 
as in \cite{Sutherland:1995aa,Beisert:2003xu,Kazakov:2004qf}. 
The solution to \eqref{eq:logbae} is uniquely specified by a set of integers $n_k$
(at least for large $L$ and small $M$).

In order to find the solution numerically by Newton's method, 
one has to specify starting points for the search. 
For generic parameters it is very difficult to give
suitable starting points that lead to convergent iterations. 
However, as we will see it is pretty simple to find good starting
points in the limit of small filling fractions $L\gg M$
where the Bethe equations \eqref{eq:logbae} 
can be solved approximately \cite{Dhar:2000aa}:

To the leading order we can drop the r.h.s.\ of \eqref{eq:logbae} and,
assuming that $u_k\sim L$ (which is true in the case we are interested in), 
we get that $u_k\simeq L/2\pi n_k$. 
We conclude that Bethe roots with the same mode number $n_k$ 
are close to each other and separated by $\oldDelta u\sim L$ 
from other groups of roots with different mode numbers.
\begin{figure}[tb]\centering
\subfloat[$L=400$]{\qquad\includegraphics[height=7.5cm]{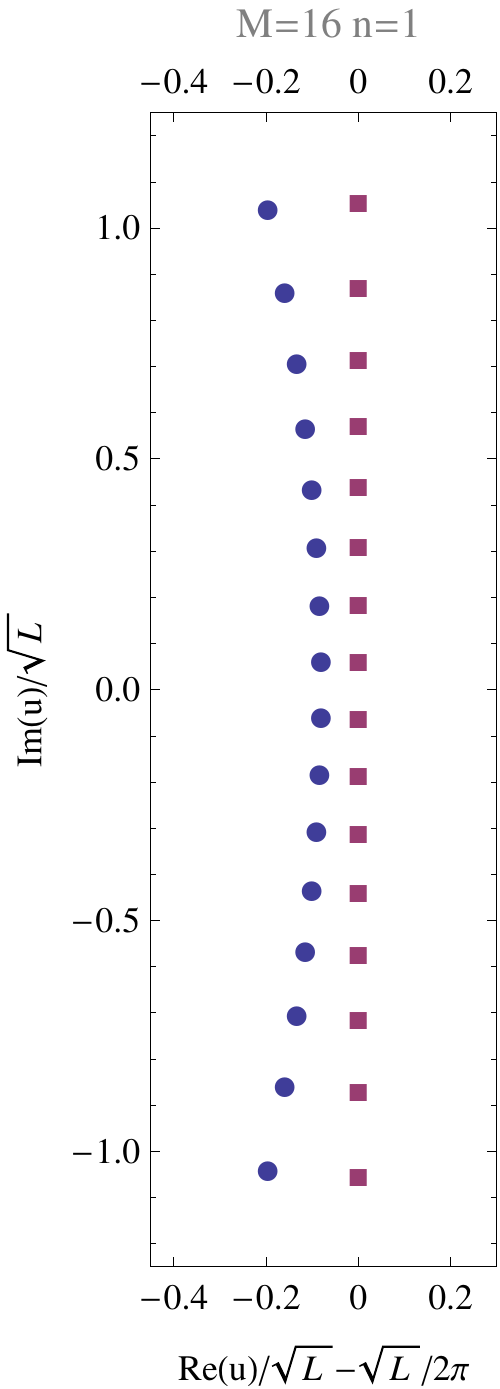}\qquad}
\subfloat[$L=1600$]{\qquad\includegraphics[height=7.5cm]{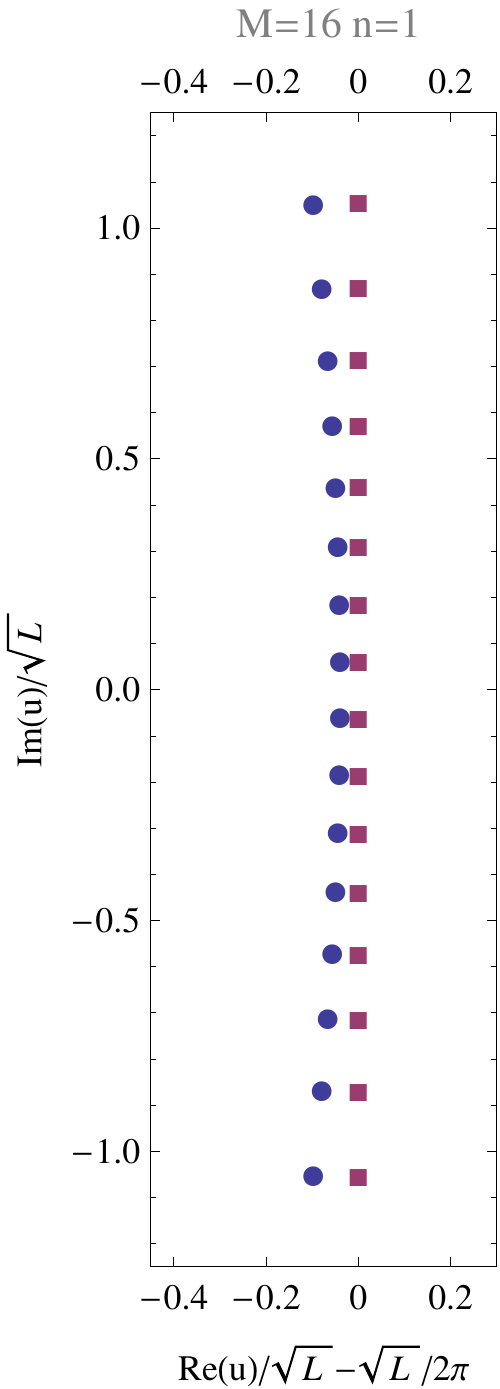}\qquad}
\caption{Approximation of a solution to the Bethe equations
\protect\eqref{eq:xxxbethe} by zeros of the Hermite polynomial for
$M=16$ roots with mode number $n=1$ and chain lengths $L=400$,
$L=1600$. Squares represent points given by 
$u_k=L/(2\pi n)+iz_k\sqrt{2L}/(2\pi n)$, where $z_k$ are zeros of Hermite polynomial
of degree $M$, while circular dots represent numerically exact Bethe
roots. This approximation provides us with a good set of starting
points.}
\label{fig:sample}
\end{figure}
For resolving the positions of the Bethe roots with the same mode number
$n$, we can
neglect terms in the r.h.s.\ of \eqref{eq:logbae} containing roots with different mode numbers
and use the ansatz
\begin{equation}
u_k=\frac{1}{2\pi n}\left(L+iz_k\sqrt{2L}+\order{1/L^0}\right) .
\label{aprox}
\end{equation}
Assuming that $n\sim 1$ and $z_k\sim 1$, we expand \eqref{eq:logbae}
and get \cite{Dhar:2000aa}
\begin{equation}
\frac{1}{z_k-z_j}-z_k=0 \,,
\end{equation}
which is the equation for the zeros of the Hermite polynomial,
$\mathrm{H}_M(z_k)=0$. Numerically exact values of Bethe roots are plotted
against their large-$L$ approximation \eqref{aprox} in
\figref{fig:sample}, and one clearly sees that the relative mismatch
is indeed smaller for larger $L$.
It is possible to continue the $1/\sqrt{L}$ expansion in a systematic
way to get a better description of Bethe roots for large $L$ and mode
number $n$ fixed. For our goal the leading approximation of the Bethe
roots by zeros of Hermite polynomials provides us with a good set of
starting points for sufficiently large $L$.

\paragraph{Iteration Procedure.}

Having starting points at hand it is almost immediate to construct a
numerical solution of the Bethe equations using Newton's method.
However, we know the starting points only for large $L$. 
To overcome this difficulty we can view the Bethe root distributions
as functions of the length $L$ while keeping the magnon number $M$ fixed.
We can solve the Bethe equations numerically, gradually varying $L$
from a sufficiently large value $L\gg M$ 
to the desired value with $L=\order{M}$.
In each step we use the numerical solution obtained
in the previous step as new starting values for smaller $L$. 
In this way we obtain a series of configurations whose filling
$\alpha=M/L$ slowly increases. Some important examples of such series
will be considered in the next subsection.

For various reasons it is favourable to use 
the Bethe equations in product form \eqref{eq:xxxbethe} 
and not in logarithmic form \eqref{eq:logbae}
for the Newton iteration.
Using this method, the mode numbers enter into the calculation only through
the starting points. Since we vary $L$ slowly, at each step we inherit
the information about the mode numbers from the previous step, and
hence the resulting configuration corresponds 
to the analytically continued curve from the regime of large $L$. 

Note that for large $L$ and fixed $M$ the roots are separated by
$\oldDelta u\sim\sqrt{L}$ and thus the stability condition is
satisfied, $|\rho|\approx 1/\oldDelta u\ll1$. However, when we decrease
$L$, at some point the separation between Bethe roots 
approaches $\oldDelta u=i$. 
The Bethe equations \eqref{eq:xxxbethe} are singular at this point
which leads to a numerical instability at small $L$.%
\footnote{In the logarithmic Bethe equations \protect\eqref{eq:logbae}
the branch cut of the logarithm furthermore introduces a discontinuity
close to the starting values.
The Newton method is thus likely not to converge
and the product form \eqref{eq:xxxbethe} is clearly preferable 
in this case.}
Therefore the step size for $L$ must be chosen sufficiently small in this regime.
For practical purposes one can even choose $L$ to take real intermediate values.
We expect the above numerical procedure to lead to a reasonably unambiguous 
final configuration if $L$ is changed in sufficiently small steps.

\paragraph{Sample Programme.}

Let us give a basic realisation of above method in \texttt{Mathematica}. 
For simplicity the programme presented in \tabref{tab:prog}
is restricted to the one-cut case.
To generalise it, it is enough to change the set of starting
points {\verb"u0"} according to the discussion in the previous section.
We tested this programme on \texttt{Mathematica} versions 5.2 and 6.0,
but it should be compatible with earlier versions as well.

\begin{table}
\footnotesize
\verb"  (*define mode number, number of roots, initial length and finial length*)"\\
\verb"n = 1; M = 12; L = 100; Lfinal = 30; "\\
\verb""\\
\verb"  (*starting points*)"\\
\verb"u0 = (L/(2 Pi) + I Sqrt[2 L]/(2 Pi n) zk /. NSolve[HermiteH[M, zk] == 0, zk])"\\
\verb""\\
\verb"  (*solve the Bethe equations while slowly decreasing the length L*)"\\
\verb"Do[u0 = Table[u[k], {k, M}] /. FindRoot[Table[((u[k] + I/2)/(u[k] - I/2))^L =="\\
\verb"   -Product[(u[k] - u[j] + I)/(u[k] - u[j] - I), {j, M}], {k, M}],"\\
\verb"   Table[{u[k], u0[[k]]}, {k, M}], WorkingPrecision -> 50, AccuracyGoal -> 20];"\\
\verb"   L-- // Print;"\\
\verb", {L - Lfinal + 1}]"\\
\verb""\\
\verb"  (*plot the result*)"\\
\verb"ListPlot[Transpose[{u0//Re,u0//Im}], PlotRange -> All, AspectRatio -> Automatic]"
\caption{\texttt{Mathematica} programme for the generation of a
finite-length numerical Bethe root distribution corresponding to a
one-cut spectral curve.}
\label{tab:prog}
\end{table}

This programme can be optimised to work much faster, especially for
the case of a large number of Bethe roots:
A major improvement can be achieved by replacing the 
built-in \verb"FindRoot" function, which is too general and inefficient, 
by a simple implementation of Newton's method. 
Furthermore the step size for $L$ can be made adaptive. 
One can even decrease $L$ in small steps during each Newton iteration
and thus avoid an excessively precise calculation 
which is merely used for new starting values.

\subsection{Some Series of Numerical Solutions}
\label{sec:numsols}

In this section we will exploit numerical solutions of the Bethe
equations obtained with our programme to test the predictions from the thermodynamic limit. We
begin with a study of the simplest one-cut solution, subsequently proceed to
more complex configurations with two cuts and consecutive mode numbers, and finish with a two-cut
solution with non-consecutive mode numbers. In the last configuration,
one of the two cuts transforms into a two-cut complex in a phase
transition, the result being a three-cut solution.

\paragraph{One-Cut Solution, $n=1$.}

\begin{figure}[tb]\centering
\resizebox{\textwidth}{!}{%
\includegraphics{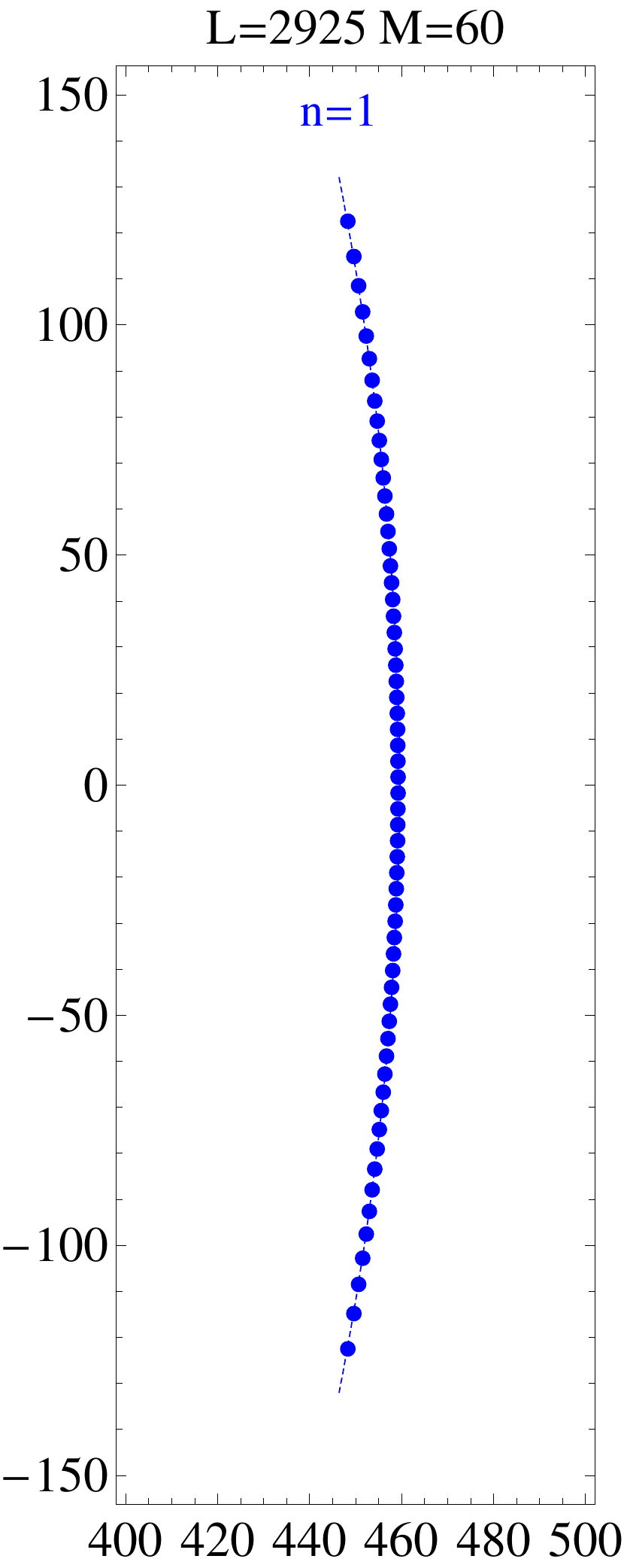}\quad%
\includegraphics{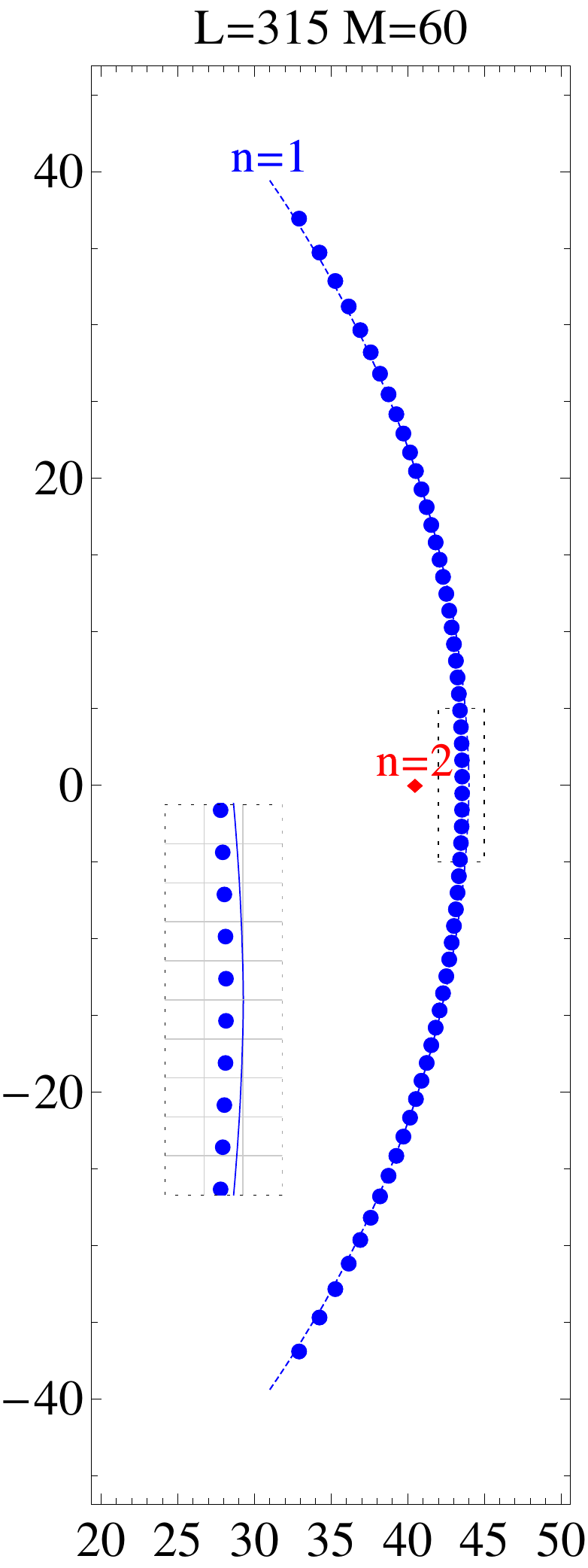}\quad%
\includegraphics{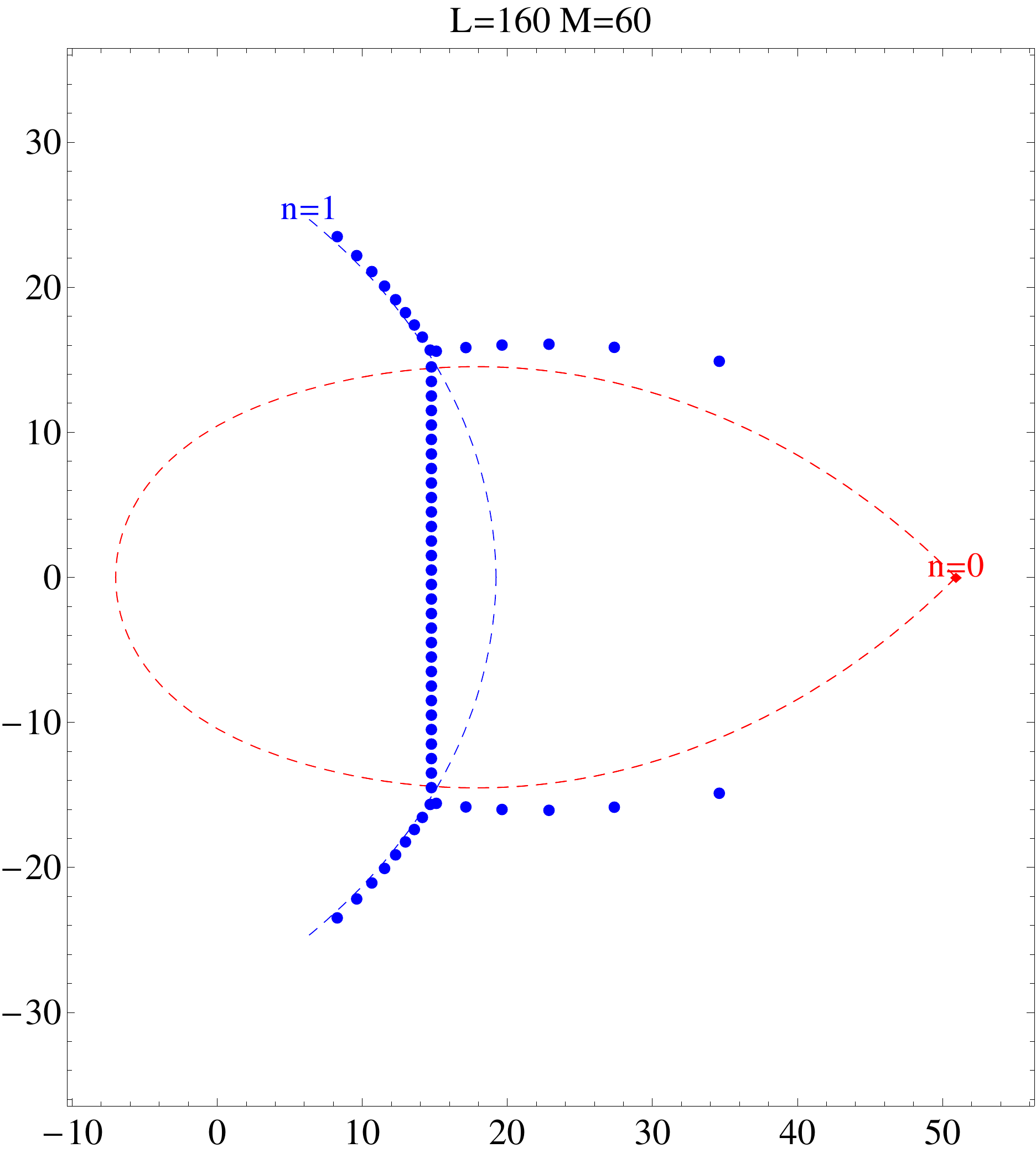}\quad%
\includegraphics{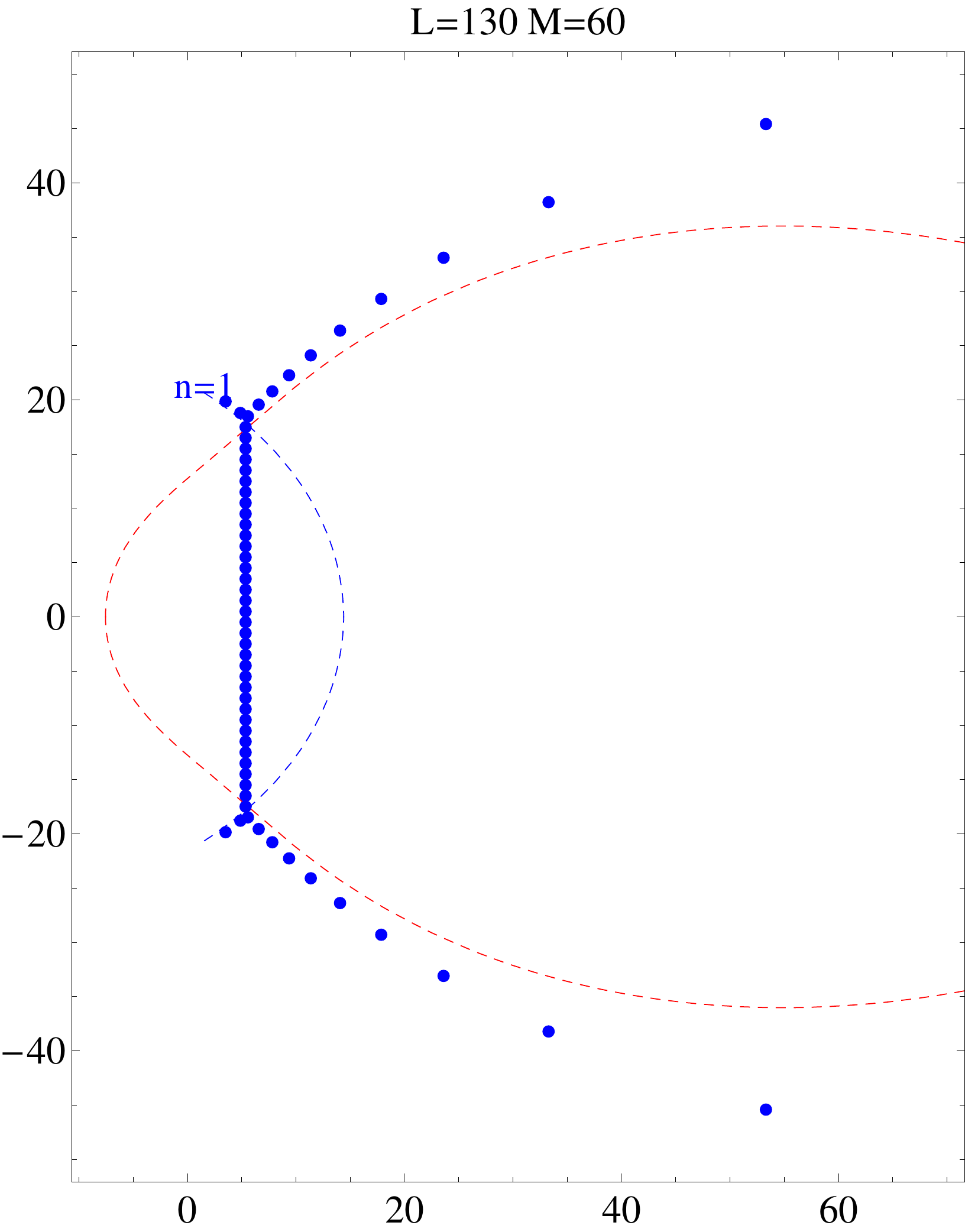}}
\caption{One-cut solutions with mode number $n=1$ and lengths
$L=2925,315,160,130$. Shown are the cut contours of the spectral curve
together with numerical solutions. When the maximum density along the cut
exceeds $|\rho|=i$, the fluctuation point with mode number $n=2$ touches the cut.
When we continue decreasing $L$, a condensate develops in the middle of
the original cut and two extra tails appear. In fact in the scaling
limit $L\sim M$ the two tails to the right are touching at the point
of the virtual excitation, cf.\ \protect\secref{sec:onecut_cond}.}
\label{fig:num1cutn1}
\end{figure}

In the thermodynamic limit this solution is discussed in
\secref{sec:onecut}. An interesting phenomenon which we
expect to arise for this configuration is the appearance of a closed
loop-cut for
filling fractions above $\alpha\indup{cond}$ (see \figref{fig:onebubblecutsb}).

The spectral curve of the one-cut solution can be discussed very
explicitly, in
particular the quasi-momentum is given by a simple algebraic formula
\eqref{eq:onecutqmom} and one can find the shape of the physical branch
cut quite easily. In the first figure of \figref{fig:num1cutn1} the
contour is shown together with the numerical solution for $M=60$ roots
and length $L=2925$. We almost do not see any deviation of the roots
from the asymptotical curve. On the next picture we considerably
increased the filling fraction to $M/L=60/315\approx 0.19$, and one can
see that the distance between roots close to the real axis is just 
slightly above $i$. This means that the filling $\alpha$ is very close
to $\alpha\indup{cond}$ and the stability condition is almost violated.
According to the discussion in \secref{sec:onecut_cond}, this implies
that the fluctuation with mode number $2$ is very close to the cut. In
the figure, we indicate it by a red diamond.

When we continue increasing the filling fraction to values above
$\alpha\indup{cond}$,
the fluctuation point passes through the cut contour and produces a
loop cut. At the
intersection point of the loop cut with the original standard cut we
expect the condensate to be formed. As is demonstrated in
\figref{fig:num1cutn1}, this is indeed the case.

We can almost reach the maximal filling $\alpha=\half$ numerically. As
discussed in \secref{sec:consecutive.global}, this
would correspond to a configuration with a condensate centred at the
origin together with two tails that extend along the imaginary axis
towards the fluctuation point, which has diverged to infinity.

Finally, we notice that the roots belonging to the two tails on the right of the
condensate do not seem to align on the contour predicted by the
spectral curve (red/dashed). In fact this is an
artefact of the finiteness of the number of roots. One can observe that the
situation becomes better when the number of roots $M$ is increased
(while
the filling fraction $M/L$ is held fixed). 
The finite-size effects become very relevant at the
tip of the loop because the density drops to zero at the
fluctuation point forming the tip. 
Close to a standard branch point the roots align on the contour more accurately, 
because the density decreases only as the square root of the distance 
from the branch point, 
whereas at the cusp of the loop the density is proportional 
to the distance itself and hence the
deviation from the spectral curve is more severe.

\paragraph{One-Cut Solution, $n=2$.}

\begin{figure}[tbp]\centering
\resizebox{\textwidth}{!}{%
\includegraphics{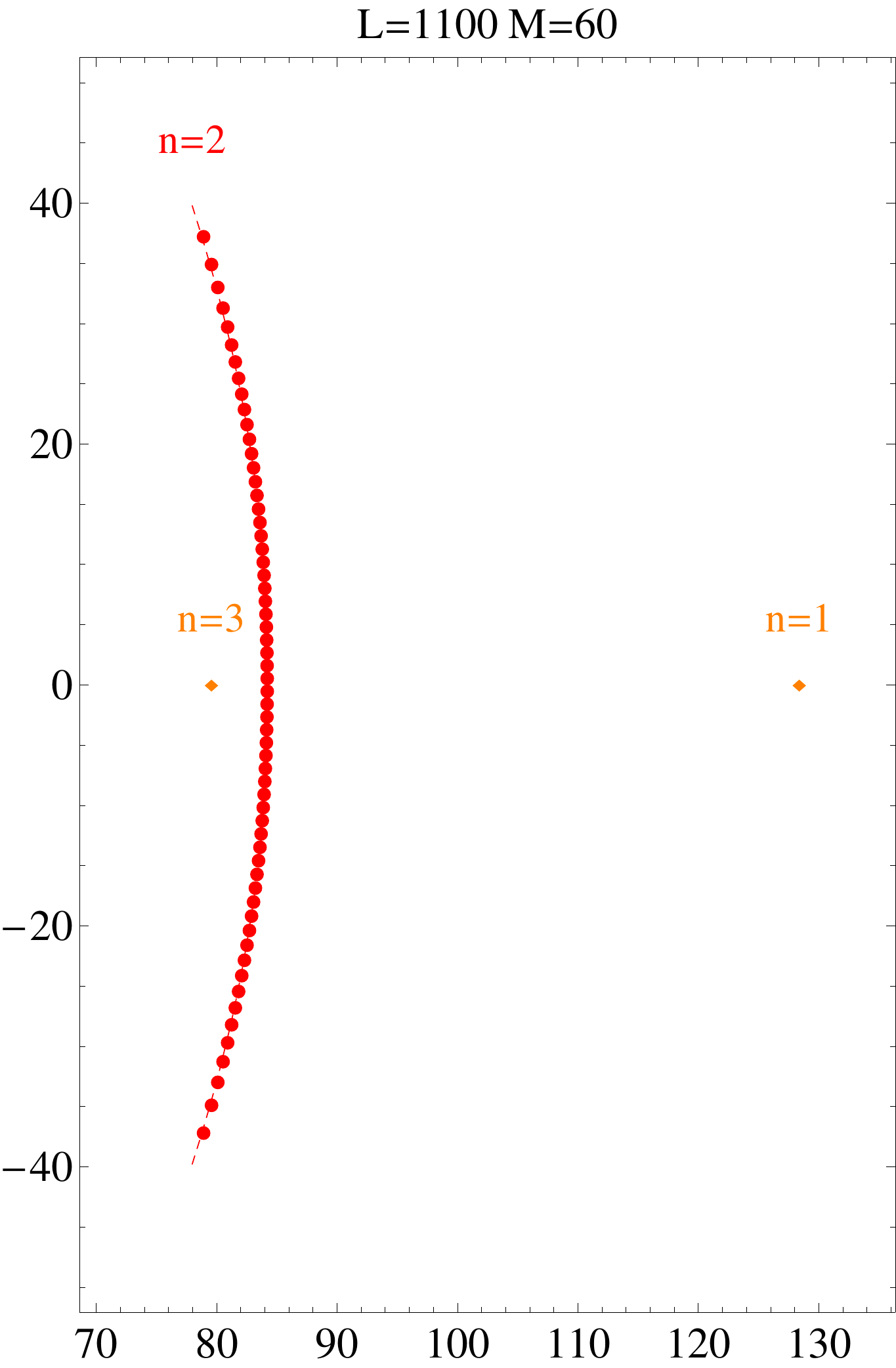}\quad%
\includegraphics{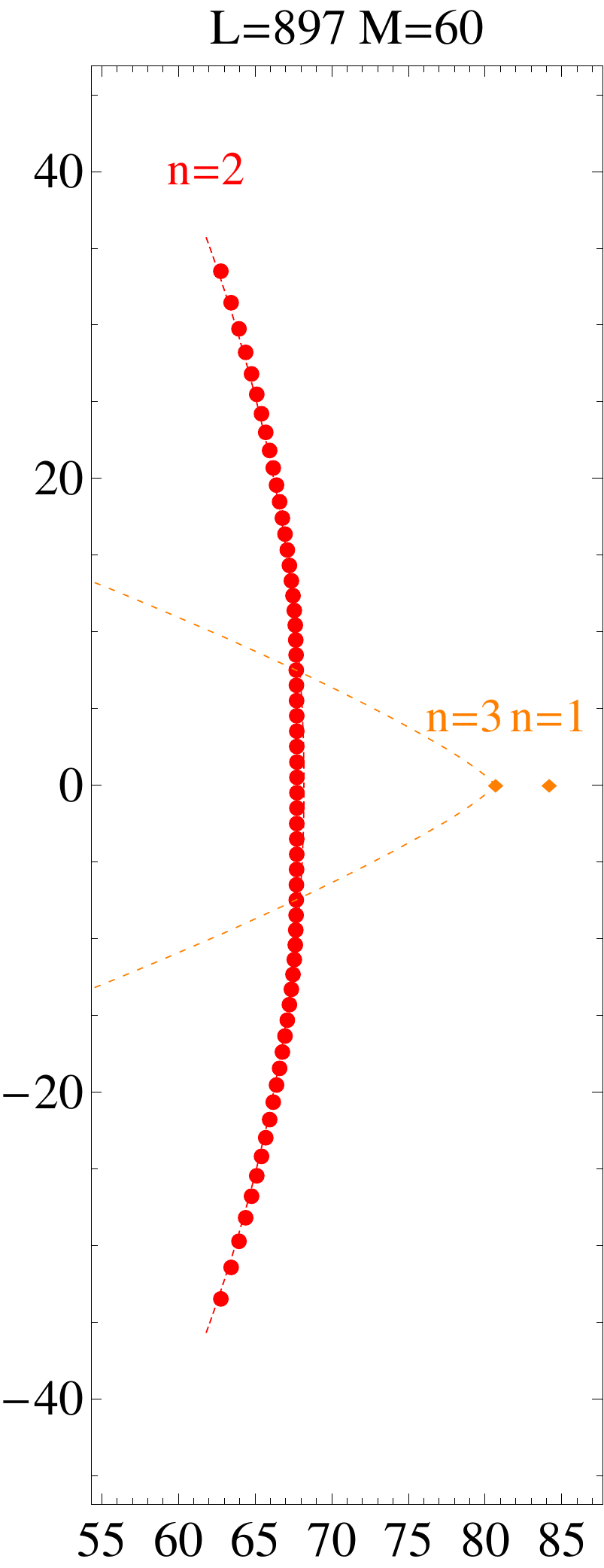}\quad%
\includegraphics{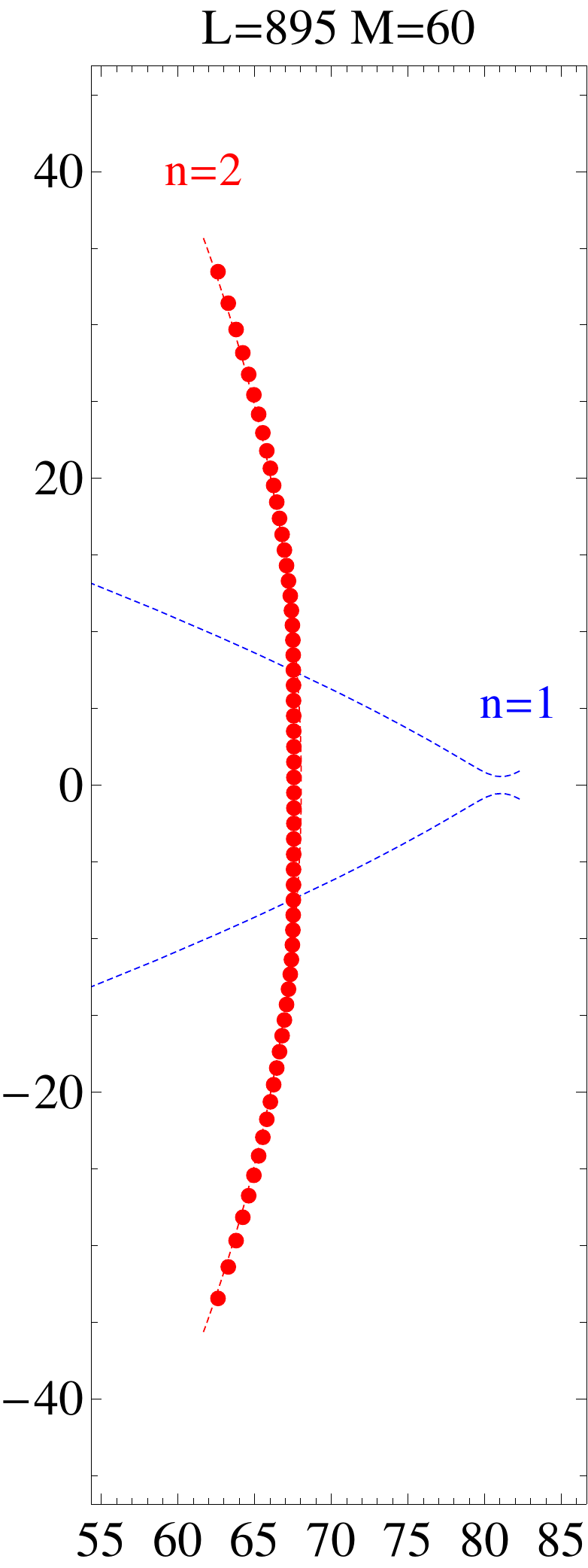}\quad%
\includegraphics{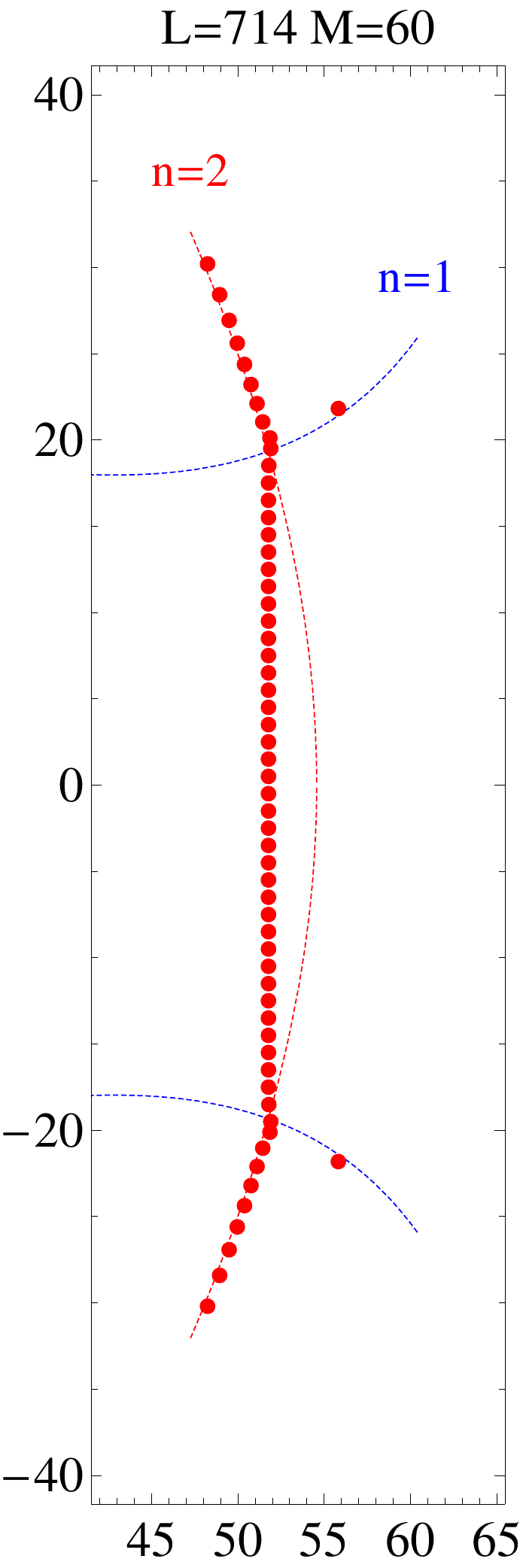}\quad%
\includegraphics{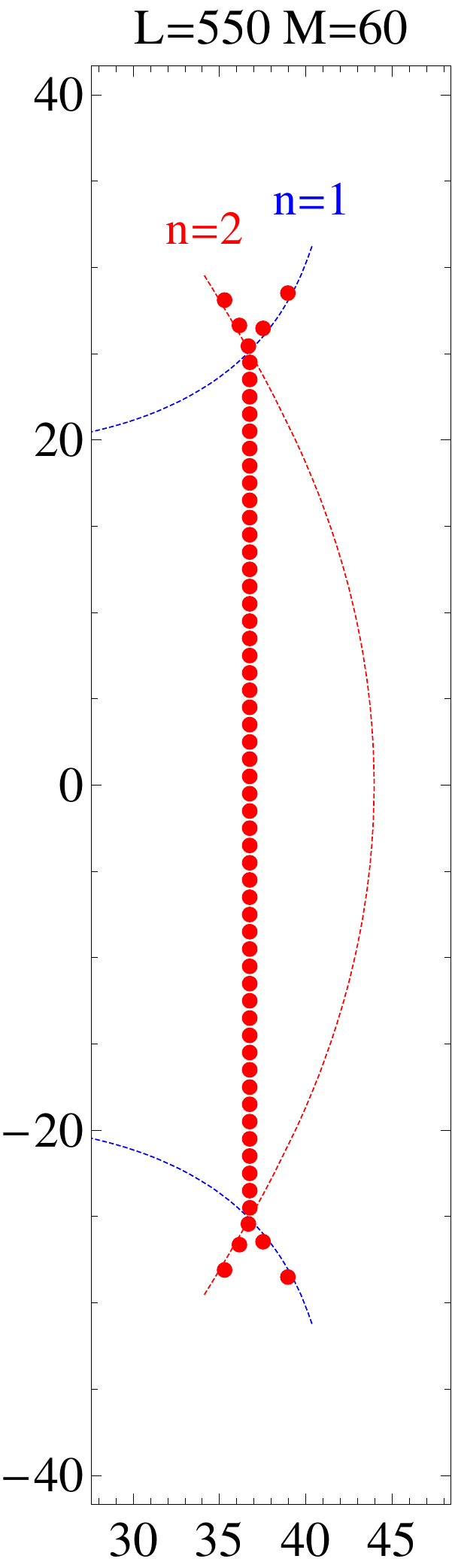}}
\caption{The one-cut configuration with mode number $n=2$ transforms
into a two-cut configuration after the collision of the two
fluctuation points with mode numbers $n=1,3$.}
\label{fig:num1cutn2}
\end{figure}

For larger mode numbers, the one-cut solution behaves similarly to the
$n=1$ case when the filling fraction is increased. However, we expect that
a configuration with $n>1$ exhibits a new type of behaviour when its filling fraction
reaches $\alpha\indup{crit}$ given by \eqref{eq:onecut.alphacrit}. As
discussed in \secref{sec:onecut_cond}, we expect that a phase transition
from one cut to two cuts occurs.

\begin{figure}[tbp]\centering
\includegraphics{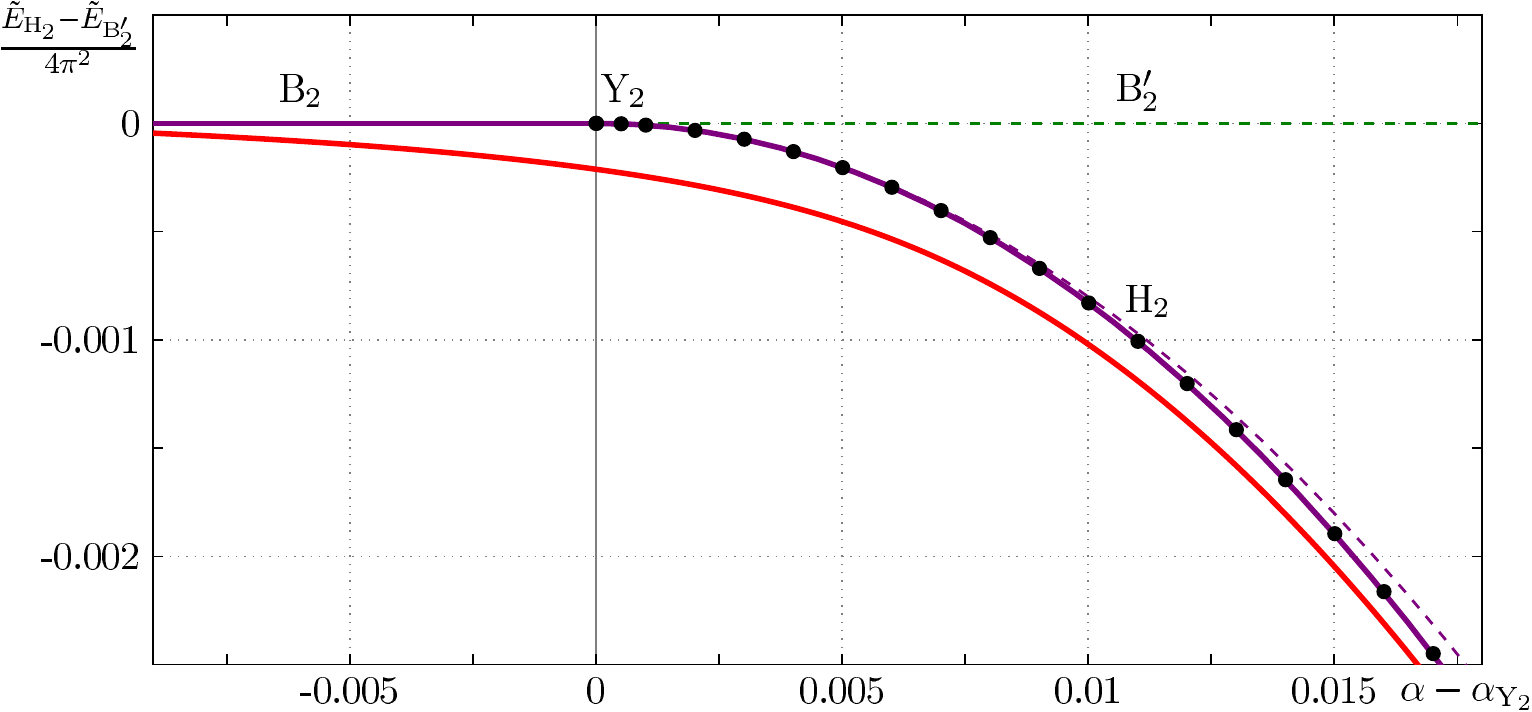}
\caption{Shown is the same diagram as in \protect\figref{fig:pointY_pt}.
In addition, the energies of a series of numerical solutions to
the discrete Bethe equations that start as a one-cut solution with
mode number $n=2$ (red/solid) are displayed. The graph shows that the
one-cut solution (line $\mathrm{B_2}$) indeed transforms
into a two-cut solution (line $\mathrm{H_2}$) when the two fluctuation
points collide (point $\mathrm{Y_2}$). The deviation from the exact
thermodynamic lines is due to the finite length and number of roots in
the numerical solutions.}
\label{fig:pointY_pt-num}
\end{figure}

In \figref{fig:num1cutn2}, we compare the distributions
of numerical Bethe roots with mode number $n=2$ for different filling fractions. We see that the
condensate forms when the nearest fluctuation from the left crosses
the cut contour at
$\alpha=\alpha\indup{cond}$, which is in full agreement with the prediction of
\tabref{tab:maxfill}.
In contrast to the $n=1$ case, there is also a fluctuation point to the
right of the cut (with mode number $n=1$), even at small fillings
$\alpha$. When we increase $\alpha$, the fluctuation point with mode
number $n=3$, which initially was on the left of the standard cut,
passes through the cut and approaches the fluctuation point with mode
number $n=1$ (second figure in \figref{fig:num1cutn2}). Beyond the
point of collision there are two possibilities -- the configuration
could continue as a one-cut solution (with a loop cut with two cusps
that could naturally be decomposed into two standard cuts), or it could
transform into a two-cut solution with lower energy as explained in
\secref{sec:onecut_cond}. We found that in our study always the second possibility
is realised, our numerical method prefers the two-cut configuration with
lower energy. It seems that this configuration is numerically more
stable. In particular we see that in the last two pictures of
\figref{fig:num1cutn2} the roots
perfectly follow the two standard contours of the two-cut spectral
curve. Accordingly, \figref{fig:pointY_pt-num} shows that the energy of
the numerical configuration after the point of collision matches the energy
of the two-cut configuration (up to deviations due to finite-size
effects), which clearly indicates that the phase
transition is indeed taking place.

\paragraph{Two-Cut Solution, $n=1,2$.}

\begin{figure}[tb]\centering
\resizebox{\textwidth}{!}{%
\includegraphics{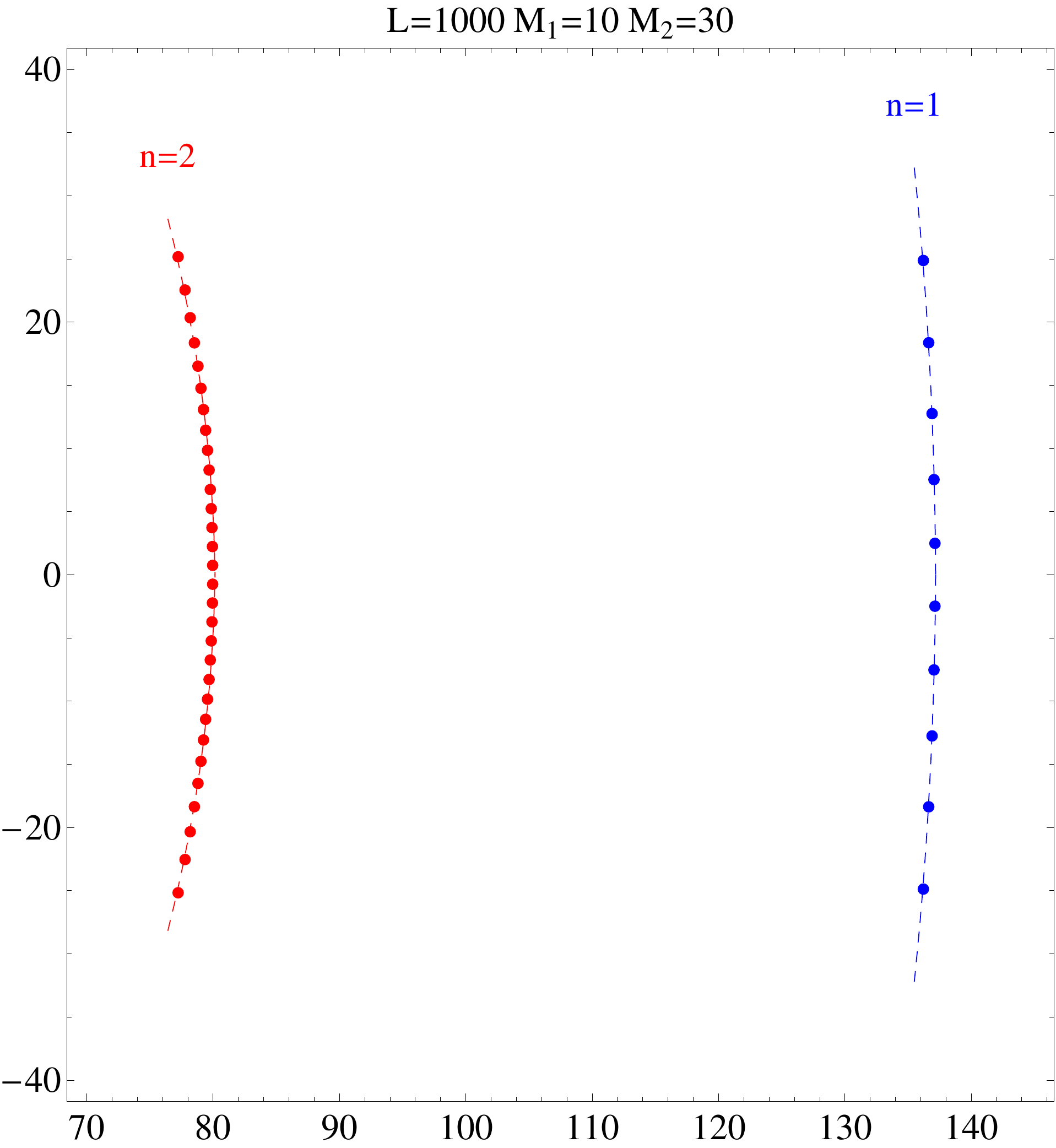}\quad%
\includegraphics{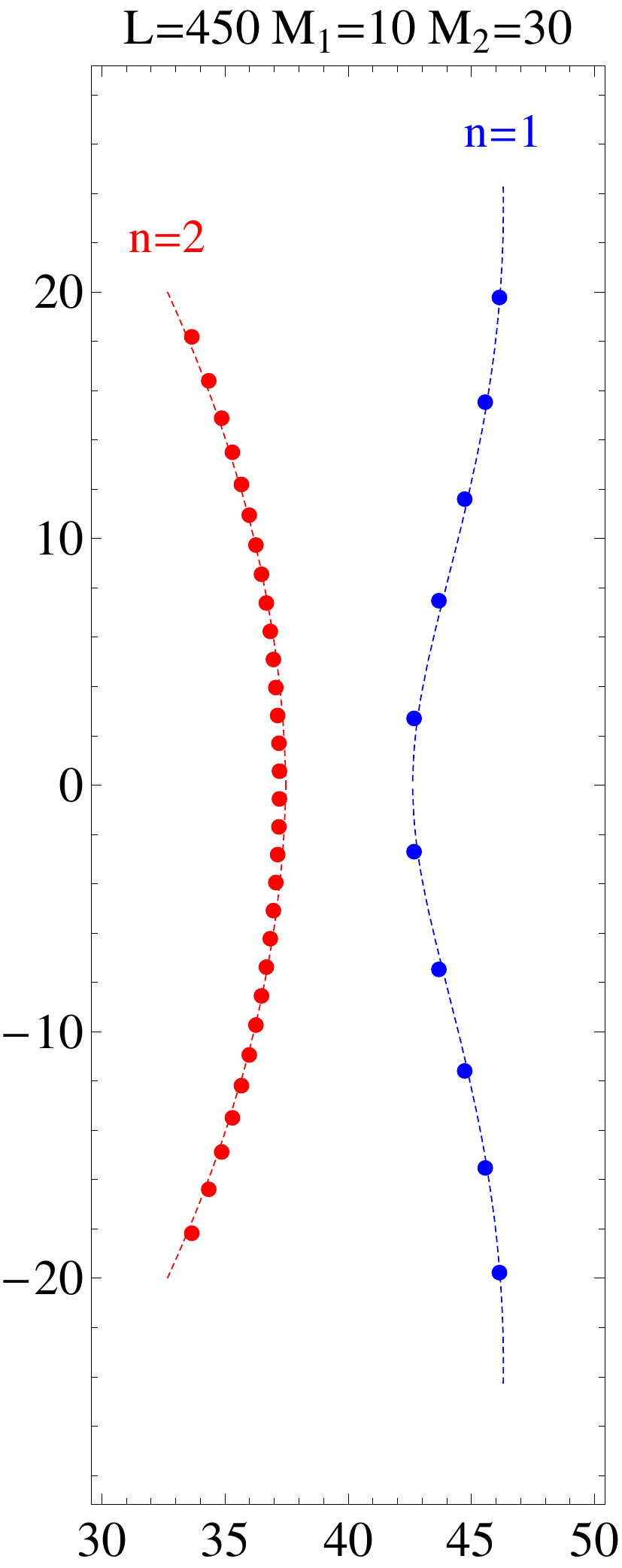}\quad%
\includegraphics{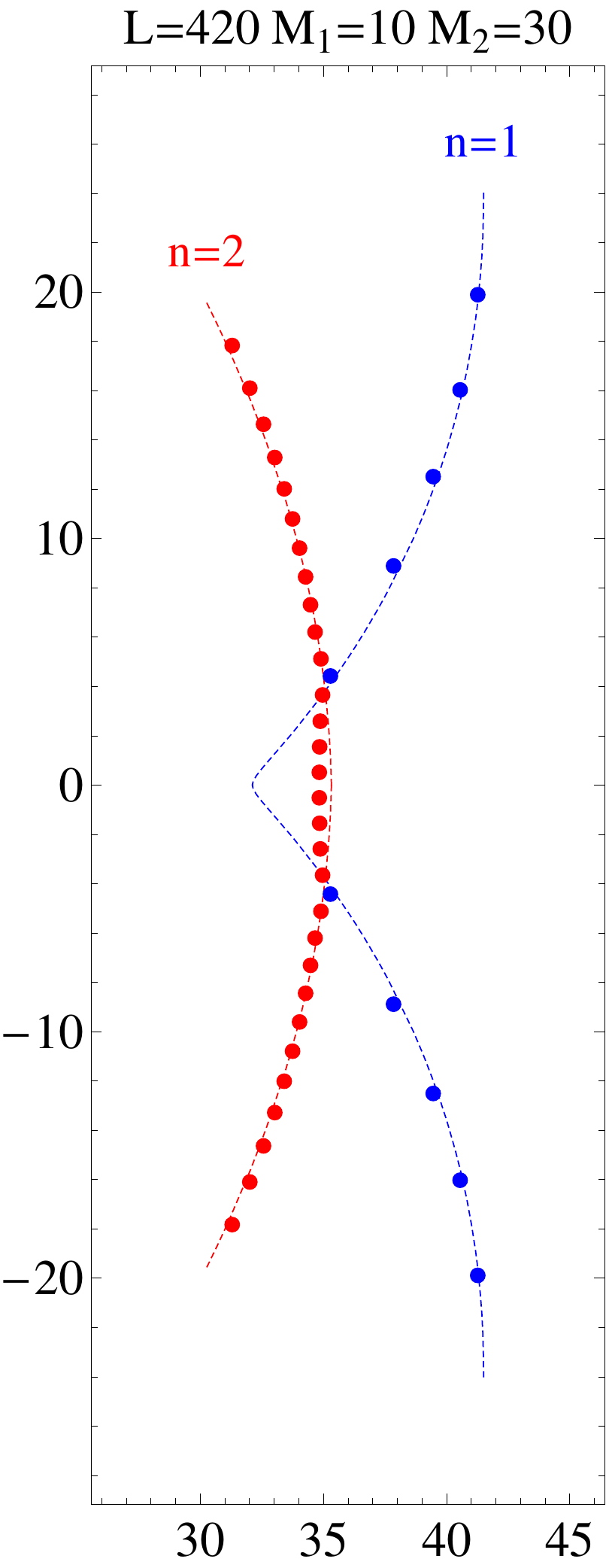}\quad%
\includegraphics{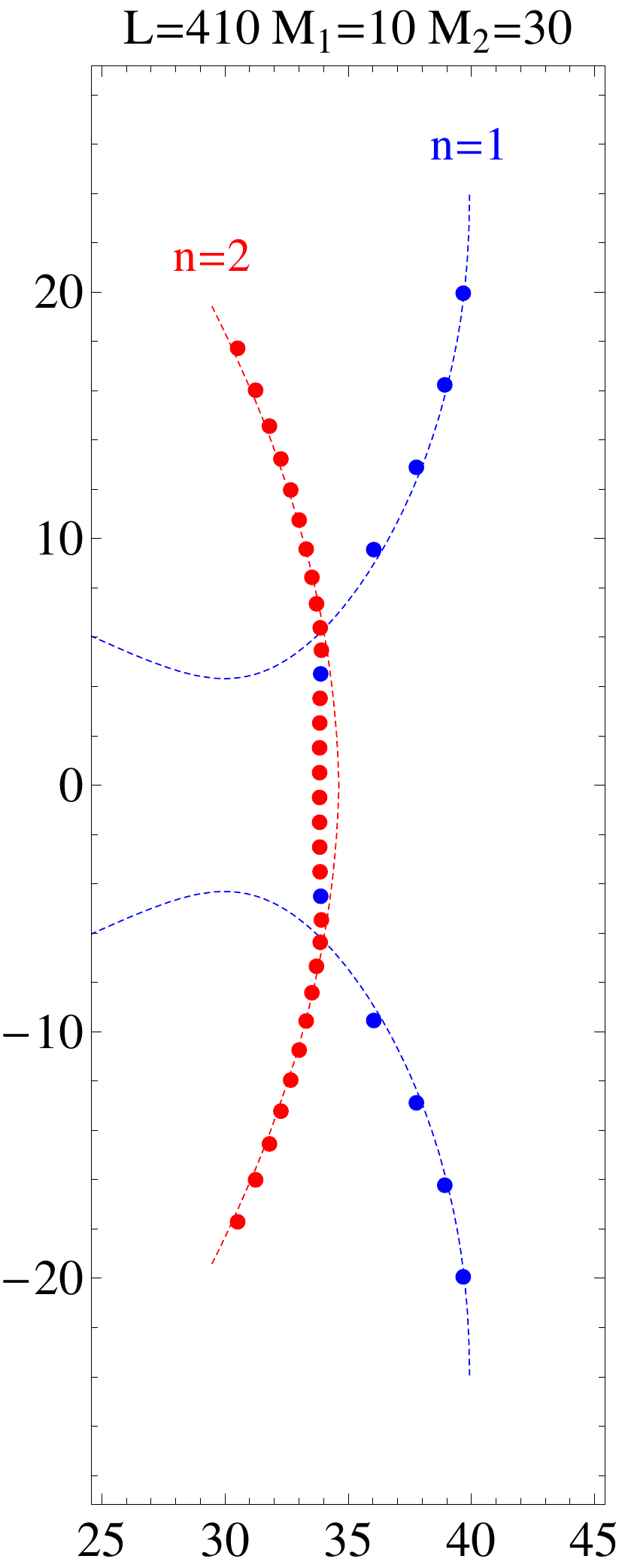}\quad%
\includegraphics{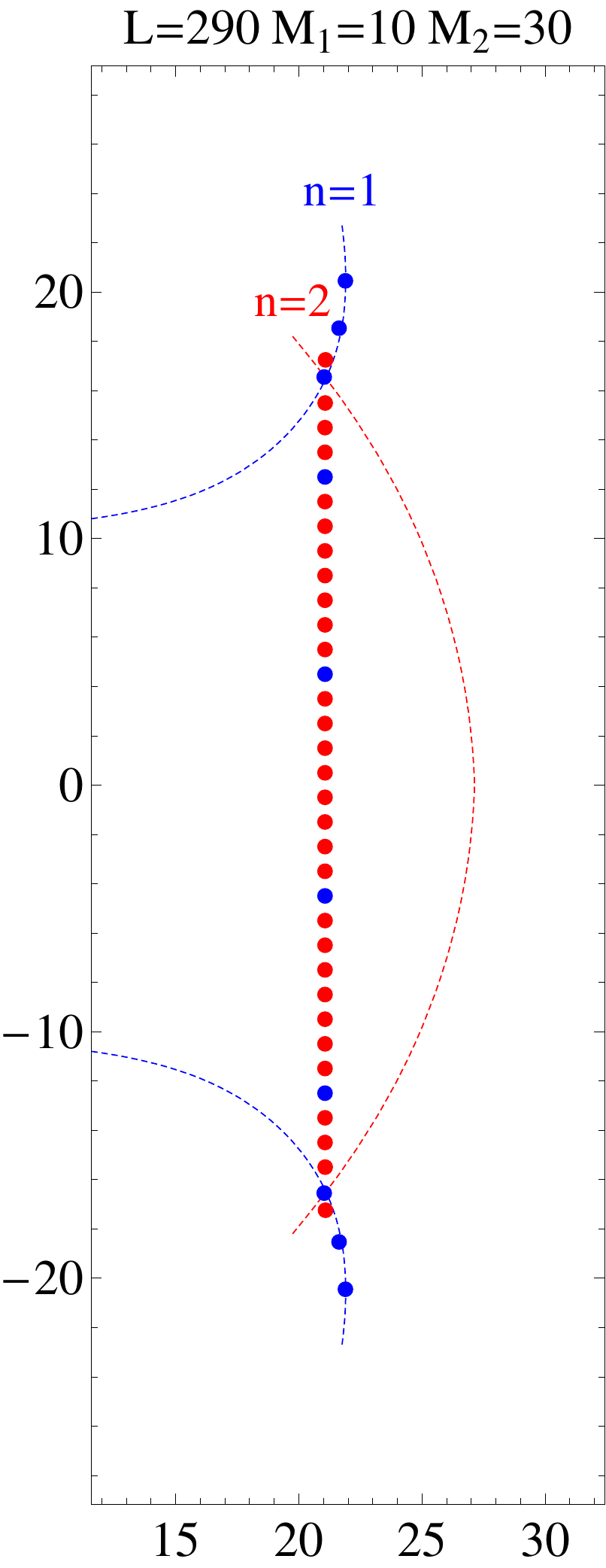}}
\caption{Two-cut solutions with consecutive mode numbers
$n=1$ (right, blue) and $n=2$ (left, red)
for different values of the length $L$. The intersection points of the two standard
contours are connected by a vertical condensate cut.}
\label{fig:num2cutn12}
\end{figure}

The spectral curve for two-cut solutions 
was constructed in \secref{sec:twocut} and is less simple than the
one-cut curve. The moduli space of two-cut
configurations is also much more complicated, for the case of
consecutive mode numbers it is described in detail
in \secref{sec:consecutive}. Numerically, however, it is not
difficult to construct a solution with an arbitrary number of cuts. We
mainly want to show here that, in accordance
with the prediction of \secref{sec:consecutive}, an intersection of the two cuts leads to
the formation of a condensate that connects the two intersection points.

In \figref{fig:num2cutn12} we compare numerical two-cut
solutions with mode numbers $n=1,2$ with the predictions from the
spectral curve. In the first two pictures the two cuts
are disjoint and hence we are in region $\mathrm{I}_{12}$ of the phase diagram
shown in \figref{fig:12_eva}. The last three configurations belong to region
$\mathrm{II}_{12}$, and we see that indeed a condensate with roots
separated by approximately $i$ connects
the intersection points of the two contours.

On the third picture, the
right (blue) $n=1$
cut does not encircle the origin of the complex plane, but it has a
cusp at
the point of intersection with the real axis. This means that this
configuration lies exactly on the line $\mathrm{G}_{12}$.
In the last configuration shown in \figref{fig:num2cutn12}, the branch
points of the $n=2$ cut (red) are almost touching the $n=1$ cut (blue)
which means that the configuration is very close to the line $\mathrm{C}_1$.

\paragraph{Two-Cut Solution, $n=1,3$.}

\begin{figure}[tb]\centering
\resizebox{\textwidth}{!}{%
\includegraphics{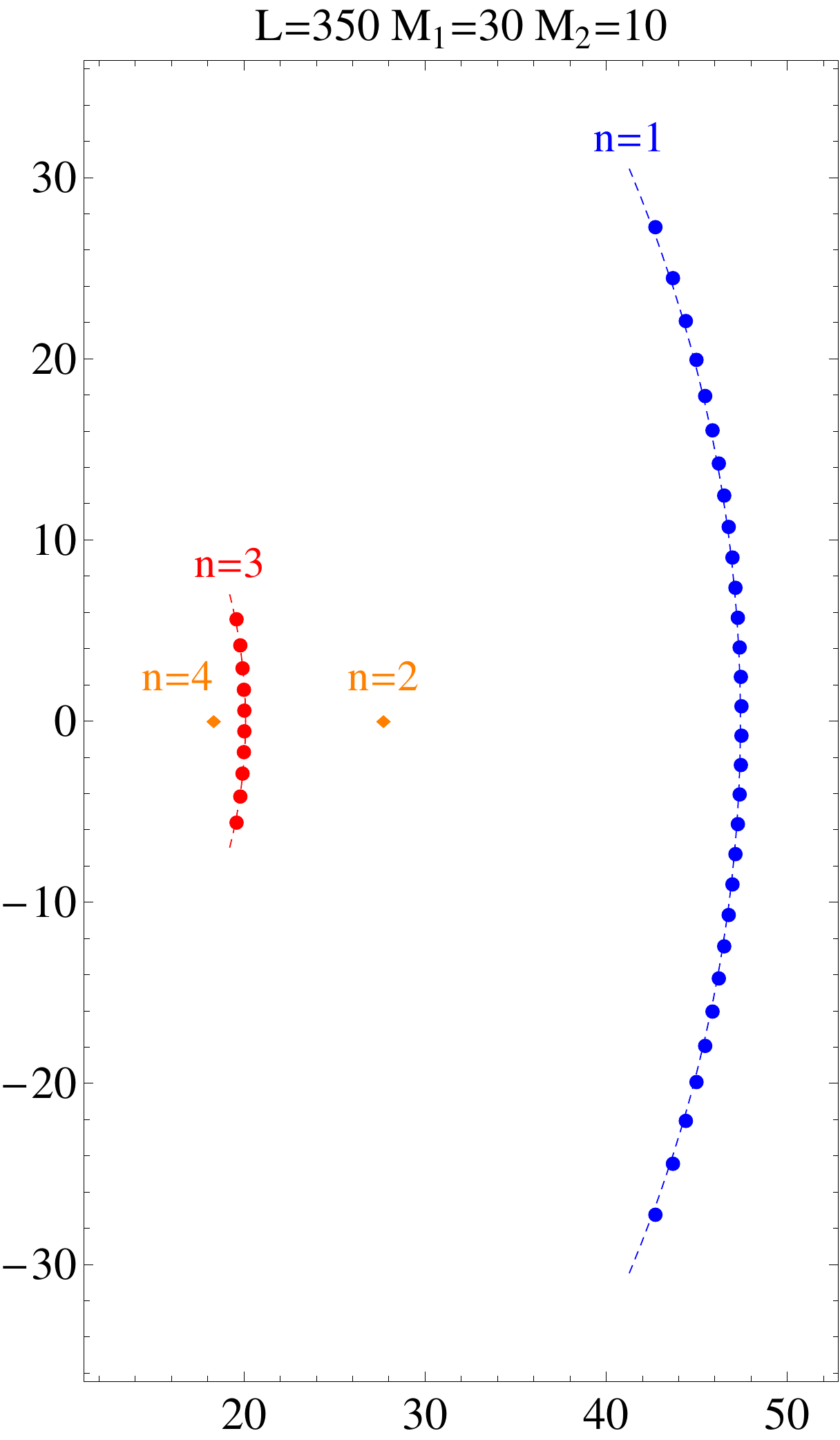}\quad%
\includegraphics{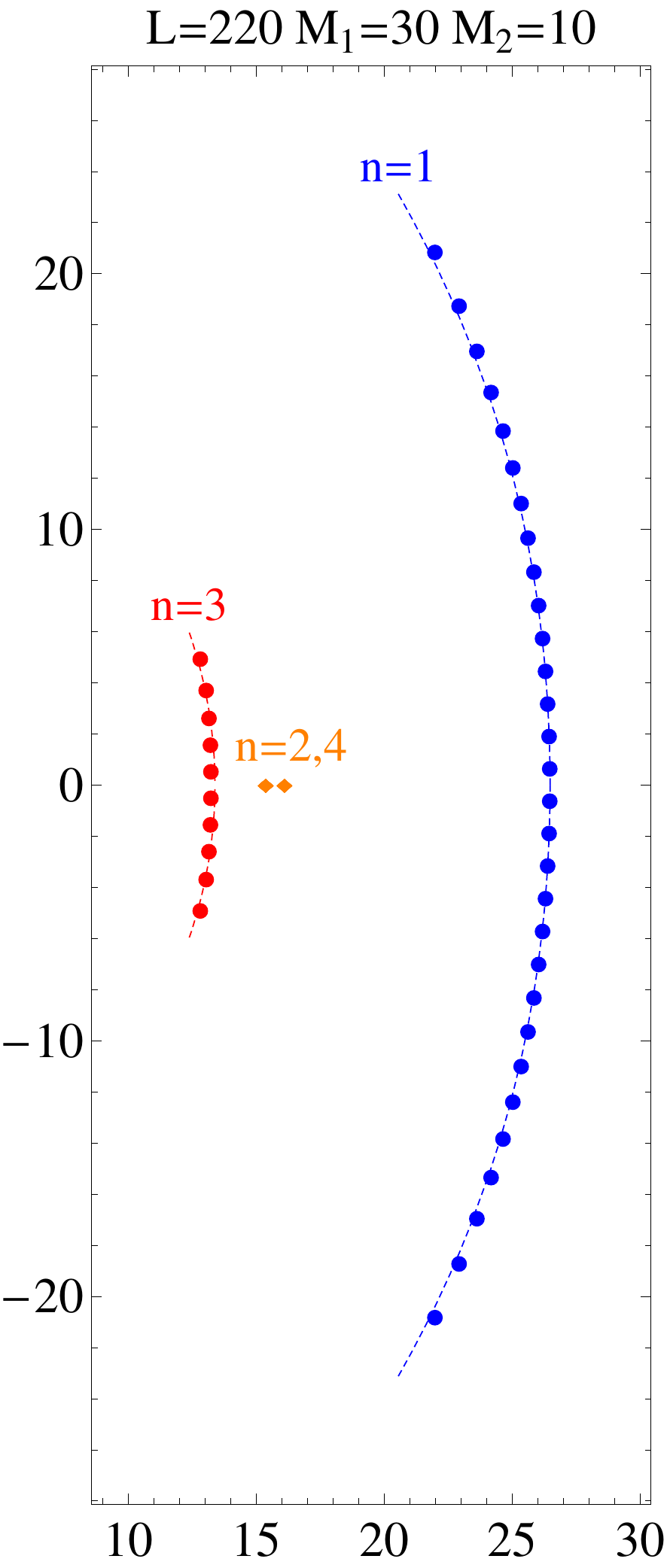}\quad%
\includegraphics{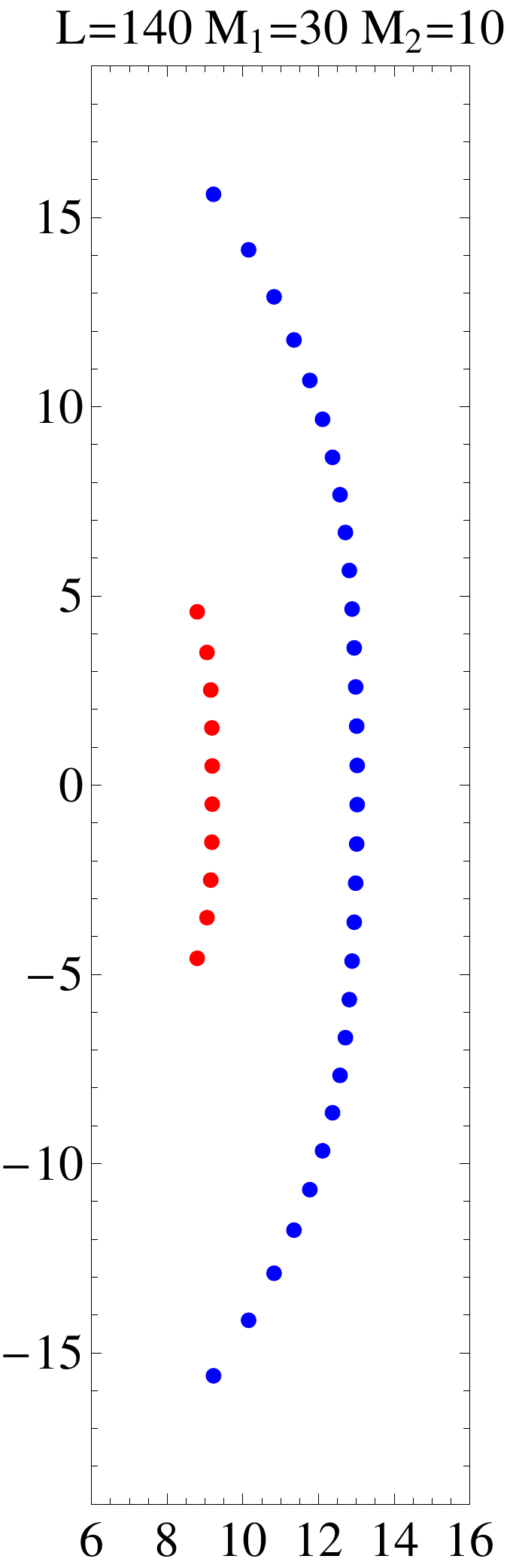}\quad%
\includegraphics{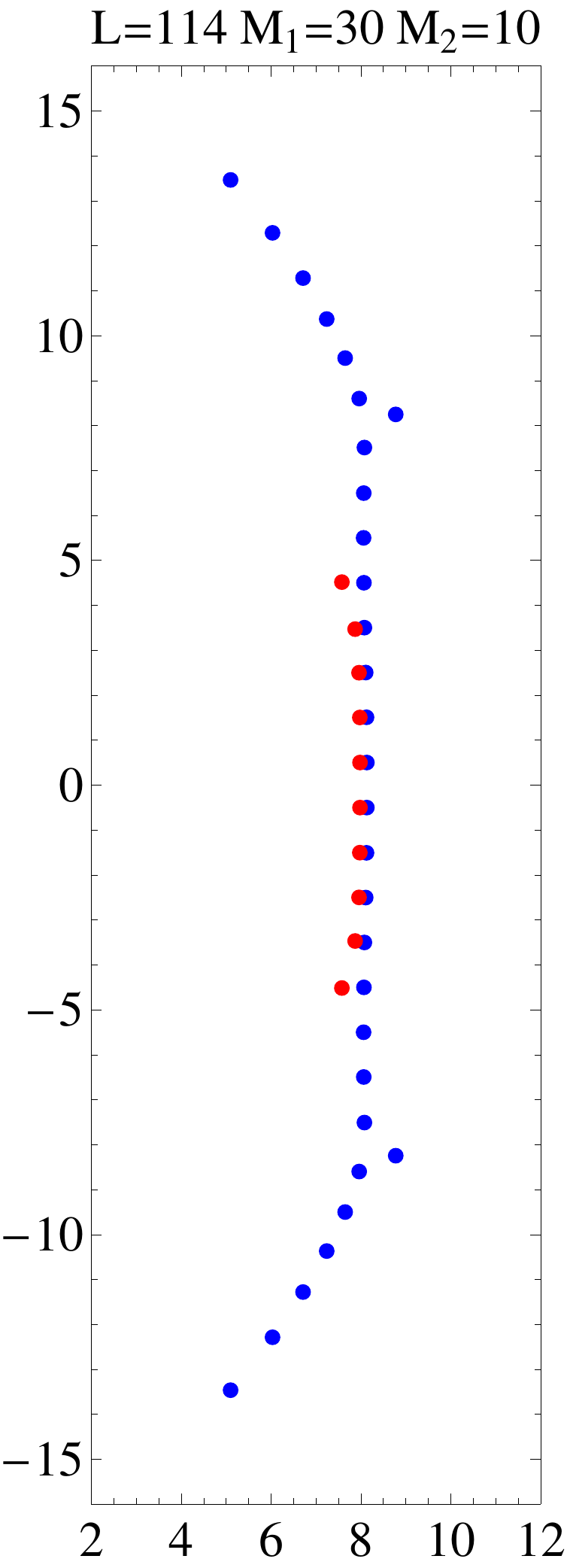}\quad%
\includegraphics{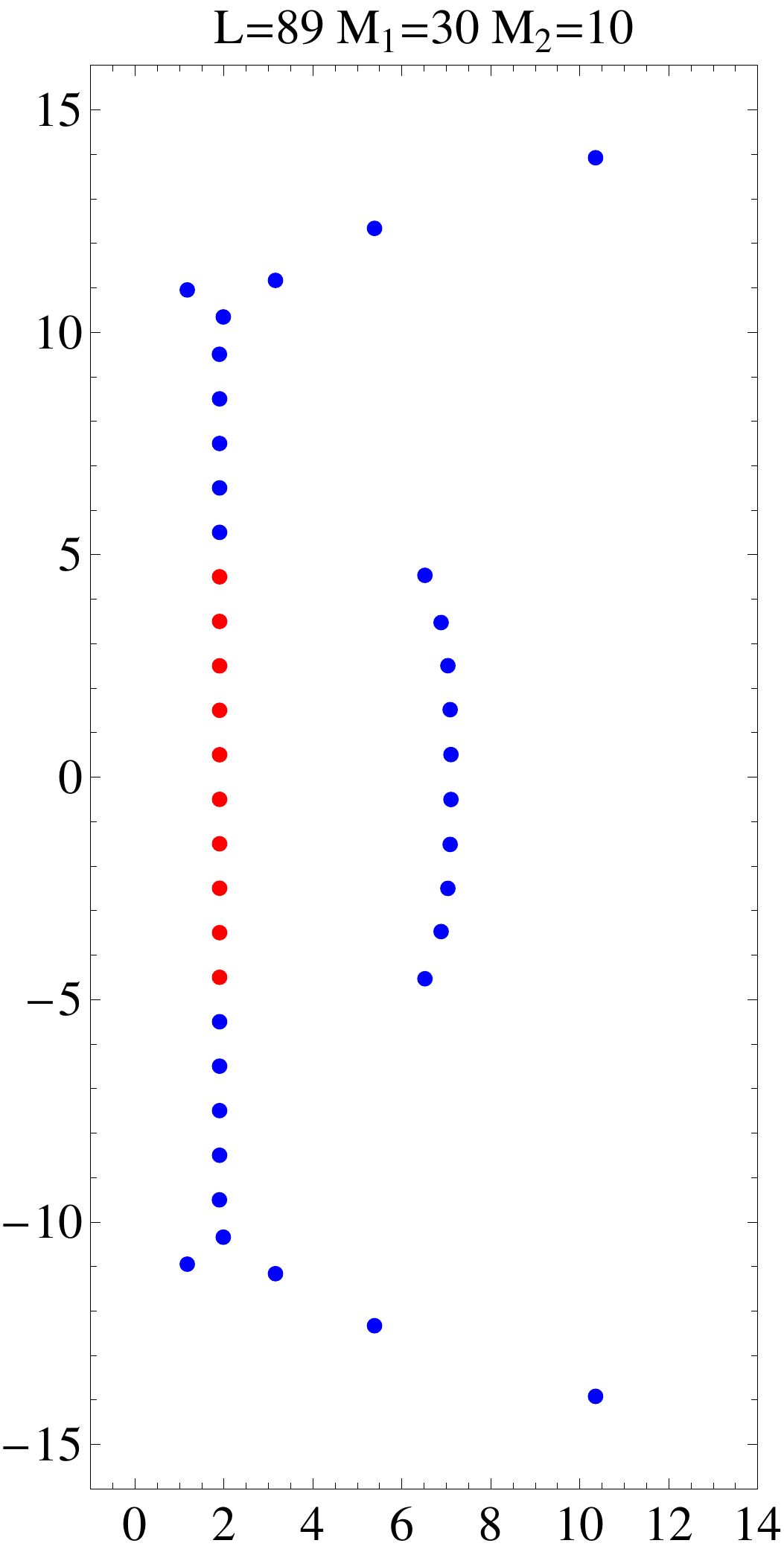}}
\caption{Two-cut solutions with non-consecutive mode numbers
$n=1$ (blue) and $n=3$ (red) for different values of the length $L$.
Both cuts develop condensates
before they intersect each other. We see that the passage of one cut
through the condensate core of the other does not affect the relative
positions of the roots inside each cut.}
\label{fig:num2cutn13}
\end{figure}

Two-cut solutions with non-consecutive mode numbers allow for even richer
properties. As is explained in
\secref{sec:13}, for some values of filling fractions
both cuts can develop condensate cores before intersecting with each other.
At higher fillings the cut with the higher mode number, which
initially lies further left, can pass
through the condensate core of the other cut. As explained in
\secref{sec:13}, this should not affect the inner structure of the cuts.

From the distributions of numerical roots for different values of the
fillings in \figref{fig:num2cutn13}, we see indeed that the
group of $n=3$ Bethe roots (left, red) reaches the condensate core of the
$n=1$ roots (right, blue) and passes through the condensate.%
\footnote{In our example, the (red) $n=3$ roots replace some of the
(blue) roots in the condensate core while the latter take the places
of the $n=3$ roots. This is merely an artefact of the numerical
procedure and has no impact on the resulting configuration.}
These numerical solutions confirm that the passage of a cut complex
through a condensate core does not affect the structure of the
individual complexes, i.e.\ their shape and their root densities.

We did not plot cut contours
for all configurations in this case, because similarly as in the
previous example, the configuration undergoes a phase transition.
Namely, this series of configurations follows the line
$P/2\pi\alpha=1.5$, which is slightly above the critical value 
$\sim 1.4$ so that we are crossing the line $\mathrm{C}_{24}$ of the
\figref{fig:13_poava}. When the two fluctuations denoted by the orange
diamonds ($n=2,4$) collide, the corresponding spectral curve transforms
into a three-cut curve, as is expected from the discussion in
\secref{sec:13}. 
Note that we have no means to plot the exact contours
for three-cut configurations and thus these lines
are missing from the figures.

\paragraph{Conclusions.}

We have presented numerical solutions for various types of 
configurations discussed in this paper, and in all cases we observe 
a perfect match with the analytical predictions of the spectral curve. 
This confirms the validity of the analysis of
the thermodynamic limit in the previous sections.

\subsection{One-Cut Solution and Finite-Size Effects}
\label{sec:numfinite}

In this paper we mainly focus on the strict thermodynamic limit. 
However, the numerical solutions containing
a finite number of Bethe roots allow us to easily analyse 
the leading correction to the thermodynamic limit.
In string theory they correspond to the semiclassical energy shift,
see \cite{Dorey:2006zj,Dorey:2006mx,Vicedo:2008jy} for a general discussion
of canonical quantisation for the spectral curve.

The energy shift was studied analytically for one-cut configurations
of the Heisenberg spin chain in \cite{Beisert:2005mq,Hernandez:2005nf,Beisert:2005bv}.
The result reads
\begin{equation}
E^{(1)}=
4\pi^2\left[\sum_{k=1}^\infty \left( k^2\sqrt{1-\frac{4
n^2\alpha(1-\alpha)}{k^2}}-k^2-2n^2\alpha(1-\alpha)\right)-n^2\alpha(1-\alpha)\right].
\label{eq:E1}
\end{equation}
Note that this result equals the sum of zero-mode energies
of the fluctuation modes in \eqref{eq:MomEngFluct}
upon zeta-function regularisation of the sum. 
The result may in principle include exponentially suppressed 
contributions, see \cite{Schafer-Nameki:2005tn,Schafer-Nameki:2006gk}.
It is however conceivable as well that these corrections are
absent in the Heisenberg/Landau--Lifshitz spectral curve
which is considerably simpler than the spectral curve 
for the principle chiral model/string theory.

The above energy shift was obtained assuming that the roots are distributed along
a single cut and it is a priori questionable 
if this result remains true for the
case $\alpha>\alpha\indup{cond}$ when the loop cut appears and the
distribution of roots is more sophisticated.
We can however use the intuition gained studying the strict thermodynamic limit:
We saw that the energy is described by the
same analytical expression \eqref{eq:onecut_EP} before and after the
condensate is formed. Therefore we expect \eqref{eq:E1} to be valid
also in the full range of $\alpha$'s.

Here we perform a numerical test of this natural guess. In
\tabref{tab:E1} we give a series of deviations of numerical values of
the energy from their asymptotical values \eqref{eq:onecut_EP} for
different numbers of roots and for two fixed filling fractions below
and above the point of condensation $\alpha\indup{cond}\approx 0.21$.
To compute $\delta E_\infty$ we fit the values of the energies from the
table for finite $M$ by a constant and inverse powers of $M$.
%
%
The value of the constant is precisely $\delta
E_\infty$ given in the table. For both filling fractions we see a
perfect match with the value given by
\eqref{eq:E1}, which shows that the correction \eqref{eq:E1} is also
valid for fillings beyond $\alpha\indup{cond}$.

\begin{table}[tb]\centering
$\begin{array}{|c||c|c|c|c|c|c|c||c|}
  \hline
  \alpha & \delta E_{10} & \delta E_{20} & \delta E_{30} & \delta E_{40}& \delta E_{50} & \delta E_{60} & \delta E_{\infty}& E^{(1)} \\
  \hline
  0.10 & \phantom{+}2.372 & \phantom{+}2.356 & \phantom{+}2.351 & \phantom{+}2.349 & \phantom{+}2.347 & \phantom{+}2.346 & \phantom{+}2.3411 & \phantom{+}2.3411 \\
  \hline
  0.33  & -2.273 & -2.472 & -2.539 & -2.573 & -2.594 & -2.608 & -2.6768 & -2.6769 \\
  \hline
\end{array}$
\caption{Deviations $\delta E_M$ of the numerical energy from its asymptotical value
for different numbers $M$ of Bethe roots and for two fixed filling
fractions $\alpha$ above and below $\alpha\indup{cond}\approx 0.21$.
$\delta E_\infty$ is the extrapolation to an infinite number of roots.
The value $E^{(1)}$ \protect\eqref{eq:E1} is given for comparison.}
\label{tab:E1}
\end{table}

A possible explanation for this could be based on the algebraic curve.
The algebraic curve appearing in the thermodynamic limit can be
modified to reproduce also the leading quantum corrections
\cite{Beisert:2005bv,Gromov:2007ky}. The same type of analysis we
have performed for the leading order hopefully can be applied for this
corrected curve as well.

\section{Conclusions and Outlook}
\label{sec:conc}

In this article we have considered the general solution
(\secref{sec:spectral_curves}) for the spectrum of the Heisenberg
ferromagnet in the thermodynamic limit by means of spectral curves.
Equivalently, we have studied the semiclassical Landau--Lifshitz model
on $S^2$, see \appref{sec:ll-model}. In particular, we have
constructed the general \emph{two-cut} (elliptic) solution
(\secref{sec:twocut}) and studied its properties
(\secref{sec:consecutive,sec:nonconsec}). This allowed us to shed
light on the issue of quantum \emph{stability} of classical solutions
(\secref{sec:onecut,sec:consecutive}) brought up in
\cite{Beisert:2005mq,Beisert:2005bv}. We have also tested our claims by
constructing some numerical solutions (\secref{sec:numerics}).
Let us summarise our findings briefly:

\medskip

\begin{figure}
  \centering
  \subfloat[$\alpha_{1,2}$ small]{\quad\includegraphics{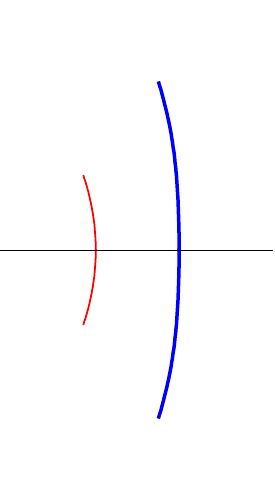}\quad\label{fig:conctwocuta}}\qquad\qquad
  \subfloat[$\alpha_{1,2}$ large]{\quad\includegraphics{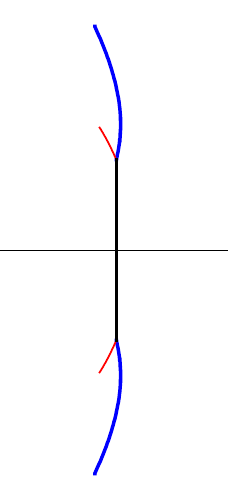}\quad\label{fig:conctwocutb}}
  \caption{Typical configurations of the two main types of two-cut
  solutions. When the fillings $\alpha_{1,2}$ are small, the two cuts
  are disjoint \protect\subref{fig:conctwocuta}. When the fillings
  increase, the cuts attract each other and eventually collide,
  forming a condensate at their common centre
  \protect\subref{fig:conctwocutb}.}
  \label{fig:conctwocut}
\end{figure}%

The two-cut solution is specified by two mode numbers $n_{1,2}$ and
two fillings $\alpha_{1,2}$. Its quasi-momentum is an elliptic
function which encodes all the relevant classical observables. In
order to quantise this solution, one needs to understand the positions
of the two branch cuts which correspond to the distribution of Bethe
roots. We have studied these in detail and we have found a rich
``phase space'' (\figref{fig:poa_vs_p,fig:13_poava}) of different
configurations depending on the fillings.

There are two main configurations, see \figref{fig:conctwocut}: When
the fillings are sufficiently small, there will be two separate branch
cuts, see \figref{fig:conctwocuta}. When the fillings are increased,
the branch cuts bend towards each other so that they eventually would
overlap. Instead of overlapping, a straight vertical \emph{condensate}
of Bethe roots with constant density forms. The configuration looks
like a Bethe string but with four curved tails of lower density, see
\figref{fig:conctwocutb}.

Classically there is no quantitative difference between the two types
of solutions, but the condensate serves several purposes for the
quantum model: It sets an upper bound to the density of Bethe roots
which is an important feature for quantum stability. In some sense,
the condensate ``hides'' some potentially dangerous distributions of
Bethe roots on an unphysical sheet of the spectral curve. Moreover,
solutions with different mode numbers (but equal $n_1-n_2$) can be
connected analytically via the condensate. For example, there is only
a \emph{single} connected phase space for solutions with consecutive
mode numbers ($n_2=n_1+1$). Finally, we have observed that the
naturally continued overall momentum $P$ never crosses the boundary of
the first Brillouin zone; it is apparently bounded by $|P|\leq \pi$
and the condensate formation plays an important role. At $P=\pm \pi$
the two-cut solution (with consecutive mode numbers) flattens out onto
the imaginary axis where it connects smoothly the two regimes with
$P>0$ and $P<0$.

\medskip

\begin{figure}[t]
  \centering
  \subfloat[$\alpha<\alpha\indup{cond}$]{\includegraphics{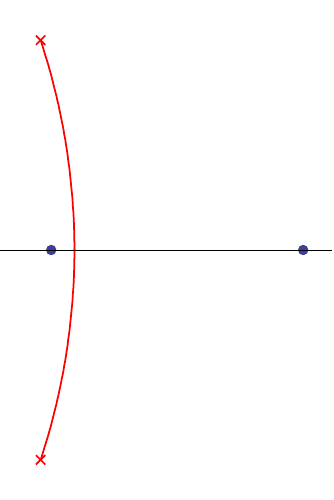}\label{fig:conconebubblecutsa}}\qquad
  \subfloat[$\alpha\indup{cond}<\alpha<\alpha\indup{crit}$]{\qquad\includegraphics{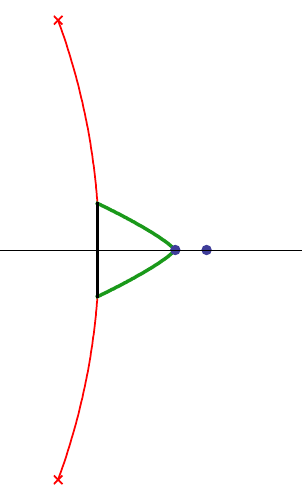}\qquad\label{fig:conconebubblecutsb}}\qquad
  \subfloat[$\alpha\indup{crit}<\alpha$]{\includegraphics{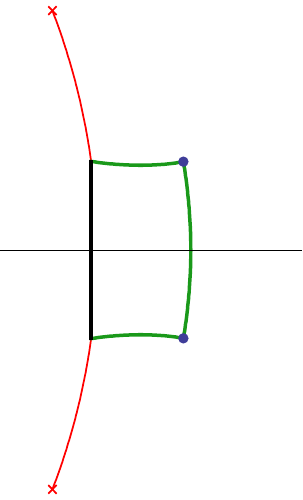}\label{fig:conconebubblecutsc}}
  \caption{The three qualitatively different types of one-cut
  solutions. For small fillings, the single cut is manifestly stable
  \protect\subref{fig:conconebubblecutsa}. At intermediate
  fillings, the cut obtains a condensate and an additional closed loop
  cut with a cusp; a potential instability is hidden in the
  condensate \protect\subref{fig:conconebubblecutsb}. When the
  filling becomes large, the classical solution acquires unstable
  fluctuation modes and the closed loop cut has two cusps
  \protect\subref{fig:conconebubblecutsc}. Nevertheless, the
  configuration is physical, albeit with non-minimal energy; see the
  discussion in the main text.}
  \label{fig:conconebubblecuts}
\end{figure}%

\begin{figure}[t]
  \centering
  \subfloat[$\alpha<\alpha\indup{cond}$]{\includegraphics{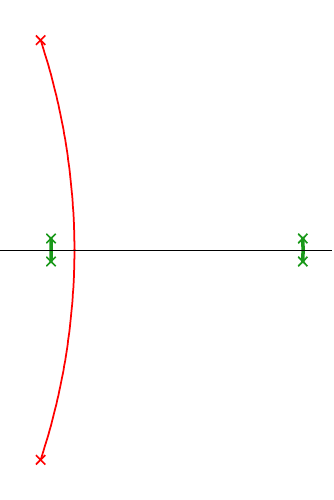}\label{fig:conconebubblecutsrega}}\qquad
  \subfloat[$\alpha\indup{cond}<\alpha<\alpha\indup{crit}$]{\qquad\includegraphics{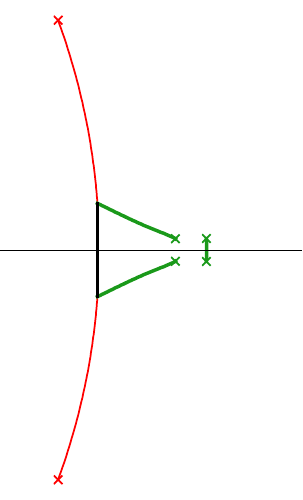}\qquad\label{fig:conconebubblecutsregb}}\qquad
  \subfloat[$\alpha\indup{crit}<\alpha$]{\includegraphics{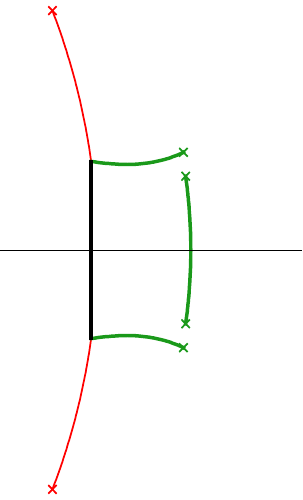}\label{fig:conconebubblecutsregc}}
  \caption{The configurations in
  \protect\figref{fig:conconebubblecutsb,fig:conconebubblecutsc}
  should be understood as degenerate cases of two- and three-cut
  solutions. This can be seen by the addition of small but macroscopic numbers
  of Bethe roots to the fluctuation points, which ``regularises''
  the configurations. The resulting configurations in
  \protect\subref{fig:conconebubblecutsregb}, \protect\subref{fig:conconebubblecutsregc}
  are perfectly well-behaved two- and three-cut solutions. Exciting
  the fluctuation points of the solution in
  \protect\figref{fig:conconebubblecutsc} lowers the energy, which is
  consistent with the (only seemingly) unstable excitation modes.}
  \label{fig:conconebubblecutsreg}
\end{figure}%

\begin{figure}[t]
  \centering
  \subfloat[$\alpha<\alpha\indup{cond}$]{\includegraphics{regularization/conconecut1reg}\label{fig:conconebubblecutsphysa}}\qquad
  \subfloat[$\alpha\indup{cond}<\alpha<\alpha\indup{crit}$]{\qquad\includegraphics{regularization/conconecut2reg}\qquad\label{fig:conconebubblecutsphysb}}\qquad
  \subfloat[$\alpha\indup{crit}<\alpha$]{\includegraphics{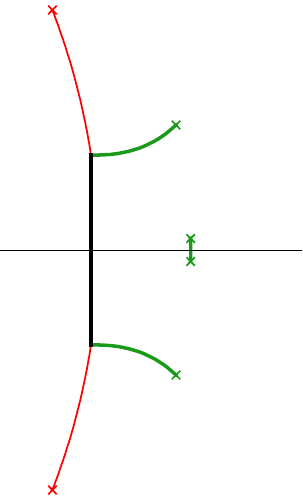}\label{fig:conconebubblecutsphysc}}
  \caption{The proper physical continuation of the one-cut solution
  beyond the critical filling $\alpha\indup{crit}$ is the two-cut
  solution, \protect\figref{fig:onecut32}. Unlike the degenerate
  three-cut solution, \protect\figref{fig:conconebubblecutsc}, this
  configuration locally minimises the energy. The phase transition
  from the one-cut to the two-cut solution occurs when the right
  fluctuation point collides with the cusp of the closed loop cut.
  Shown here is the regularised version of this process, where the
  fluctuation points carry a small but finite number of excitations.}
  \label{fig:conconebubblecutsphys}
\end{figure}%

Concerning quantum stability we find that an instability does not
actually put a threat to the quantisation of some classical solution;
the instability is rather caused by an unsuitable way of looking at
the classical solution. This can be observed conveniently by means of
the one-cut solution (\figref{fig:conconebubblecuts}) Here there are
three main classes of cut configurations (which serve as subclasses
for the two-cut solution):

For small filling, there is just a simple cut, see
\figref{fig:conconebubblecutsa}, in which case the solution is
perfectly stable. For intermediate filling the naive one-cut solution
violates the stability criterion but without any observable flaws like
unstable fluctuation modes. In this case, the physical cut
configuration develops a condensate and a loop with a cusp, see
\figref{fig:conconebubblecutsb}, which achieves to bypass the
stability criterion; it is nevertheless semiclassically equivalent to
the one-cut solution. For large filling, the naive one-cut solution
acquires an unstable fluctuation mode which appears to render the
solution unphysical. The physical cut configuration now has a
condensate and a loop with two cusps, see
\figref{fig:conconebubblecutsc}, which gives a clue of how to
understand the instability:

The latter two dressed configurations should be interpreted as
singular limits of two-cut or three-cut solutions, respectively, see
\figref{fig:conconebubblecutsreg}. In the naive one-cut configuration,
a fluctuation would break apart the loop at a cusp by a tiny amount.
This is fine when the cusp is on the real axis (for intermediate
filling), but away from the real axis (for large filling) it leads to
complex momenta and thus an instability. Conversely, fluctuations
of the regularised three-cut configuration would merely change the
filling of either the two-cut complex or the third cut; in either case
the configuration remains symmetrical with respect to the real axis,
and there is no physical instability. Therefore the classical one-cut
state with large filling does have a corresponding quantum state. This
state however does not minimise the energy locally in accordance with
the apparent instability. The energy is lowered by taking away Bethe
roots from the third cut and moving them onto the two-cut complex. At
the end of this process the third cut has evaporated and only the
two-cut complex remains which is a genuine two-cut solution
(\figref{fig:conconebubblecutsphysc,fig:onecut32}).

This leads to the conclusion that the one-cut solution undergoes a
phase transition at the point which separates intermediate and large
filling: When the one-cut solution is analytically continued from
small to large filling one obtains a degenerate three-cut solution.
However in the vicinity of the latter there exists a two-cut solution
with slightly smaller energy. It is useful to consider this solution
the proper physical continuation of the one-cut curve. We find that at
the phase transition point the second derivative of the energy is
discontinuous. An exact expression for the discontinuity has been
found on the basis of numerical data; an analytic confirmation of its
form would be desirable. Technically, the phase transition turns out
to be equivalent to the one observed in 2D QCD on $S^2$
\cite{Douglas:1993ii} or equivalently a particular matrix model. In
\appref{sec:DKtrans} we summarise the relevant functions and
illustrate the phase transition for that model. Note that the apparent
instability and the phase transition are merely artefacts of the
thermodynamic limit in accordance with general results of statistical
physics; they are resolved naturally at finite size, i.e.\ by the
quantum theory: At finite $L$ there exists no precise notion of
continuous cuts and the distribution of Bethe roots onto these cuts is
often ambiguous. Thus two cuts never actually end on the very same
point which was an essential assumption for the instability and the
phase transition. In particular we show that numerically exact
solutions with finite $L$ are smoothly interpolating between two phases
and there is no real instability when the condensate forms.

\medskip

We hope to have presented a fairly complete picture of the Heisenberg
ferromagnet in the thermodynamic limit, its quantum fluctuations and
the issue of stability. However there are several further questions
which may be addressed in future work. One of them would be how
precisely the third cut interacts with the two-cut complex in
solutions of the type shown in
\figref{fig:conconebubblecutsc,fig:conconebubblecutsregc}. An
investigation of this issue would require an explicit expression for
the (hyperelliptic) three-cut quasi-momentum. Further, it would be
interesting to construct the general elliptic solution to the
Landau-Lifshitz model which includes
the solutions obtained in \cite{Kruczenski:2003gt,Dimov:2004qv}.
It should be equivalent to the two-cut solution
presented here and the circular solution \eqref{eq:ll.class.sol}
should transform into the elliptic solution in the same way the
two-cut solutions emerges from the one-cut solution at large fillings.
Another question would be how the thermodynamic solutions of the
Heisenberg magnet behave under the influence of a magnetic field and
how this would change the picture of the phase space.

Concerning the AdS/CFT correspondence, a generalisation of the
results to the $\grp{SU}(2)$ principal chiral model describing the
dual string theory on $\Reals\times\mathrm{S}^3$ would be desirable.
As discussed in \cite{Kazakov:2004qf},
the solution to this model is given by a spectral curve
that reduces to the Heisenberg spectral curve in a certain limit.
Because of their strong structural similarity,
we expect that stability and loop-formation
work in qualitatively the same fashion.
The phase space of general two-cut solutions
(or elliptic string states)
is likely to be similar as well,
but new features may appear due to the
more elaborate singularity structure on the curve.
Therefore it is worth obtaining and discussing the general
two-cut curve for the principal chiral model.
Some special cases of two-cut solutions have appeared 
in \cite{Beisert:2004hm,Vicedo:2007rp}.
Analogously one could find the corresponding elliptic
solution of the equations of motion explicitly,
e.g.\ using the results of \cite{Dorey:2006zj,Dorey:2006mx}.
This solution may be related to the solutions studied in 
\cite{Arutyunov:2003uj,Arutyunov:2003za}.

Finally, it remains an open question whether a similar study of the
thermodynamic phase space can be carried out for spin chains which have
different symmetry groups, quantum-deformed symmetry, or for which the
spins transform in different representations. For example, studying the
spectrum for symmetry groups of higher rank requires working with
spectral curves with more than two sheets, which complicates the
analysis substantially. Moreover for more general symmetry groups the
algebraic curves are quite sensitive to different boundary conditions
of the spin chain \cite{Gromov:2007ky}, thus making the analysis even
more complicated (and interesting).

\subsection*{Acknowledgements}

We would like to thank
V.\ Bazhanov,
V.\ Kazakov,
M.\ Staudacher and
P.\ Vieira
for interesting discussions.

The work of N.G.\ was partially supported by
a French Government PhD fellowship,
by RSGSS-1124.2003.2
and by RFFI project grant 06-02-16786.
N.G.\ was also partly supported by ANR grant
INT-AdS/CFT (contract ANR36ADSCSTZ).
N.G.\ thanks the AEI Potsdam, where part
of this work was done, for the hospitality during the visit.


\appendix

\section{The Landau--Lifshitz Model}
\label{sec:ll-model}

The long-wavelength limit of the Heisenberg
ferromagnet is equivalent to the Landau--Lifshitz model on
$\mathrm{S}^2$ \cite{Fradkin:1991aa,Kruczenski:2003gt},
see also \cite{Tseytlin:2004xa} for a review 
of Landau--Lifshitz models in connection with the AdS/CFT correspondence.
Here we shall discuss the simplest circular solution
and its fluctuation spectrum which corresponds
to the one-cut solutions discussed in \secref{sec:onecut}.

\subsection{Definitions}

The Landau--Lifshitz model on $S^2$
is a classical two-dimensional sigma model with two fields
$\theta(\sigma,\tau)$ and $\phi(\sigma,\tau)$
which are the standard angular coordinates on $S^2$.
We identify the site $k$
of the corresponding Heisenberg chain
with the coordinate $\sigma=2\pi k/L$
of the sigma model.
The expectation value of the spin projection $S_z$
at that site corresponds to
the value $\half\cos\theta$ at $\sigma$.

In analogy to the closed spin chain
we take the world sheet to be closed,
i.e.\ the coordinate $\sigma$ is
identified with $\sigma+2\pi$:
Consequently, the embedding coordinates $\theta$ and $\phi$
(more precisely the equivalence classes describing points on $S^2$)
must be $2\pi$-periodic in $\sigma$. This leads to the
following periodicity conditions
\[
\theta(\sigma+2\pi,\tau) \in \pm \theta(\sigma,\tau)+2\pi\Integers \,, \qquad
\phi(\sigma+2\pi,\tau) \in \phi(\sigma,\tau)+2\pi\Integers+\half \pi(1\mp 1).
\]

The Lagrangian reads
\[\label{eq:LLLag}
\mathcal{L} = -\frac{L}{4\pi}\,\dot\phi \cos\theta
-\frac{\pi}{2L} \bigbrk{\theta^{\prime\,2} +\phi^{\prime\,2}\sin^2\theta },
\]
where the constants $L/4\pi$ and $\pi/2L$
are adjusted to match with our conventions for the Heisenberg chain.
In fact the first term in the Lagrangian
originates from a topological Wess--Zumino term in three dimensions
and for a fully consistent treatment
one should make the replacement
\[\label{eq:LLWZ}
\dot\phi \cos\theta
\to
\int_0^1\lrbrk{\frac{\ud\phi}{\ud\tau}\, \frac{\ud\cos\theta}{\ud r}
-
\frac{\ud\phi}{\ud r}\, \frac{\ud\cos\theta}{\ud\tau} }\ud r.
\]

The equations of motion following from the Lagrangian are
\begin{align}
  \dot\theta &= \left( \frac{2\pi}{L} \right)^2
        \bigbrk{2\theta'\phi'\cos\theta +\phi''\sin\theta } , \nln
  \dot\phi &= \left( \frac{2\pi}{L} \right)^2
        \bigbrk{\phi^{\prime\,2}\cos\theta -\theta''\csc\theta } .
\end{align}
Taking the derivative of $\mathcal{L}$ with respect to $\dot\theta$
and $\dot\phi$ yields the conjugate momentum densities
\[
\pi_\theta = 0 \,, \qquad
\pi_\phi = -\frac{L}{4\pi} \cos\theta .
\]

The Lagrangian is homogeneous and therefore the total energy and total momentum
are conserved.
The rescaled total energy and a suitably shifted total momentum
read
\<
\tilde{E} \eq EL
= L \int \bigl( \pi_\theta \dot\theta + \pi_\phi \dot\phi -
\mathcal{L} \bigr) \ud\sigma
= \frac{\pi}{2} \int \bigl( \theta^{\prime\,2} +\phi^{\prime\,2}\sin^2\theta
\bigr) \ud\sigma ,
\nln
P \eq
\frac{1}{2}\int \phi' \,\ud\sigma
+
\frac{2\pi}{L}\int \bigbrk{\pi_\theta \theta'+\pi_\phi \phi'}\,\ud\sigma
= \frac{1}{2} \int \bigl( 1-\cos\theta \bigr) \phi'
\,\ud\sigma .
\>
Furthermore the Lagrangian is invariant under shifts
$\phi\mapsto\phi+\varepsilon$ which yields the conserved total spin
$S=L(\half-\alpha)$ with the filling
\[
\alpha =
\frac{1}{2}+\frac{1}{L}
\int \pi_\phi\,\ud\sigma
=
\frac{1}{4\pi} \int \bigl( 1-\cos\theta \bigr) \,\ud
\sigma .
\]
%

\subsection{Circular Solution}

The simplest non-trivial solution of the Landau--Lifshitz model
is the circular string at constant
latitude $\theta_0$ with $n$ windings
\cite{Kazakov:2004qf}
\[
  \theta(\sigma,\tau) = \theta_0, \qquad
  \phi(\sigma,\tau) = n\sigma + (2\pi/L)^2
  n^2 \tau \cos\theta_0.
\label{eq:ll.class.sol}
\]
It has the same charges as the one-cut solution of the Heisenberg
magnet \eqref{eq:onecut_EP}
\[\label{eq:circularcharges}
\alpha = \half \bigl( 1-\cos\theta_0 \bigr) , \qquad
P = 2\pi n\alpha, \qquad
\tilde{E} = 4\pi^2 n^2\alpha (1-\alpha) .
\]

The momentum $P$ and the filling $\alpha$ are defined such that they both
vanish for the vacuum solution at $\theta_0=0$
\[
  \alpha = P = \tilde{E} = 0 .
\]

\subsection{Fluctuations}

We would like to obtain the spectrum of fluctuations around this solution.
For the case of $\alpha=\half$ this problem was solved in \cite{Minahan:2005mx}
which coincides with the ultra-relativistic limit of spinning strings
discussed in \cite{Frolov:2003tu}.
To this end, fluctuation modes with Fourier mode $k$
and frequency $\omega_k$ are added to the background solution
with the following ansatz
\[
\delta\theta =
\frac{a}{\sqrt{L}}\,
\cos\bigbrk{k\sigma+\omega_k \tau+c}
,
\qquad
\delta\phi =
\frac{b}{\sqrt{L}}\,
\sin\bigbrk{k\sigma+\omega_k \tau+d}.
\]
From the equations of motion it follows immediately that $d=c$
as well as
\[
\omega_k=\lrbrk{\frac{2\pi}{L}}^2
\lrbrk{2nk\cos\theta +\frac{b}{a}\,k^2\sin\theta }
=\lrbrk{\frac{2\pi}{L}}^2
\lrbrk{2nk\cos\theta+\frac{a}{b}\,\frac{k^2-n^2\sin^2\theta}{\sin\theta} } .
\]
This fixes the frequency and ratio of amplitudes
\<
\omega_k \eq
\lrbrk{\frac{2\pi}{L}}^2
\lrbrk{
2nk(1-2\alpha)+k^2
\sqrt{1-\frac{4n^2\alpha(1-\alpha)}{k^2}}},
\nln
\frac{b}{a}\eq
\frac{1}{2\sqrt{\alpha(1-\alpha)}}
\sqrt{1-\frac{4n^2\alpha(1-\alpha)}{k^2}}\,.
\>

Now it turns out that the
total energy, total momentum and filling
are not yet completely fixed.
They still depend on the value of $a$
and curiously also on the second variation of the
$\sigma$-independent mode%
\footnote{This feature is
related to the fact that
this mode is already macroscopically
filled in the classical solution.
Essentially $f$ corresponds
to changing $\alpha$ in the classical solution
by a microscopic amount.}
\[
\delta^2\theta=\frac{f}{L}\,.
\]
We shall parametrise our ignorance by two constants $N_k$ and $N_0$
as follows
\<
a\eq\sqrt{2N_k}\bigg/\sqrt[\scriptstyle 4]{1-\frac{4n^2\alpha(1-\alpha)}{k^2}}\,,
\nln
b\eq
\frac{\sqrt{N_k}}{\sqrt{2}\sqrt{\alpha(1-\alpha)}}
\sqrt[\scriptstyle 4]{1-\frac{4n^2\alpha(1-\alpha)}{k^2}}\,,
\nln
f \eq
\frac{1}{\sqrt{\alpha(1-\alpha)}}
\lrbrk{N_0+N_k-N_k\frac{1-2\alpha}{2}
\bigg/\sqrt{1-\frac{4n^2\alpha(1-\alpha)}{k^2}}}.
\>
This leads to the following shift in the observables
\<
\delta^2\alpha \eq
\frac{N_0+N_k}{L}\,,
\nln
\delta^2 P \eq
N_0\,\frac{2\pi n}{L}
+
N_k\,\frac{2\pi (n+k)}{L}\,,
\\\nn
\delta^2\tilde{E} \eq
N_0\frac{4\pi^2 n^2(1-2\alpha)}{L}
+
N_k\frac{4\pi^2}{L}  \lrbrk{
n(n+2k)(1-2\alpha)
+k^2\sqrt{1-\frac{4n^2\alpha(1-\alpha)}{k^2}}
}.
\>
The excitations enumerated by $N_0$ and $N_k$ carry
each one quantum $1/L$ of filling
equivalent to one unit of spin.
They have mode numbers $n$ and $n+k$, respectively.%
\footnote{The shift by $n$ w.r.t.\ the Fourier
mode in $\delta\theta,\delta\phi$ is
induced by the linear dependence of $\phi$ on $\sigma$.}
Their energies can be read off from the above expression.
It is not hard to see that a quantum of mode $n$ essentially just
increases $\alpha$ by $1/L$ in the classical solution,
cf.\ \eqref{eq:circularcharges}.
The charges for mode $n+k$ are
in complete agreement
with the fluctuations in the
ferromagnetic Heisenberg magnet
\eqref{eq:MomEngFluct}.
A careful semiclassical analysis actually shows
that the numbers $N_0$ and $N_k$ must be integers.%
\footnote{Note that our Lagrange function
evaluated on the classical solution is not zero
and does depend on the filling $\alpha$.
Unfortunately, this obscures the correct phase space normalisation
in connection with the shift $\delta^2\theta$.
If one uses the Wess--Zumino term \eqref{eq:LLWZ} instead,
the classical Lagrange function is zero
and the above problem should be absent.}

%


The classical solution \eqref{eq:ll.class.sol} is unstable if some
of the oscillator frequencies $\omega_k$ are complex.
Obviously, this is the case if
\[
2n\sqrt{\alpha(1-\alpha)} > 1 \,.
\]
The transition from stable to unstable classical solutions exactly
happens at the critical density $\alpha\indup{crit}$
\eqref{eq:onecut.alphacrit}, where the one-cut solution of the
Heisenberg magnet transforms into a two-cut solution, as discussed in
\secref{sec:cons_points}.

\section{The Douglas--Kazakov Transition}
\label{sec:DKtrans}

Here we shall review the Douglas--Kazakov transition \cite{Douglas:1993ii}
for QCD on $S^2$ on a technical level.
It mimics the transition between our one-cut and two-cut
solutions, albeit in a somewhat simplified and
more transparent fashion.

\subsection{Definitions}

The description of the partition function in the large-$N$ limit of
QCD on $S^2$ uses a density function $\rho(h)$ with support on the
real axis. It must obey certain integral equations which will not be
needed in detail here.

In order to extract the relevant observables it is convenient
to introduce the continuation $u(h)$ of $\rho(h)$
into the complex plane.
The function $u(h)$ has a branch cut on the real axis
such that the real/principal part equals the density
\[
\rho(h)=\half u(h+i\epsilon)+\half u(h-i\epsilon).
\]
The function $u(h)$ is the analog of our quasi-momentum $p(x)$ and
its derivative $u'(h)$ is an algebraic curve.
Apart from the branch cut, the only
singularity is at $h=\infty$ with
\[\label{eq:DKexp}
u(h)=\frac{1}{2\pi i}\lrbrk{-Ah+\frac{2}{h}+\frac{4}{h^3}\,(F'+\sfrac{1}{24})+\order{1/h^5}}\,.
\]
This expansion also defines the area $A$ and the derivative of the free energy $F'$.
It is clear that these quantities play similar roles as
the total filling $\alpha$, momentum $P$ and energy $\tilde E$.

\begin{figure}\centering
\subfloat[rational regime, $A<A\indup{crit}$]{\includegraphics[height=4.5cm]{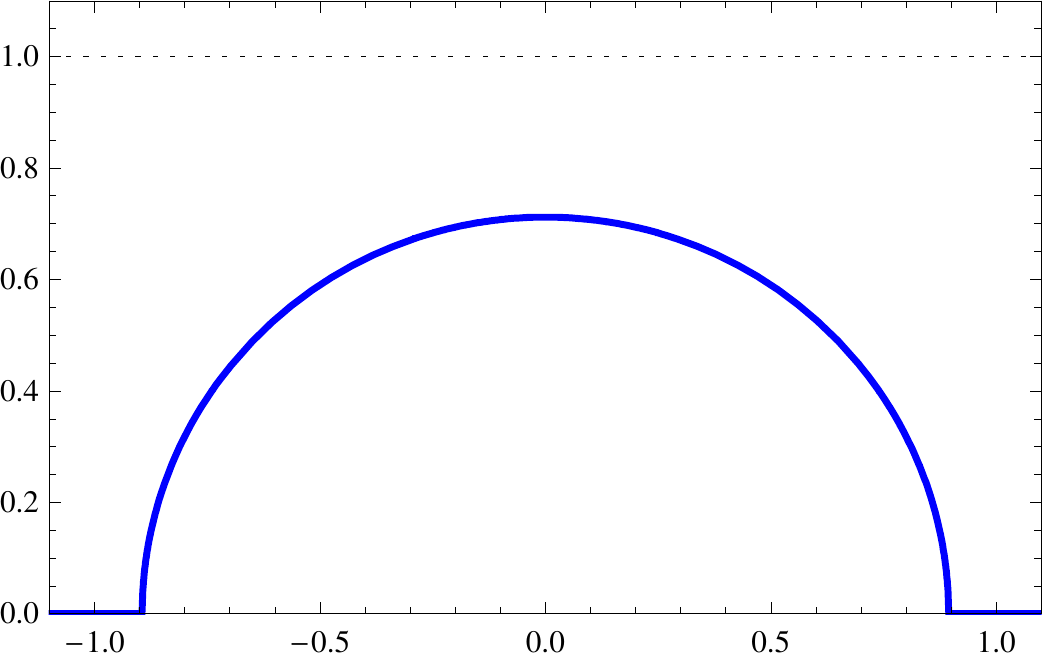}\label{fig:DKDensRat}}
\hfill
\subfloat[elliptic regime, $A>A\indup{crit}$]{\includegraphics[height=4.5cm]{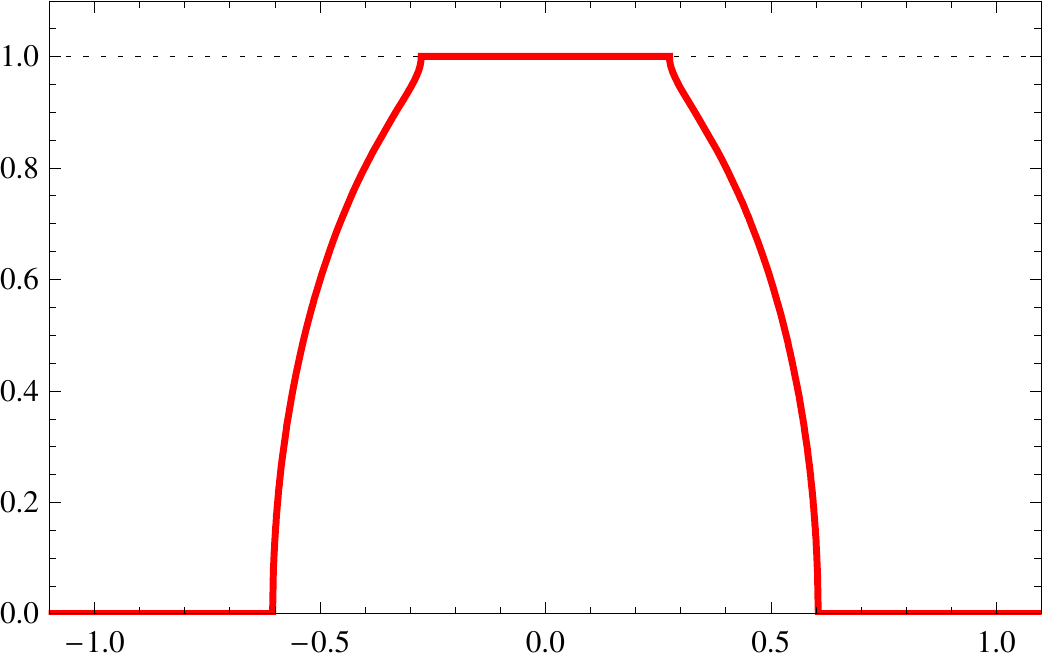}\label{fig:DKDensEll}}
\caption{Density function $\rho(h)$
in the respective regimes.}
\label{fig:DKDens}
\end{figure}

\subsection{Rational Regime}

The simplest solution has one branch cut
at $[-h\indup{max},+h\indup{max}]$.
The function $u(h)$ consequently is of semi-circle form
\[
u(h)=\frac{1}{2\pi}\sqrt{A(4-Ah^2)}\,,
\]
with $h\indup{max}=2/\sqrt{A}$,
see \figref{fig:DKDensRat}.

The resulting free energy from \eqref{eq:DKexp} is simply
\cite{Rusakov:1992uf}
\[
F'=\frac{1}{2A}-\frac{1}{24}\,.
\]

This function is physically correct for $A\approx 0$. However,
it was noticed that it does not have the correct behaviour at $A\to\infty$.
For instance $F'\to-\frac{1}{24}$ but not $F'\to 0$.

\subsection{Elliptic Regime}

On the technical level, the pathology can be attributed
to a density which grows beyond a threshold of
\[\rho\indup{max}=1.\]
A density with $\rho(h)>1$ is unphysical.
Indeed for $A>A\indup{crit}$ with
\[
A\indup{crit}=\pi^2
\]
the maximum density $\rho(0)=\sqrt{A}/\pi$
exceeds the maximum allowed density $\rho(0)>\rho\indup{max}$.

This problem is resolved by inserting a
condensate with constant density $\rho(h)=1$ on the interval $[-b,b]$
\cite{Douglas:1993ii}.
The density then falls off to zero on the intervals
$[-a,-b]$ and $[b,a]$.
This leads to the two-cut solution,
cf.\ \figref{fig:DKDensEll}
\[
u(h)=\frac{2}{\pi ah}\sqrt{(a^2-h^2)(h^2-b^2)}\,\ellPi(b^2/h^2\mathpunct{|}b^2/a^2).
\]

The expansion \eqref{eq:DKexp} at $h\to\infty$ leads to
\<
A\eq 4\ellK(q)\bigbrk{2\ellE(q)-(1-q)\ellK(q)},
\nln
F'\eq\frac{4(1+q)\ellE(q)-(1+3q)(1-q)\ellK(q)}{24\bigbrk{2\ellE(q)-(1-q)\ellK(q)}^3}-\frac{1}{24}\,,
\>
where
\[
a=\frac{1}{2\ellE(q)-(1-q)\ellK(q)}\,,\qquad
b=\frac{\sqrt{q}}{2\ellE(q)-(1-q)\ellK(q)}\,.
\]
As usual this implicitly defines $F'$ as a function of $A$.

Now for $A\to\infty$ one finds
\[
F'=e^{-A/2}+(\half A^2-3A+1)e^{-A}+(\sfrac{1}{2}A^4-\sfrac{16}{3}A^3+14A^2-8A+4)e^{-3A/2}+\order{e^{-2A}}
\]
in agreement with physical expectations
\cite{Gross:1992tu,Gross:1993hu,Gross:1993yt}.

\subsection{Transition}

\begin{figure}\centering
\subfloat[$F'(A)$]{\includegraphics[height=4.5cm]{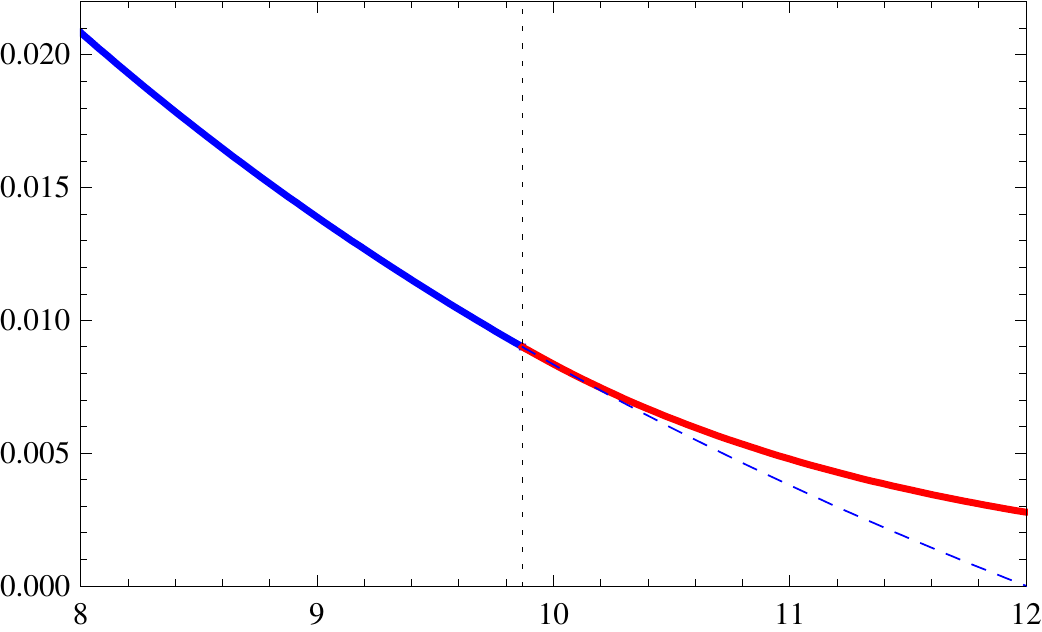}\label{fig:DKTransF}}
\hfill
\subfloat[Comparison of $\rho(h)$]{\includegraphics[height=4.5cm]{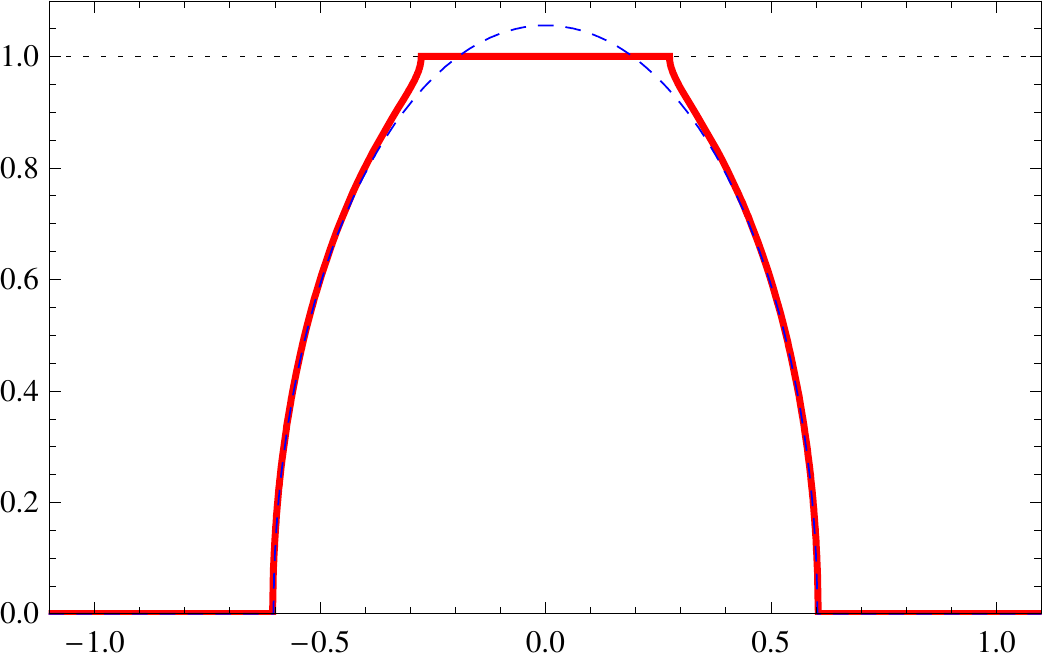}\label{fig:DKComp}}
\caption{Transition of the derivative of the free energy $F'$ at
$A=A\indup{crit}=\pi^2$
and comparison of density functions slightly above $A\indup{crit}$.}
\label{fig:DKTrans}
\end{figure}

We can now compare the two regimes.
The expansion of the derivative of the free energy
around the critical point $A=A\indup{crit}=\pi^2$ yields in the two cases
\<
F'\indup{rat}\eq\frac{1}{2\pi^2}-\frac{1}{24}
-\frac{A-A\indup{crit}}{2\pi^4}
+\frac{(A-A\indup{crit})^2}{2\pi^6}
-\frac{(A-A\indup{crit})^3}{2\pi^8}
+\ldots\,,
\nln
F'\indup{ell}\eq\frac{1}{2\pi^2}-\frac{1}{24}
-\frac{A-A\indup{crit}}{2\pi^4}
+\frac{3(A-A\indup{crit})^2}{2\pi^6}
-\frac{13(A-A\indup{crit})^3}{4\pi^8}
+\ldots\,,
\>
see \figref{fig:DKTransF}.
The deviation is therefore of second order
\[
\oldDelta F'=
+\frac{(A-A\indup{crit})^2}{\pi^6}
-\frac{11(A-A\indup{crit})^3}{4\pi^8}
+\ldots
\]
implying a third-order transition in the free energy $F$.

In fact the function $u(h)$ deviates only quadratically
at a generic point away from the branch points,
see \figref{fig:DKComp}.
It can be expected that the same holds
whenever an elliptic curve degenerated into a rational curve.
This leads to the conclusion that all quantities
which can be read off directly from the curve
(e.g.\ in an expansion)
will deviate at least quadratically.

\section{Integration of the Partial Filling Fractions}
\label{app:int_alphas}

Analytically integrating the partial filling fractions $\alpha_1$,
$\alpha_2$ of the two-cut solution \eqref{eq:p_def} is possible
because they can be expressed as integrals over A-cycles
\eqref{eq:fillingcurve} of the derivative $\tilde{p}_0'(x)$, which is
an algebraic function:
\<
  \alpha_{1,2} \eq - \frac{1}{2\pi i} \oint _{\contour{A}_{1,2}}
  p(x) \, \ud x
\nln\eq
 - \frac{1}{2\pi i} \oint _{\contour{A}_{1,2}} \tilde{p}_0
  \circ \mu (x) + C \, \ud x
\nln\eq
 - \frac{1}{2\pi i} \oint _{\contour{A}_{1,2}} \bigl(
  (\tilde{p}_0 + C) \circ \mu  \bigr) (x) \, \ud x
\nln\eq
 - \frac{1}{2\pi i} \oint _{\mu(\contour{A}_{1,2})} \left(
  \tilde{p}_0 + C \right) (y) (\mu^{-1})' (y) \, \ud y
\nln\eq
 \frac{1}{2\pi i} \oint _{\mu(\contour{A}_{1,2})} \tilde{p}_0'
  (x) \mu^{-1}(x) \, \ud x
\nln\eq
   \frac{1}{2\pi i} \oint _{\mu(\contour{A}_{1,2})}
  \sqrt{\frac{a_0^2 (b_0^2 - x^2)}{b_0^2 (a_0^2 - x^2)}}
\nl
   \cdot \frac{\oldDelta n \left( b_0^2 (sx-u)^2
  \ellE(q) - (a_0^2 s - u x) (b_0^2 s - u x) \ellK(q) \right)}{a_0
  (rx-t) (sx-u) (b_0^2 - x^2)} \, \ud x \,.
\>
In the following, this integral will be evaluated for the case $0 <
a_0 < b_0 < \infty$. The result can then be analytically continued to
the case of complex $a_0$ and $b_0 = \bar{a}_0$. For $0 < a_0 < b_0 <
\infty$ the contours $\contour{A}_{1,2}$ encircle the real intervals
$I_1 := [a_0, b_0]$, $I_2 := [-b_0, -a_0]$. Except for an overall
sign, the integrand $\tilde{p}_0' (x) \mu^{-1} (x)$ has the same values
above and below the intervals $I_{1,2}$. Therefore, the integrals
along the contours $\contour{A}_{1,2}$ can be expressed as integrals
over $I_{1,2}$:
\[
  \oint \limits _{\mu(\contour{A}_{1,2})} \tilde{p}_0' (x) \mu^{-1}(x)
  \, \ud x
  = 2 \lim _{\varepsilon \rightarrow 0} \int \limits _{I_{1,2} -
  i \varepsilon} \tilde{p}_0' (x) \mu^{-1} (x) \, \ud x \,.
\]
Since
\[
  \Im \left( \frac{b_0^2 - (x-i\varepsilon)^2}{a_0^2 - (x -
  i\varepsilon)^2} \right) \xrightarrow {\varepsilon \rightarrow 0}
  \begin{cases}
    < 0 & \text{for $x \in I_1$} \\
    > 0 & \text{for $x \in I_2$}
  \end{cases} \,,
\]
the square root in
the integrand on the right hand side becomes
\[
  \sqrt{\frac{b_0^2 - (x - i \varepsilon)^2}{a_0^2 - (x -
  i\varepsilon)^2}} \xrightarrow {\varepsilon \rightarrow 0}
  \begin{Bmatrix}-i & (x \in I_1) \\ i & (x \in I_2)\end{Bmatrix}
  \sqrt{\frac{b_0^2 - x^2}{x^2 - a_0^2}} \,.
\]
The partial filling fraction integrals can now be written as
\[
  \alpha_{1,2} = \pm \frac{\oldDelta n}{\pi b_0} \int _{I_{1,2}}
  \frac{1}{\sqrt{(x^2 - a_0^2) (b_0^2 - x^2)}} \left( A +
  \frac{A_{t/r}}{x - \frac{t}{r}} + \frac{A_{u/s}}{x - \frac{u}{s}}
  \right) \ud x \,,
\label{eq:fieses_integral}
\]
where $\alpha_1$ has the positive and $\alpha_2$ the negative sign.
The constants are
\begin{align}
  A & = \frac{u^2 \ellK(q) - b_0^2 s^2 \ellE(q)}{r s} \,, \nn\\
  A_{t/r} & = \frac{(a_0^2 r s - t u) (b_0^2 r s - t u) \ellK(q) - b_0^2
  (st-ru)^2 \ellE(q)}{r^2 (st-ru)} \,, \nn\\
  A_{u/s} & = \frac{- (a_0^2 s^2 - u^2) (b_0^2 s^2 - u^2) \ellK(q)}{s^2
  (st-ru)} \,.
\end{align}
The integral \eqref{eq:fieses_integral} can be performed term by term.
The first term is the simplest:
\[
  \int _{I_{1,2}} \frac{1}{\sqrt{(x^2 - a_0^2) (b_0^2 - x^2)}} \,
  \ud x = \frac{1}{b_0} \ellK(q) \,.
\]
The second and third term are of the same form, but the results for
either contour differ by a sign:
\begin{align}
  \int _{I_1} & \frac{1}{\sqrt{(x^2 - a_0^2) (b_0^2 -
  x^2)}} \frac{1}{x-c} \, \ud x
\nln
  & = \int _{a_0} ^{b_0} \frac{x+c}{\sqrt{(x^2 - a_0^2) (b_0^2 - x^2)}
  (x^2 - c^2)} \, \ud x
\nln
  & = \frac{\pi}{2 \sqrt{(a_0^2 - c^2) (b_0^2 - c^2)}} -
  \frac{c}{b_0 (c^2 - a_0^2)} \left( \frac{c^2 - a^2}{c^2} \ellK(q) +
  \frac{a_0^2}{c^2} \elltPi(c) \right) ,
\nln
  \int _{I_2} & \frac{1}{\sqrt{(x^2 - a_0^2) (b_0^2 -
  x^2)}} \frac{1}{x-c} \, \ud x
\nln
  & = - \frac{\pi}{2 \sqrt{(a_0^2 - c^2) (b_0^2 - c^2)}} -
  \frac{c}{b_0 (c^2 - a_0^2)} \left( \frac{c^2 - a^2}{c^2} \ellK(q) +
  \frac{a_0^2}{c^2} \elltPi(c) \right) .
\end{align}
For obtaining these two forms, the integrals
\begin{align}
  \int _{a_0} ^{b_0} \frac{x}{\sqrt{(x^2 - a_0^2) (b_0^2 - x^2)}
  (x^2 - c^2)} \, \ud x
  & = \int _{a_0^2} ^{b_0^2} \frac{1}{2 \sqrt{(t - a_0^2) (b_0^2 - t)}
  (t - c^2)} \, \ud t
\nln
  & = \frac{\pi}{2 \sqrt{(a_0^2 - c^2) (b_0^2 - c^2)}}
\end{align}
and
\[
  \int _{a_0} ^{b_0}  \frac{c}{\sqrt{(x^2 - a_0^2) (b_0^2 - x^2)}
  (x^2 - c^2)} \, \ud x
  = - \frac{c}{b_0 (c^2 - a_0^2)} \left(
  \frac{c^2 - a^2}{c^2} \ellK(q) + \frac{a_0^2}{c^2} \elltPi(c)
  \right)
\]
were used. The latter can be found in \cite{Byrd:1971aa}, formula 217.12.
Putting it all together and writing the occurring square roots in the
form \eqref{eq:rootform} in order to select the branch of the square
root consistently, one arrives at the result \eqref{eq:alpha12}.

\section{Integral of the Two-Cut Quasi-Momentum}
\label{app:int_density}

In order to find the physical cut contours for a given two-cut
quasi-momentum $p(x)$, one can use the fact that expression
\eqref{eq:rho_int} must be real on the physical contours. That
expression contains the integral of the quasi-momentum, which will be
given here. Direct integration yields
\begin{align}
  \int^y & \mu^{-1}(z) \tilde{p}_0'(z) \, \ud z =
\nln
  & - \frac{i a_0 \oldDelta n}{r s (b_0^2 - y^2)} \sqrt{\frac{-1}{a_0^2}}
  \sqrt{\frac{a_0^2 (b_0^2 - y^2)}{b_0^2 (a_0^2 - y^2)}} \sqrt{1 -
  \frac{y^2}{a_0^2}} \sqrt{1 - \frac{y^2}{b_0^2}}
\nln
  & \mspace{100mu} \cdot \left( b_0^2 s^2 \ellE(q) - u^2 \ellK(q)
  \right) \ellF\left( i \arsinh\left(
  {\sqrt{\frac{-1}{a_0^2}}\, y} \right)\mathpunct{\Bigg|} 1-q \right)
\nln
  & + \frac{4 \oldDelta n\, a_0}{b_0^2} \,
  \frac{\displaystyle\sqrt{\frac{a_0 (b_0 +
  y)}{(a_0 - b_0) (y - a_0)}} \sqrt{\frac{a_0 (b_0 - y)}{(a_0 + b_0)
  (a_0 - y)}} }{\displaystyle \sqrt{\frac{(a_0 - b_0) (a_0 + y)}{(a_0 + b_0) (a_0
  - y)}} \sqrt{\frac{a_0^2 (b_0^2 - y^2)}{b_0^2 (a_0^2 - y^2)}}}
\nln
  & \mspace{20mu} \cdot \Biggl( \frac{b_0^2 (s t - r u)^2 \ellE(q) -
  (a_0^2 r s - t u) (b_0^2 r s - t u) \ellK(q)}{(a_0^2 r^2 - t^2) (s
  t - r u)} \ellPi_{r,t}(y)
\nln
  & \mspace{50mu} + \frac{b_0^2 s^2 - u^2}{s t - r u} \ellK(q)
  \ellPi_{s,u}(y)
\nln
  & \mspace{50mu} - \frac{b_0^2 s (s t - r u) \ellE(q) + (a_0^2 r s
  u - t u^2 + a_0 r (u^2 - b_0^2 s^2)) \ellK(q)}{2 a_0 r s (a_0 r -
  t)}
\nln
  & \mspace{140mu} \cdot \ellF\left( \arcsin
  \sqrt{\frac{(a_0 - b_0) (a_0 + y)}{(a_0 + b_0) (a_0 -
  y)}}\,\mathpunct{\Bigg|} \frac{(a_0 + b_0)^2}{(a_0 - b_0)^2} \right) \Biggr)
  \,,
\label{eq:monsterintegral}
\end{align}
where
\[
  \ellPi_{A,B}(x) := \ellPi\left(
  \frac{(a_0 + b_0) (a_0 A - B)}
       {(a_0 -   b_0) (a_0 A + B)}; \arcsin \sqrt{\frac{(a_0 -
  b_0) (a_0 + x)}{(a_0 + b_0) (a_0 - x)}}\, \mathpunct{\Bigg|}
  \frac{(a_0 + b_0)^2}{(a_0 - b_0)^2} \right)
\]
and where $\ellF(z\mathpunct{|}q)$ is the incomplete elliptic integral of the
first kind:
\[
  \ellF(z\mathpunct{|}q) = \int _0 ^{\sin z} \frac{1}{\sqrt{(1-t^2)(1-qt^2)}} \,\ud t
  \,.
\]
Using the integral \eqref{eq:monsterintegral}, one can obtain a closed
expression for the integral $\Lambda(x)$ of the two-cut quasi-momentum
$p(x)$ \eqref{eq:p_def}, which can be expressed as
\<
  \Lambda(x) \eq \int^x p(y)\,\ud y
  = xp(x) -\int^x yp'(y)\,\ud y
\nln
  \eq xp(x) -\int^x y\tilde{p}_0'(\mu(y))\mu'(y)\,\ud y
  = xp(x) -\int^{\mu(x)} \mu^{-1}(z)\tilde{p}_0'(z)\,\ud z \,.
\>

\bibliographystyle{nb}
\bibliography{twocut}

\end{document}